\documentclass[%
 reprint,
 amsmath,
 amssymb,
 aps,
 longbibliography,
]{revtex4-1}

\usepackage{graphicx}% Include figure files
\usepackage{dcolumn}% Align table columns on decimal point
\usepackage{bm}% bold math
\usepackage{hyperref}
\usepackage{braket}
\newcommand*{\affmark}[1][*]{\textsuperscript{#1}}

\begin{document}

%\preprint{APS/123-QED}

\title{Transmission lines and resonators based on quantum Hall plasmonics: \\
electromagnetic field, attenuation and coupling to qubits}

\author{S. Bosco\affmark[1,3]}
\email {bosco@physik.rwth-aachen.de}
\author{D. P. DiVincenzo\affmark[1,2,3]}
 \email{d.divincenzo@fz-juelich.de}
\affiliation{
\affmark[1]Institute for Quantum Information, RWTH Aachen University,                                
  D-52056
  Aachen,                              
  Germany
}

\affiliation{
  \affmark[2]Peter Gr\"{u}nberg Institute, Theoretical Nanoelectronics,
    Forschungszentrum J\"{u}lich,
  D-52425
  J\"{u}lich,
  Germany
}

\affiliation{
\affmark[3]J\"{u}lich-Aachen Research Alliance (JARA),
    Fundamentals of Future Information Technologies,
  D-52425
  J\"{u}lich,
  Germany
}

\date{\today}

\begin{abstract}
%Quantum Hall effect edge states  have some characteristic features that makes them be useful to measure and control solid state qubits.
%For example, their high voltage to current ratio and their dissipationless nature can be exploited to manufacture low loss microwave transmission lines and resonators with a characteristic impedance of the order of the quantum of resistance.
%The high value of impedance guarantees a high voltage per photon and for this reason high impedance resonators can prove useful to obtain larger values of coupling to qubits with a small charge dipole, e.g. spin qubits.
%
Quantum Hall edge states  have some characteristic features that can prove useful to measure and control solid state qubits.
For example, their high voltage to current ratio and their dissipationless nature can be exploited to manufacture low-loss microwave transmission lines and resonators with a characteristic impedance of the order of the quantum of resistance $h/e^2\sim 25\mathrm{k\Omega}$.
The high value of the impedance guarantees that the voltage per photon is high and for this reason high impedance resonators can be exploited to obtain larger values of coupling to systems with a small charge dipole, e.g. spin qubits.
In this paper, we provide a microscopic analysis of the physics of quantum Hall effect devices capacitively coupled to external electrodes. The electrical current  in these devices is carried by edge magnetoplasmonic excitations and by using a semiclassical model, valid for a wide range of quantum Hall materials, we discuss the spatial profile of the electromagnetic field in a variety of   situations of interest.
Also, we perform a numerical analysis to estimate the lifetime of these excitations and, from the numerics, we extrapolate a simple fitting formula which quantifies the $Q$ factor in quantum Hall resonators.
We then explore the possibility of reaching  the strong photon-qubit coupling regime, where the strength of the interaction is higher than the losses in the system.
We compute the Coulomb coupling strength between the edge magnetoplasmons and singlet-triplet qubits, and we obtain values of  the coupling parameter of the order $100\mathrm{MHz}$; comparing these values to the estimated  attenuation in the resonator, we find that for realistic qubit designs the coupling can indeed be strong.
% Although our analysis is restricted to singlet-triplet qubits, we believe that the qualitative features discussed here can be extended to a wider class of qubits.
%
%We analyze the possibility of implementing high impedance transmission lines and resonators by exploiting edge states in a quantum Hall system.\\
%Objectives: \\
%1) Analysis of the electromagnetic fields.\\
%2) Quantitatively analyze the coupling to exchange Qbits, focusing on charge coupling to singlet-triplet states.  \\
%3) Semi-quantitative analysis of dissipation. \\
%3) Comparison with Quantum Anomalous Hall effect and Topological insulators. (here or separate paper only about plasmons there? )\\

\end{abstract}

\pacs{Valid PACS appear here}% PACS, the Physics and Astronomy
                             % Classification Scheme.
%Use showkeys class option if keyword
                              %display desired
\maketitle

%\tableofcontents

\section{\label{sec:intro}Introduction}

Since its first discovery \cite{QHEVK}, the quantum Hall (QH) effect has captured the attention of researchers because of its fascinating physics and for its possible real-world applications \cite{QuantumHallGirvin,quantumHallReviewVK}.
A key feature which makes the QH effect so special is that over a wide range of magnetic field values and electronic densities, the bulk of the 2-dimensional material is insulating, but a net electrical current can still flow.
This current is carried by extended states localized at the edge of the sample, whose existence is guaranteed by a topological argument valid as long as the bulk has a mobility gap \cite{TKNN,LaughlinTheory}.

These edge states have several interesting properties, which make them appealing in different branches of applied science.
In particular, each of these states provides a dissipationless conduction channel in DC with a quantized value of conductance $e^2/h$; this quantity can be measured with an extremely high precision (about a part per billion) and for this reason is now used in metrology to define the electrical resistance standard \cite{metrologyQHE}.
Another intriguing feature of these states is their chirality. 
In the context of quantum computing, the chiral and lossless nature of these edge states was invoked to propose them as a candidate for  one-way processing of quantum information \cite{Stace1}.
A different possibility is to exploit the unidirectional motion of the QH states to manufacture passive low loss non-reciprocal devices such as gyrators and circulators \cite{Viola-DiVincenzo,Wick-patent,Placke,Bosco,Reilly}, that are broadly used for manipulation of qubits and noise reduction.
The advantage of using the QH effect compared to other passive implementations of non-reciprocal devices \cite{Stace2,Koch-Girvin} is that QH effect devices provide better scalability performances \cite{Viola-DiVincenzo} and they are naturally compatible with externally applied magnetic fields, which makes them appealing for semiconductor qubits.\\

%In the context of quantum computation, the dissipationless nature of these states was invoked to propose QH  as a candidate to process quantum information \cite{Stace1}.
%
%For example, in metrology, the quantized value of the off-diagonal resistance $\rho_q=h/e^2$ can be measured with an extremely high precision (about of a part per billion) that the QH effect is  now used as the electrical resistance standard.

%In the context of quantum information processing, there have been proposals to exploit the interesting physics of the quantum Hall effect  
%The states localized at the edges of Quantum Hall (QH) systems have several interesting properties that can prove useful to process quantum information \cite{Stace1}. 

%For example, their chirality can be exploited to implement scalable non-reciprocal devices such as gyrators and circulators \cite{Viola-DiVincenzo,Placke,Reilly} that are broadly used  for manipulation of qubits and noise reduction. Other passive implementations are also possible \cite{Stace2}.

Materials in the QH regime have another interesting property, which was sometimes overlooked, that is they exhibit a large voltage drop between opposite edges when a low current is applied.
This high voltage to current ratio is related to the large value of the quantum of resistance, $h/e^2\sim 25\mathrm{k\Omega}$; for this reason, it was  pointed out that the QH effect can be exploited to manufacture low-loss transmission lines and resonators with a high characteristic impedance \cite{HITLpt1}.
The characteristic impedance of these devices was estimated to be proportional to the resisitance quantum, and so orders of magnitude higher than the typical value $\sim 50\mathrm{\Omega}$ of microwave circuits \cite{Pozar}.
There has been a growing interest in high impedance transmission lines in the quantum information community \cite{Scarlino,Landig,Doherty2,Doherty1,Benito2} and different implementations have been proposed \cite{Manucharyan, Masluk, Annunziata, Santavicca, Niepce, Hagmann, Portier,Burke,Prober2}.
In fact, the excitations in devices with a large characteristic impedance have a high electric field: this property can enhance the electrostatic coupling between the photon and the qubit.
This enhancement is particularly attractive  for semiconductor based quantum computing, where the charge dipole of the qubits  can be low, and it can be exploited  for the challenging task of reaching the strong coupling regime, where the photon-qubit interaction strength is higher than the losses in the system \cite{Scarlino,Landig,Doherty2,Doherty1,Benito3,Benito-exp,Petta}. \\

In this paper, we focus mostly on this aspect and we analyze QH effect transmission lines and resonators.
Our goal is twofold. 
On one hand, we provide a microscopic analysis of the electromagnetic field in these devices; on the other hand, we estimate the strength of the Coulomb interactions of the QH edge states with semiconductor spin qubits and we discuss the possibility of achieving the strong coupling regime.

We restrict our analysis to QH devices that are capacitively coupled to external electrodes \cite{Viola-DiVincenzo,Wick-patent}: this coupling scheme allows to manufacture low-loss devices working in the microwave domain, in contrast to Ohmic coupling, which always causes a high intrinsic contact resistance, degrading the performance \cite{Wick,Girvin}.
The electrical current flowing in capacitively coupled devices is carried by low energy and long wavelength plasmonic excitations localized at the edge of the QH material; these excitations are usually called edge magnetoplasmons (EMPs).
The physics of EMPs has been studied in depth in a variety of different cases  \cite{Volkov,JohnsonVignale,Glazman,Glazman2,Rudner,Mahoney,Glattli,Circuit_EMP,MacDonald_plasmon,QEMP,Mikhailov,Thouless_plasmon}. We use here a semiclassical model that captures the main features of these excitations and we adapt it to describe actual devices, such as the ones in  \cite{HITLpt1}.
In particular, we study in detail the electromagnetic field propagating in these devices, with a particular focus on the effect of the  metal electrodes and of the externally applied AC voltage sources. We also consider the effect of the Coulomb drag between EMPs propagating at different edges of a nanowire.
With this analysis, we are able to justify the model of EMP propagation used in  \cite{HITLpt1}, and to quantify its phenomenological parameters.

Although our focus here is only on 2-dimensional electron gasses in the integer QH regime, i.e. where the mobility gap in the bulk is opened by the application of a quantizing perpendicular magnetic field, with a few straightforward modifications, the results presented in this paper can be extended to a wider range of QH materials, including graphene and quantum anomalous Hall materials.

Also, to gain insight into the possibility of achieving strong photon-qubit coupling, we extend the EMP model to capture the dissipation in a QH resonator due to a finite real-valued bulk conductivity. 
By  fitting our numerical results to a simple expression  inspired by  \cite{Volkov}, we provide an analytic  formula to quantify the quality ($Q$) factor in QH resonators: we find  $Q\sim 10^3$ for commonly measured values of diagonal conductivity in the integer QH effect \cite{Stromer_Res,Briggs_Res} and in state-of-the-art anomalous QH materials \cite{QAH-exp,QAH-exp2}.\\

We then direct our attention to the electrostatic interactions between EMPs and qubits; our analysis is restricted for simplicity to singlet-triplet (ST) qubits \cite{Levy}.
We examine two possible ways of coupling the qubit to the QH resonator, namely via the coupling to the gradient of the electric field of the resonator \cite{Doherty2} and via the coupling to the  electric field of the resonator, averaged over the qubit area, which has to be mediated by an externally applied electric field \cite{Doherty1}.
We find that the effective interaction Hamiltonian is longitudinal  \cite{Susanne1,Susanne2,Blais_longit}, and that the  strength of the interaction term obtained for the two mechanisms is comparable and can be of the order $100\mathrm{MHz}$ for realistic qubit designs. 
Interestingly, the tunability of the second coupling mechanism via an external electric field can be used to switch on and off the photon-qubit interaction, potentially allowing for on demand control of the individual coupling terms if several qubits are coupled to the same resonator.

Using our estimation of the $Q$ factor of the resonator, which we believe is the limiting attenuation factor in these systems, we find that the ratio between the photon-qubit interaction strength and the inverse lifetime of the EMPs can be higher than one.
In particular, for realistic qubit designs, we find that this ratio can be higher than $\sim 30$; this value is at least an order of magnitude larger than has been measured in recent experiments where the strong coupling regime was reached  \cite{Scarlino,Landig,Benito-exp}.

Although our analysis is restricted to a single type of qubit, we believe that our conclusions can be extended to a wider class of semiconductor spin qubits, such as single electron qubits in a magnetic field gradient and three-electron spin qubits \cite{3electronsQ}.\\

The paper is organized as follows.
In Sec. \ref{sec:EMP}, we discuss the semiclassical model of the EMPs.
In Subsec. \ref{sec:EMP-half-plane}, we introduce a simple approximation scheme, which allows to find a solution for the electromagnetic field that is accurate sufficiently far from the edge of QH material; we then use this solution to describe the physics of capacititively coupled QH effect devices and to justify the treatment used in \cite{HITLpt1}. 
In Subsec. \ref{sec:field_distribution}, we present a more detailed calculation which captures also the behavior of the electromagnetic field near the edge. In Subsec. \ref{sec:dissipation-field}, we discuss the corrections to our model due to dissipation and we quantify the $Q$ factor in QH resonators.
In Sec. \ref{sec:coupling}, we analyze the electrostatic coupling between the resonator and a ST qubit. We compute the susceptibility of the qubit to an electric field and to its gradient and we use these results to obtain simple approximate formulas that capture the dependence of the coupling strength on the qubit design parameters. The range of validity of these formulas is examined by comparing them to a more rigorous calculation based on the explicit computation of the Hartree interaction integral. We conclude the paper by discussing in Sec. \ref{sec:strong-coupl} the possibility of reaching the strong photon-qubit coupling limit.
%
%
%In semi-classical model and readapt it to describe the physics of capacitively coupled QH devices. justify the phenomenological model introduced in high-impedance transmission lines introduced in Ref. \cite{HITLpt1}.  
%To do so, we begin by introducing a simple approximation scheme which captures the behavior of the field far from the edge of the quantum Hall material. We then compare with a more detailed calculation which describe near field effects.
%We discuss also dissipation  and quantum corrections.
%In Sec \ref{sec:coupling}, we quantify the Coulomb coupling to singlet-triplet qubits and we discuss the possibility of achieving strong photon-qubit coupling.

\section{\label{sec:EMP}Edge magnetoplasmons}

%Our main goal here is to justify Eq.(\ref{eq:motion-emp-local-velocity-app}) and to estimate the EMP velocities.
%We will also comment on the spatial distribution of the potential in different cases of interest and on the lifetime of the excitations. 

%\subsection{\label{sec:classical-model-EMP} Classical model}
%In this paper, we focus only on quantum Hall (QH) effect devices that are capacitively coupled to external electrodes \cite{Viola-DiVincenzo,Wick-patent}: this coupling scheme allows to manufacture low loss devices, in contrast to Ohmic coupling, which always causes a high intrisic contact resistance degrading the performances \cite{Wick,Girvin}.
%
%The current flowing in these kind of the devices is carried by low energy and long wavelength plasmonic excitations localized at the edge of the QH material, which are usually referred to as edge magnetoplasmons (EMPs).

The main features of the dynamics of the EMPs are captured by a model based on the following system of partial differential equations in the frequency domain \cite{Volkov, JohnsonVignale,Glazman}
\begin{subequations}
\label{eq:classical-system-eq}
\begin{align}
    \label{eq:continuity-eq}
    i\omega\rho (\textbf{r},\omega )&=-\nabla_{\textbf{r}}\pmb{.}\textbf{j}(\textbf{r},\omega ),\\
    \label{eq:inverted-poisson-eq}
    V(\textbf{r},z,\omega )&=V_a(\textbf{r},z,\omega)+\int d\textbf{r}' G\left(\textbf{r},\textbf{r}',z\right)\rho \left(\textbf{r}',\omega \right),\\
    \label{eq:conducivity-definition}
    \textbf{j}(\textbf{r},\omega )&=-\underline{\sigma }(\textbf{r},\omega )\cdot\nabla_{\textbf{r}} V(\textbf{r},0,\omega ).
\end{align}
\end{subequations}

These equations relate the excess charge density $\rho$, the screened potential $V$ \cite{Vignale} and the current density $\textbf{j}$ in the $(x,y)$ plane; here, $\textbf{r}=(x,y)$ and $\nabla_{\textbf{r}}$ is the 2-dimensional nabla operator in the $(x,y)$ plane.
The continuity equation (\ref{eq:continuity-eq})  imposes the conservation of charge in the QH material. The screened potential $V$  is modeled  by the inverted Poisson equation (\ref{eq:inverted-poisson-eq}) with appropriate boundary conditions. It accounts for the external driving voltage applied at the metal electrodes and for the self-consistent rearrangement of charge due to Coulomb interactions. In particular, the Coulomb interactions are captured by the electrostatic Green's function $G$, obtained mathematically by grounding all the driving electrodes, and the effect of the external potentials is captured by the function $V_a$, which is the particular solution of the Laplace equation required to  fix the potential of the electrodes to the appropriate time-dependent value; $V_a$ also accounts for fringing fields.
The peculiar physics of the Hall materials enters in this model through the microscopic Ohm's law (\ref{eq:conducivity-definition}) via a non-reciprocal conductivity tensor
\begin{equation}
\label{eq:conducitivy-convention}
\underline{\sigma}=\left(\begin{array}{cc}
\sigma_{xx} & \sigma_{xy} \\
-\sigma_{xy} & \sigma_{xx}
\end{array}
\right).
\end{equation}
Although we focus only on integer QH effect in 2-dimensional electron gasses, the solution presented here can be modified to describe a wide range of Hall responses, such as in graphene and in anomalous QH materials \cite{Glattli,Rudner,Mahoney}.

Note that in Eq. (\ref{eq:conducivity-definition}),  we equated the electric field to the gradient of the scalar potential: this equality holds only in the electro-quasi static approximation \cite{EQS}, where the electric field is assumed to be approximately irrotational, i.e. $\nabla_{\textbf{r},z}\times\textbf{E}\approx 0$ ($\nabla_{\textbf{r},z}$ is the 3-dimensional nabla operator).
%In QH droplets, this approximation is justified by the fact that the characteristic scale of the EMP velocity is reduced with respect to the speed of light (in the medium) by the fine structure constant $\alpha\approx 1/137$ \cite{QEMP}.
This approximation is justified in the low-frequency limit when the electric energy is high compared to the magnetic energy, or, analogously, when the speed of diffusion of the electric charge $v_E\sim |\textbf{E}|/|\textbf{B}|$ is much lower than the speed of diffusion of the electric current $\sim c_S^2/v_E$, with $c_S$ being the speed of light in the medium, see e.g. Sec. 3 of \cite{Poynt_V}.
This condition is usually not met in conventional microwave transmission lines, where $v_E$ and $c_S$ are comparable, but it holds in QH droplets because $v_E\sim c_S\alpha$ \cite{QEMP}, with $\alpha$ being the fine structure constant $\alpha\approx 1/137$.
Additionally, since we restrict our analysis to low frequencies (compared to the bulk mobility gap), we   neglect retardation effects and take the DC limit of the conductivity tensor $\underline{\sigma}(\textbf{r},\omega\rightarrow 0)=\underline{\sigma}(\textbf{r})$.

The model presented so far is purely classical.
To analyze the physics of the edge excitations, we now evaluate the conductivity tensor in the QH limit, i.e. $\sigma_{xx}=0$, and we find the semiclassical relation 
\begin{equation}
\label{eq:classical-rho-V-relation}
 i\omega\rho (\textbf{r},\omega )=-\left(\nabla_{\textbf{r},z}\sigma_{xy}(\textbf{r})\right)\cdot\left(e_z\times \nabla_{\textbf{r},z} V(\textbf{r},0,\omega )\right).
\end{equation}
between the charge density and the screened potential.

%
%When the QH limit is taken, i.e. $\sigma_{xx}=0$, the charge density and the screened potential are related by
%\begin{equation}
%\label{eq:classical-rho-V-relation}
% i\omega\rho (\textbf{r},\omega )=-\left(\nabla_{\textbf{r},z}\sigma_{xy}(\textbf{r})\right).\left(e_z\times \nabla_{\textbf{r},z} V(\textbf{r},0,\omega )\right).
%\end{equation}

We use this equation to study different cases and to analyze the spatial profile of the electromagnetic fields and the attenuation of the EMPs. 
We begin by proposing a simple approximate solution of the equation of motion (\ref{eq:classical-rho-V-relation}) based on the introduction of a phenomenological length $l$, which physically characterizes the width of the EMP charge density.
This approximation gives a good qualitative description of the physics of the problem when $|\textbf{r}|\gg l$ and it can be used to analyze a variety of situations.
In this paper, we refer to the limits $|\textbf{r}|\gg l$ and $|\textbf{r}|\approx l$ respectively as far- and near-field; the definition of far-field limit here differs from the conventional electromagnetic definition, where $|\textbf{r}|$ is compared to the wavelength.
A more rigorous solution of Eq. (\ref{eq:classical-rho-V-relation}) capturing also near-field corrections is provided in Sec. \ref{sec:field_distribution}.
The attenuation of the EMPs caused by a finite diagonal conductivity $\sigma_{xx}$ is discussed in Sec. \ref{sec:dissipation-field}.

\subsection{Far-field analysis \label{sec:EMP-half-plane}}

\subsubsection{EMPs in the half-plane \label{sec:emp-half}}
In this section, we consider a conductivity profile varying abruptly from zero to the bulk value and we model the spatial dependence of the Hall conductivity by a step function constant in the $y$-direction and with support in $x>0$, i.e.
\begin{equation}
\label{eq:cond-heaviside}
\sigma_{xy}(\textbf{r})=\sigma_{xy} \Theta(x).
\end{equation}
Here, $\sigma_{xy}=\nu e^2/h$ is the  QH conductivity and $\nu$ is the filling factor.
For now, we also neglect the effect of nearby metal electrodes and of driving potentials, and we look for self-consistent excitations at the edge of the half-plane, i.e. $V_a=0$.

A closely related problem was solved analytically by Volkov and Mikhailov \cite{Volkov} by using the Wiener-Hopf decomposition.
However, the solution provided there is quite complicated and, most importantly, it crucially relies on the presence of a frequency dependent complex-valued diagonal conductivity $\sigma_{xx}(\omega)$.
Here, we propose instead a simpler approach that still captures the main features of the EMPs in the far-field limit.

Using the conductivity tensor in (\ref{eq:cond-heaviside}), Eq. (\ref{eq:classical-rho-V-relation})  reduces to an integro-differential equation for the charge density. 
%, and since for a straight edge, the interaction kernel preserves translational invariance in $y$-direction,
By Fourier transforming the translational invariant $y$-coordinate and introducing the corresponding momentum $q$, we obtain
\begin{equation}
\label{eq:FT-self-consistent-charge-straight-line}
\omega\rho (x,q,\omega)=2\pi q \sigma_{xy} \delta(x) \int d x' G_0\left(x-x',q,0\right)\rho \left(x',q,\omega\right),
\end{equation}
where the function
\begin{equation}
\label{eq:G-nogates}
G_0(x,q,z)=\frac{1}{4\pi^2\epsilon_S}K_0\left(\left|q \right|\sqrt{x^2+z^2}\right),
\end{equation}
is the Fourier transform of $G_0(\textbf{r},\textbf{r}',z)$ in $y-y'$ and $\epsilon_S$ is the average dielectric constant of the medium; $K_0$ is the modified Bessel function of the second kind.
In this paper, we use the index 0 to label the electrostatic Green's function $G$ obtained in free space, i.e. without including the effect of metal gates.

From Eq. (\ref{eq:FT-self-consistent-charge-straight-line}), it follows that the excess charge density is proportional to $\delta(x)$; this proportionality however leads to an unphysical divergence of the integral kernel, which is related to the well-known electrostatic instability of a 1-dimensional line of charge \cite{Macdonald-Rice}. 
This divergence can be dealt with by including a finite and complex-valued $\sigma_{xx}$ \cite{Volkov}: in this case, the excess charge density spreads into the bulk with a penetration length dependent on $\mathrm{Im}(\sigma_{xx})$ and the Coulomb interactions are regularized.
In this section, however, we focus on another approach to  circumvent this problem, which allows for a simpler solution: we add a phenomenological length $l$, below which the interactions in the $x$ direction are cut-off, i.e. $\lim_{x\rightarrow 0}G_0\left(x,q,0\right)\approx G_0\left(l,q,0\right)$.
With this approximation, the eigenfrequency of the EMP is
\begin{equation}
\label{eq:eigenfreq-classical-cutoff}
\omega\approx q v_0(q),
\end{equation}
with the momentum dependent velocity
\begin{equation}
\label{eq:inter_edge_v}
v_0(q)= 2\pi\sigma_{xy} G_0\left(l,q,0\right)= 2 v_p K_0(\left|q \right| l),
\end{equation}
and with a characteristic velocity
\begin{equation}
\label{vp-scale}
v_p=\frac{\sigma_{xy}}{4\pi\epsilon_S}=\frac{c \alpha}{2\pi \epsilon_S^*}\nu.
\end{equation}
Here, $c$ is the speed of light in vacuum, $\alpha\approx 1/137$ is the fine structure constant and $\epsilon_S^*$ is the dimensionless  dielectric constant of the medium; the definition of $v_p$ differs from the one used in \cite{QEMP} by a factor $\nu$.
Note the presence of a familiar $\log(\left|q \right|)$ divergence for long wavelengths \cite{Volkov,Glazman}.

A more rigorous treatment of Eq. (\ref{eq:classical-rho-V-relation}), not relying on the introduction of an ad-hoc lengthscale to cut-off the Coulomb interactions, is postponed to   Sec. \ref{sec:field_distribution}, where we consider a smoother conductivity profile, varying from zero to the bulk value in a finite length $l'$.
%In this paper, we focus on another approach to circumvent this problem, valid also in the DC quantum Hall limit $\sigma_{xx}=0$, i.e. . 
%In Appendix \ref{sec:field_distribution}, we present a more rigorous treatment of Eq. (\ref{eq:classical-rho-V-relation}), where we include a characteristic length $l'$ over which the conductivity tensor changes from zero to the bulk value.
Including a length $l'$ in the calculations is another well-known procedure to avoid the divergence of $G_0$ and this procedure works also in the DC quantum Hall limit $\sigma_{xx}=0$ \cite{Glazman,Glazman2}. 
In atomically defined edges, $l'$ is proportional to the magnetic length $l_B=\sqrt{\hbar/(eB)}$ and this approach is consistent also with quantum mechanical calculations \cite{QEMP,Thouless_plasmon,MacDonald_plasmon} up to a quantum correction discussed in Appendix \ref{sec:quantum-field}.
We anticipate that the EMP eigenfrequency obtained for a smoother conductivity profile, given in Eq. (\ref{eq:eigen-freq-classical-lw}), coincides with Eq. (\ref{eq:eigenfreq-classical-cutoff}) in the long wavelength limit $l,l'\ll 1/q$ if we consider $l=c_0 l'$, with $c_0$ being a constant of order 1 dependent on the precise spatial profile of the conductivity. For example, for the conductivity profile in Eq. (\ref{eq:conductivity-erf}), we obtain $c_0\approx0.53$.\\
%Comparing to the more rigorous solution of Eq. (\ref{eq:classical-rho-V-relation}) presented in Appendix \ref{sec:field_distribution}, we find that this approach gives a good estimation of the eigenfrequency when the $l$ is identified with the characteristic length over which the conductivity changes from zero to the bulk value.
%far-field limit, where $x\gg l$  $l$ 
%In particular, we find that $l$ is 
%
% to avoid the divergence in a dissipationless model with $\sigma_{xx}=0$, we include a small characteristic length $l'$ over which the conductivity tensor changes from zero to the bulk value; for sharp edges, this characteristic length is of the order of the magnetic length $l_B\equiv \sqrt{\hbar/(eB)}$ up to a quantum correction discussed in Appendix \ref{sec:quantum-field}.
%  $l$ depends on the characteristic length over which the conductivity changes from zero to the bulk value; for a sharp edge $l$ is proportional to the magnetic length $l_B\equiv \sqrt{\hbar/(eB)}$, with a proportionality constant of order 1 which depends on the details of . For example, for a ga of the edge and of for . 

%Also, the phenomenological length $l$ can be modified to  capture the effect of dissipation, as shown in Sec. \ref{sec:dissipation}.
%By including a small and real $\sigma_{xx}$  in the long wavelength limit, Eqs. (\ref{eq:eigenfreq-classical-cutoff}) and  (\ref{eq:inter_edge_v}) remain approximately the same, but $l$ becomes a complex number, whose imaginary part parametrizes the lifetime of the EMP.\\

Using our approximation, the spatial variation of the charge, potential and current density reduce to
\begin{subequations}
\label{eq:CVC}
\begin{align}
    \label{eq:charge}
    \rho(\textbf{r})&\approx \rho_0 \delta(x)e^{i q y}/(2\pi),\\
    \label{eq:voltage}
    V(\textbf{r},z)&\approx \rho_0 e^{i q y}G_0\left(x,q,z\right),\\
    \label{eq:current}
    \textbf{j}(\textbf{r})&\approx \rho_0 \sigma_{xy}  e^{i q y} \Theta(x)\left(\begin{array}{cc}
-i q G_0\left(x,q,0\right) \\
\partial_x G_0\left(x,q,0\right)
\end{array}
\right),
\end{align}
\end{subequations}
with $\rho_0$ being a constant of units charge per meter.
These results are in agreement with the asymptotic far-field limit of the solution of Volkov and Mikhailov \cite{Volkov} and the one presented in Sec. \ref{sec:fields-classical}.

The transverse component of the current density $j_{x}$ is small compared to $j_y$, and the potential and the current density decay into the bulk of the material on a scale $1/q$, which is generally quite long.
This behavior is quite different from conventional conductors, where the skin depth is often negligible, and it is  related to the fact that QH materials are a novel form of insulator, and so the electric field is unscreened in the bulk.
This also implies that even if the excess charge is localized at the edge, the current density is quite broadly distributed in the material, making QH devices quite unique.
Note that the electromagnetic waves traveling in this setup are not TEM modes, but more complicated hybrid TE-TM modes, with a finite component in the direction of propagation. A detailed calculation of the electric and magnetic field valid also in the near-field limit is presented in Sec. \ref{sec:fields-classical}.\\

From a microwave engineering perspective, the complicated structure of the fields means that the choice of the reference potential is not unique, and so the definition of the characteristic impedance $Z_0$ of the device can vary \cite{Pozar}.
For example, the characteristic impedance can be defined from the microwave $S$-parameters \cite{HITLpt1, QEMP, Bosco} by setting it equal to the values of the impedance of the external circuit that minimizes reflection at the electrodes. Using this approach in capacitively coupled QH devices, one obtains  \cite{HITLpt1}
\begin{equation}
\label{eq:characteristic-impedance}
Z_0=\frac{1}{2\sigma_{xy}}.
\end{equation} 
To verify the validity of this approach, we now compare this result to the alternative definition for $Z_0$:
\begin{equation}
P= \frac{1}{2}Z_0 I_c^2,
\end{equation}
which relates the average power flow $P$ to the amplitude of the conduction current $I_c$.

The total conduction current at position $y$ can be found by integrating the current density in the direction of propagation over a circular cross section $\mathcal{C}$ of radius $R \rightarrow\infty$.
The integration leads to
\begin{equation}
\label{eq:cond-current}
I_c(y)\approx \frac{v_0(q)}{2\pi}\rho_0 e^{i q y};
\end{equation}
to avoid the divergence of the integral at $x\rightarrow 0$, we use again the cut-off length $l$ and we restrict the  domain of integration to $[l,\infty)$. 
Note that at the EMP propagation frequency, the conduction current at any point in $y$ is compensated for by a displacement current $I_d(y)\approx -\omega\rho_0 e^{i q y}/(2\pi q)=-I_c(y)$.

%Since the travelling waves in this set up are in general not TEM modes, but more some complicated hybrid TE-TM modes, the choice of reference potential is not unique, hence the definition of the characteristic impedance can vary.
In the electro-quasi static approximation, the power flow in the infinite circular cross section $\mathcal{C}$ is given by (see e.g. Sec. 11 of \cite{Poynt_V})
\begin{equation}
\label{eq:power-flow}
P = \frac{1}{2} \int_{\mathcal{C}} V \left( \textbf{j}^*-i\omega\epsilon_s \textbf{E}^* \right) d\textbf{S} 
   \approx \frac{I_c^2}{4\sigma_{xy}},
\end{equation}
leading to $Z_0=1/(2\sigma_{xy})$, in  agreement with the $S$-parameter definition.
Also, this result coincides in the long wavelength limit with the characteristic impedance computed with the near-field solution,  see Eq. (\ref{eq:power-near}) in Sec. \ref{sec:fields-classical}.
%The characteristic impedance found with this procedure is in perfect agreement with the values shown in Sec. \ref{sec:transmission-line}; thus we can confirm that the impedance is of order of the quantum of resistance.

%For gated devices, defining the impedance in this way becomes increasingly complicated and therefore in these cases an alternative definition of the impedance can be given by considering the microwave scattering parameters and check for what external impedance one gets perfect transmission.

%We now use this EMP model to analyze the behavior of QH devices.

\subsubsection{Quantum Hall effect devices \label{sec:EMP-gated}}

In this section, we present a way to model the response of a QH droplet capacitively coupled to external electrodes.
A phenomenological model of these devices \cite{Viola-DiVincenzo,HITLpt1,Reilly} relies on the chiral equation of motion for the EMP charge density along the edge
\begin{equation}
\label{eq:motion-emp-local-velocity-app}
i\omega\rho(y,\omega )=  \partial_y\left(v(y) \rho(y,\omega)\right)+\sigma_{xy} \partial_y V_a(y,\omega),
\end{equation}
and on the relation between $\rho$ and the current in the $i$th electrode
\begin{equation}
\label{eq:current-ith-electrode}
I_i(\omega)=-i\omega\int_{\mathcal{R}_i}dy\rho(y,\omega ).
\end{equation}

The velocity $v(y)$ and the driving term $V_a(y)$ are both functions of the position along the perimeter of the droplet, parametrized by $y$. For simplicity, $v(y)$ and $V_a(y)$ are often approximated by piecewise functions, and so the EMPs  propagate at a constant velocity in the regions $\mathcal{R}_i$ coupled to the $i$th electrode and are boosted by the applied voltage in a narrow region at the boundary of $\mathcal{R}_i$.\\

\begin{figure}
\includegraphics[width=0.45\textwidth]{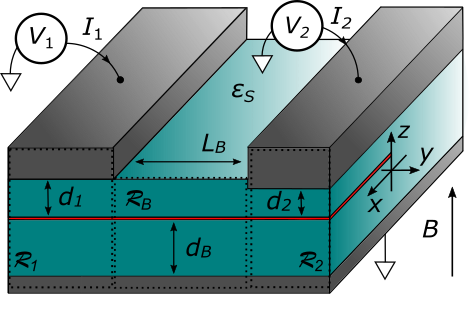}
\caption{\label{fig:3gates-device} Cross-section of a gated quantum Hall device. A QH material (red line) is coupled capacitively to a back gate at distance $d_B$ and to two top gates placed at a distance $d_{1,2}$ and separated by a distance $L_B$ in the $y$-direction. The back gate is grounded while the $i$th top gate is driven by a voltage $V_i$, measured with respect to ground. For simplicity, all the gates extend indefinitely in the $x$-direction, while the QH material occupies only the half-plane $x>0$.
The $y$-direction is divided into the three regions $\mathcal{R}_i$ characterized by the different gating configuration; $\mathcal{R}_{1,2}$  extends to $y\rightarrow \mp \infty$, respectively. The dielectric background is assumed to have a homogeneous and isotropic dielectric constant $\epsilon_S$.
An homogeneous magnetic field $B$ is applied in the $z$-direction.  }
\end{figure}

The main goal of this section is to discuss the validity of this model and to characterize the EMP velocities.
To do so, we analyze the simple configuration shown in Fig. \ref{fig:3gates-device}, which gives us valuable insight into the coupling between the edge excitations of a QH droplet and external electrodes.
We consider a grounded back gate and two top gates respectively at a distance $d_B$, and $d_{1,2}$ from the QH material in the $z$-plane. 
The top gates $1,2$ are placed at position $y<0$ and $y>L_B$, respectively, and they are driven by external time-dependent potentials $V_{1,2}(\omega)$. The discussion here can then be generalized to setups with more electrodes. All the electrodes are assumed to be perfectly conducting.

In this configuration and in the QH regime, the EMP dynamics is captured by Eq. (\ref{eq:classical-rho-V-relation}), but inverting the Poisson equation and finding the screened potential $V$ becomes a challenging task.
In fact, the two top gates break the translational invariance of the system in the $y$-direction, so that $G\left(l,y-y',0\right)\rightarrow G\left(l,y,y',0\right)$, and this causes momentum mixing in the $y$-direction. 
Also, in this situation, the screening potential includes a driving term $V_a$, which guarantees that the value of $V$ at the boundaries matches the time-dependent applied voltage, see Eq. (\ref{eq:inverted-poisson-eq}).

%$V_a$ modifies the self-consistent equation (\ref{eq:FT-self-consistent-charge-straight-line}) by the addition of the term $q \sigma_{xy}\delta(x)V_a(x,q,0,\omega)$  on the right hand side.

%To simplify the problem, we use a local approximation for the Green's function.
%To this aim, we divide the $y$-direction into three different regions characterized by the presence (or absence) of a top electrode.

To obtain an approximate equation of motion for the EMP charge which resembles Eq. (\ref{eq:motion-emp-local-velocity-app}), we divide the $y$-direction into three different regions $\mathcal{R}_i$, with $i=(1,2,B)$, as shown in Fig. \ref{fig:3gates-device}.
The total excess charge density $\rho$ can then be decomposed into a sum of densities $\rho_i$ with support only in $\mathcal{R}_i$, i.e. $\rho(x,y,\omega)=\sum_i \delta(x)\rho_i(y,\omega)$, and the equation of motion in real space reduces to
\begin{multline}
\label{eq:eq-motion-parts}
i\omega\rho_i(y,\omega )=\sigma_{xy} \partial_y \left( V_{a,i}(0,y,0,\omega)+  \vphantom{\sum_j}\right.  \\
\left.\sum_j \int d y' G_{ij}\left(l,y,y',0\right)\rho_j\left(y',\omega \right)\right).
\end{multline}
Here, $G_{ij}\left(l,y,y',0\right)=G\left(l,y\in\mathcal{R}_i,y'\in \mathcal{R}_j ,0\right)$ relates the charge densities of the $i$th and $j$th regions, $V_{a,i}(0,y,0,\omega)=V_{a}(0,y\in\mathcal{R}_i,0,\omega)$, and we introduced again the small cut-off length $l$ required for the integrand to be finite. \\

To simplify the problem, we now use a local approximation for the Green's function $G$, valid for smooth excitations characterized by a wavelength $1/q$ in the $y$-direction satisfying $q d_i\ll 1$ and when $d_i/L_B\ll 1$.
In this local approximation, we keep  in the integral on the right hand side of Eq. (\ref{eq:eq-motion-parts}) only the terms that couple the charge densities in the same region, i.e. $G_{ij}\approx\delta_{ij}G_i$. 

Although an exact computation of $G_{i}$ is still challenging, the limiting behavior of these functions is known.
In particular, far from the boundaries of $\mathcal{R}_i$, $G_i$ can be approximately assumed to be translational invariant, and given by
\begin{equation}
\label{eq:G-function-app}
G_i(l,q,0)=\frac{1}{4\pi^2\epsilon_s}\int dk \frac{e^{i k l}}{\sqrt{k^2+q^2}} f_i\left(\sqrt{k^2+q^2}\right),
\end{equation}
where
\begin{subequations}
\begin{flalign}
f_{1,2}(s)&=\left(\coth\left(s d_{1,2} \right)+\coth\left(s d_B\right)\right)^{-1},\\
f_{B}(s)&=\frac{1}{2}\left(1-e^{-2s d_B}\right).
\end{flalign}
\end{subequations}

In the long wavelength limit, $q l\ll 1$, these results can be used to  further simplify Eq. (\ref{eq:eq-motion-parts}) by approximating
\begin{equation}
\label{eq:estimation-local-G}
G_i(l,y,y',0)\approx 2\pi \delta(y-y') G_i(l,q\rightarrow 0,0).
\end{equation}
%
%\begin{multline}
%\label{eq:G-function-app}
%G_i(l,q,0)=\frac{1}{4\pi^2\epsilon_s}\int dk \frac{e^{i k l}}{\sqrt{k^2+q^2}} \times\\
%\left(\coth\left((d_T)_i\sqrt{k^2+q^2}\right)+\coth\left(d_B\sqrt{k^2+q^2}\right)\right)^{-1},
%\end{multline}
%with $(d_T)_i=(d_1,d_2,\infty)_i$ being the distance of the top gate in $\mathcal{R}_i$.
%In the longwavelength limit, the functions $G_i$ and $V_{a,i}$ can be estimated by using the long wavelength limit $q l\rightarrow 0$ of Eqs. (\ref{eq:G-function-app}) and (\ref{eq:Va_app}), respectively.
Note that the presence of one or more metal electrodes in every region is required to regularize the $\log(\left|q \right|)$ singularity of the EMP velocity \cite{Volkov} and, consequently, to guarantee that $G_i(l,q,0)$ is finite when $q  l\rightarrow 0$.\\

To gain insight into the driving term $V_a$, let us neglect the capacitive cross-talk between the two top electrodes. This approximation holds when $d_i\ll L_B$  and it allows to decouple the effects of the voltages $V_{1,2}(\omega)$ applied to the top gates; the analysis of the capacitive coupling between the electrodes can be done a posteriori, see e.g. \cite{Bosco,Placke,HITLpt1, Reilly}.
In this case, we obtain that well-inside $\mathcal{R}_{1,2}$ the field $V_{a,(1,2)}$ (evaluated at the position of the EMP $x=z=0$) is approximately homogeneous  and  is related to $V_{1,2}(\omega)$ by
\begin{equation}
\label{eq:Va-longwl-full}
V_{a,(1,2)}(0,y,0,\omega)\approx V_{1,2}(\omega) \frac{d_B}{d_B+d_{1,2}}.
\end{equation}

\begin{figure}
\includegraphics[width=0.45\textwidth]{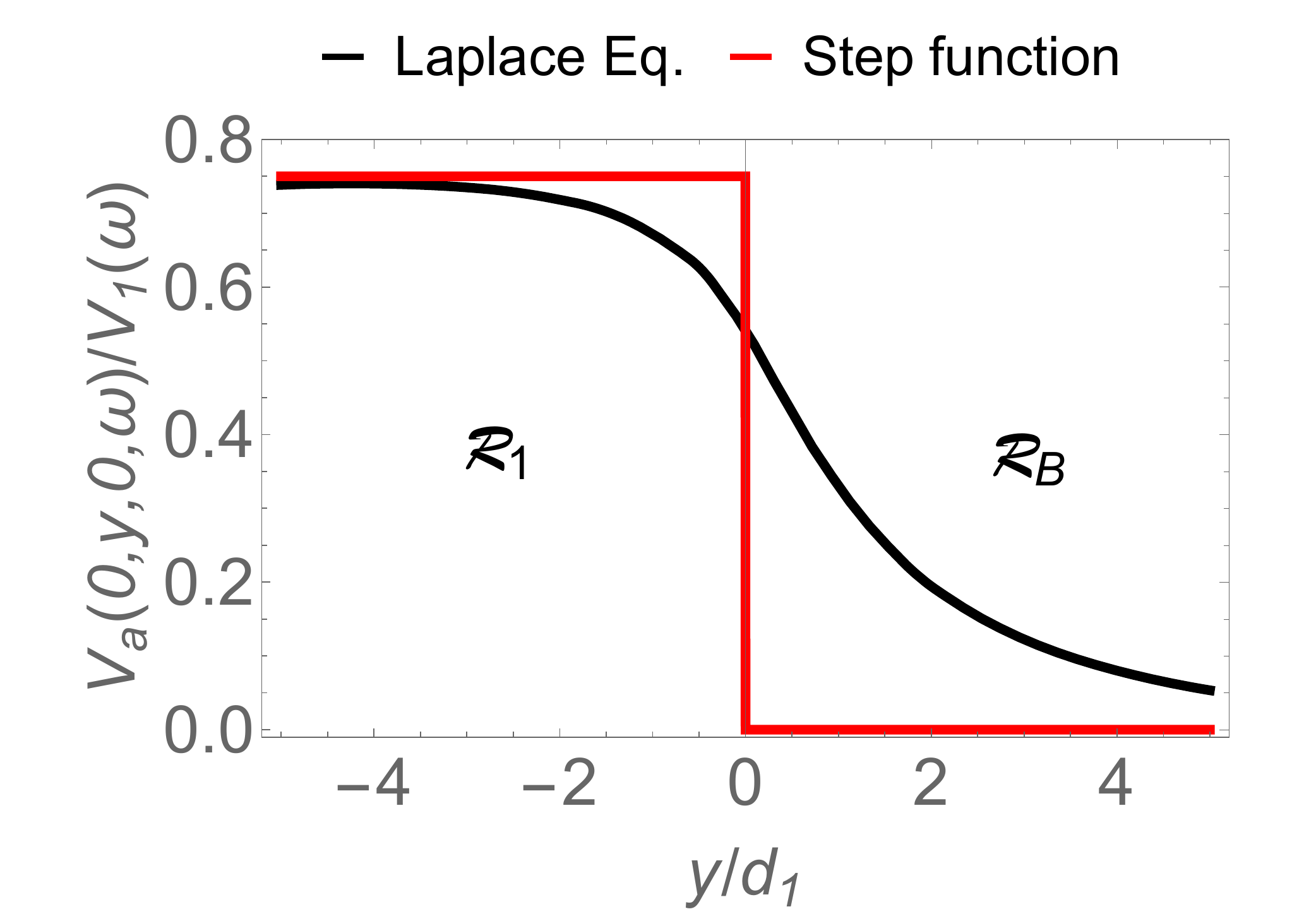}
\caption{\label{fig:Va_app} Driving term $V_a(0,y,0,\omega)$ as a function of $y$. We show  a comparison between the numerical solution of the Laplace equation for the electrostatic configuration in Fig.  \ref{fig:3gates-device} (black line) and the step function approximation discussed in the text (red line). The value of  $V_a(0,y,0,\omega)$ in $\mathcal{R}_1$ used for the step function  is defined in Eq. (\ref{eq:Va-longwl-full}). 
For the plot, we fix the ratio $d_B/d_1=3$ and neglect the effect of the second top electrode; the latter approximation is justified when the two electrodes are far away from each other $d_{1,2}\ll L_B$.
}
\end{figure}

Approaching the edge of $\mathcal{R}_{1,2}$, the value of $V_a$ in $x=z=0$ decreases, and it vanishes  in  $\mathcal{R}_{B}$ at a distance $\sim d_{1,2}$ from the boundary.
Because the driving voltage enters the equation of motion (\ref{eq:eq-motion-parts}) via $\partial_y V_a$, the applied potential does not influence the plasmon dynamics inside $\mathcal{R}_i$, but it accelerates the EMPs at the edge of $\mathcal{R}_i$.
When $d_i\ll L_B$ and in the long wavelength limit, one can neglect the fringing effects and approximate $V_a$ by using step functions, in agreement with the treatment presented in \cite{QEMP}.
To illustrate this approximation, we find $V_a$ by solving numerically the Laplace equation  in the electrostatic configuration shown in Fig. \ref{fig:3gates-device}. In Fig. \ref{fig:Va_app}, we show the solution $V_a$ evaluated at $x=z=0$ close to the boundary of $\mathcal{R}_1$ and compare it with the step function approximation obtained by neglecting fringing fields. \\

Let us now focus on the limit $d_{1,2}\ll d_B$, which models the response of the devices in Refs. \cite{Viola-DiVincenzo, Reilly, Mahoney,HITLpt1, Bosco}: in this case, we find that Eq. (\ref{eq:motion-emp-local-velocity-app}) holds. In particular, we obtain the piecewise equation of motion,
\begin{equation}
\label{eq:eq-motion-linearized-gate}
i\omega\rho_i(y,\omega )\approx v_i \partial_y \rho_i\left(y,\omega \right),
\end{equation}
% \begin{equation}
%\label{eq:eq-motion-linearized-gate}
%i\omega\rho_i(y,\omega )\approx v_i \partial_y \rho_i\left(y,\omega \right)+\sigma_{xy} \partial_y V_{a,i}(y,\omega),
%\end{equation}
with velocities
\begin{equation}
\label{vel-i}
v_i=2\pi\sigma_{xy}G_i(l,q\rightarrow 0,0)=v_p\log\left(1+\frac{4d^2_i}{l^2}\right),
\end{equation}
and $v_p$ defined in Eq. (\ref{vp-scale}).
Also, as expected, if we neglect the fringing fields, the applied voltages $V_{1,2}$ enter in the dynamics of the EMPs only via the matching conditions at the boundaries between adjacent regions $\mathcal{R}_i$. In particular, we find that at the edge of $\mathcal{R}_{1,2}$ $\rho_i$ satisfies 
\begin{equation}
\label{eq:matching-cond}
v_B\rho_B-v_{1,2}\rho_{1,2}= \sigma_{xy} V_{1,2}.
\end{equation}

Here, we used the simplified form of Eqs. (\ref{eq:G-function-app}) and (\ref{eq:Va-longwl-full}), valid for $d_{1,2}\ll d_B$:
\begin{subequations}
\begin{flalign}
\label{eq:approximate-GF}
G_i(l,q,0)&\approx\frac{K_0(\left|q\right| l)-K_0(\left|q\right|\sqrt{l^2+4d_i^2})}{4\pi^2\epsilon_s},\\
V_{a,(1,2)}(0,y,0,\omega)&\approx V_{1,2}(\omega).
\end{flalign}
\end{subequations}

%Also, in this setup, the relation between EMP charge and current is given by Eq. (\ref{eq:current-ith-electrode}).

If we consider a side gate instead of a top gate, the EMP velocities in Eq. (\ref{vel-i}) modify as $v_i^{\text{SG}}=2 v_p\log(\left|2d_i/l-1\right|)$; the two situations are quantitatively different only when $d_i$ comparable to the cut-off length $l$.\\

We now analyze a more general situation, where $d_{1,2}$ and $d_B$ are comparable. In this case, one obtains an equation of motion similar to Eq. (\ref{eq:eq-motion-linearized-gate}), but with different EMP velocities and matching conditions.
In particular, the velocities are now proportional to the long wavelength limit of Eq. (\ref{eq:G-function-app}), and the voltages $V_{1,2}$ in the right hand side of the matching conditions (\ref{eq:matching-cond})  acquire an additional proportionality constant $\frac{d_B}{d_B+d_{1,2}}$ dependent on the distance of the QH material from both gates, see Eq. (\ref{eq:Va-longwl-full}).

Also, to characterize the response of a QH device, one needs to compute the current flowing in the top electrodes, which is generally given by the integral of the displacement current, i.e. by the time variation of the surface charge localized at the top gates ($\mathcal{\textit{\textbf{R}}}_{1,2}$ indicates the 2-dimensional area of the top electrodes),
\begin{equation}
I_{1,2}(\omega)=i\omega \epsilon_S \int_{\mathcal{\textit{\textbf{R}}}_{1,2}} d \textbf{r} \int d\textbf{r}' \frac{\partial}{\partial z}G(\textbf{r},\textbf{r}',d_{1,2})\rho(\textbf{r}',\omega).
\label{eq:current-displacement}
\end{equation}
By using the same approximations discussed above, when $d_{1,2}\sim d_B$, this integral reduces  to
\begin{equation}
I_{1,2}(\omega)\approx -i\omega \frac{d_B}{d_B+d_{1,2}} \int_{\mathcal{R}_{1,2}} dy \rho_{1,2}(y,\omega).
\label{eq:current-displacement_simplified}
\end{equation}
%Note that this result agrees with the simple Eq. (\ref{eq:current-ith-electrode}) when $d_{1,2}\ll d_B$.

The prefactor $\frac{d_B}{d_B+d_{1,2}}$ in Eq. (\ref{eq:current-displacement_simplified}) is the same one that modifies the driving voltage in the matching conditions at the boundaries of $\mathcal{R}_{1,2}$.
The physical origin of this term is qualitatively understood by introducing the capacitance per unit area $c_i=\epsilon_s/d_i$, which parametrizes the electrostatic coupling of the QH material to the $i$th metal gate.
In $\mathcal{R}_{1,2}$, there are two capacitances $c_{1,2}$ and $c_B$ in series that connect the top gate to ground: only a fraction $\frac{c_B^{-1}}{c_B^{-1}+c_{1,2}^{-1}}=\frac{d_B}{d_B+d_{1,2}}$ of the total applied voltage reaches the QH material, and, conversely, only a fraction $\frac{c_B^{-1}}{c_B^{-1}+c_{1,2}^{-1}}$ of the current flowing the  QH material reaches the top gates.

In the literature, the EMP velocities are sometimes related to a capacitance per unit length $\tilde{c}$, which quantifies the Coulomb coupling of the electrode to QH edge state \cite{Viola-DiVincenzo,Glattli,Circuit_EMP}.
In the configuration examined here, the velocity in $\mathcal{R}_{1,2}$ can also be  roughly estimated by considering the effect of two parallel capacitors $\tilde{c}$ by approximating Eq. (\ref{eq:G-function-app}) as
\begin{equation}
G_{1,2}(l,q\rightarrow 0,0)\approx \frac{1}{2\pi}\frac{1}{\tilde{c}_B+\tilde{c}_{1,2}},
\end{equation}
where $\tilde{c}_i=4\pi\epsilon_s/\log(1+(2d_i/l)^2)$ are obtained from the $q\rightarrow 0$ limit of Eq. (\ref{eq:approximate-GF}).

Note that the two capacitances $c_i$ and $\tilde{c}_i$ are qualitatively different quantities: the former is the usual parallel plate capacitance (per unit area) which characterizes the electrostatic coupling between 2-dimensional charged planes, while the latter characterizes the Coulomb interactions (per unit length) between a 2-dimensional electrode and a (quasi) 1-dimensional line of charge.\\

In  \cite{HITLpt1}, the qualitative difference  between $c$ and $\tilde{c}$ is neglected. This is a reasonable estimation only in a fully local capacitance approximation, which is appropriate for smooth edges where $l'\gg d_i$ ($l'$ quantifies the broadening of $\sigma_{xy}(\textbf{r})$ into the bulk) \cite{Volkov,JohnsonVignale}.
In this situation, one can assume that the Green's function in the inverted Poisson Eq. (\ref{eq:inverted-poisson-eq}) is local in both $x$ and $y$ and can be  approximated in $\mathcal{R}_{1,2}$ as
\begin{equation}
G_{1,2}(\textbf{r},\textbf{r}',0)\approx \frac{\delta(\textbf{r}-\textbf{r}')}{\epsilon_s}\frac{d_B d_{1,2}}{d_B+d_{1,2}}=\frac{\delta(\textbf{r}-\textbf{r}')}{c_B+c_{1,2}}.
\end{equation}
%\begin{equation}
%G_{1,2}(\textbf{r},\textbf{r}',0)\approx \frac{\delta(x)}{2\pi\epsilon_s}\frac{d_B d_{1,2}}{d_B+d_{1,2}}=\frac{\delta(x)}{2\pi(c_B+c_{1,2})}.
%\end{equation}
With this simplification, and assuming a linear profile of $\sigma_{xy}(\textbf{r})$ at the edge as in \cite{JohnsonVignale}, one obtains the value of EMP velocity $v_{1,2}=\sigma_{xy}/(l'(c_B+c_{1,2}))$ used in \cite{HITLpt1}.
For sharp QH edges, however,  the difference between $c_i$ and $\tilde{c}_i$ is not negligible, and so the substitutions in Eq. (27) of  \cite{HITLpt1} have to be adjusted as
\begin{subequations}
\begin{flalign}
\tau_{1,2} &\rightarrow \tilde{\tau}_{1,2}\equiv L_{1,2} \frac{\tilde{c}_B+\tilde{c}_{1,2}}{\sigma_{xy}},\\
\sigma_{xy} & \rightarrow \sigma_{xy}\left(\frac{c_B^{-1}}{c_B^{-1}+c_{i}^{-1}}\right)\left(\frac{c_B^{-1}}{c_B^{-1}+c_{j}^{-1}}\right).
\end{flalign}
\end{subequations}

\subsubsection{Nanowires and Coulomb drag \label{sec:EMP-nanowire}}

\begin{figure}
\includegraphics[width=0.4\textwidth]{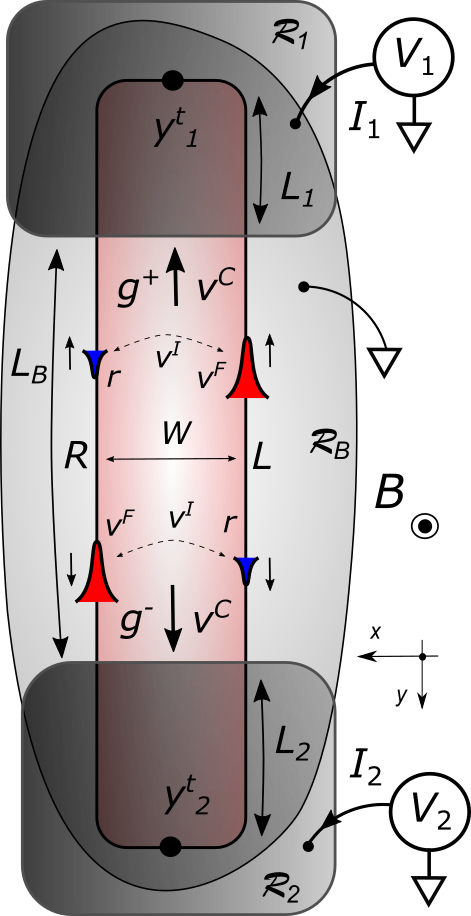}
\caption{\label{fig:TL-NW} Top view of a QH nanowire of width $W$. The nanowire is drawn in light red, while the gray areas indicate the position of the metal electrodes. The vertical cross-section of the device is shown in Fig. \ref{fig:3gates-device} and the $y$-direction is divided into  three regions $\mathcal{R}_i$ characterized by a different gating structure. The length of the nanowire is $y^t_2-y^t_1=L_1+L_2+L_B$, where $y^t_i$ are the end points of the QH material in the $y$-direction and $L_i$ is the length of $\mathcal{R}_i$. The intra- and inter-edge Coulomb couplings are parametrized by the velocities $v^{F,I}$, respectively. The inter-edge component $v^I$ causes Coulomb drag. We sketch schematically the charge profile in the $x$-direction of the two EMP eigenmodes $g^{\pm}$, moving with velocities $\pm v^{C}=\pm \sqrt{(v^F)^2-(v^I)^2}$, respectively. A positive charge (red) localized mostly at one edge drags a $r$ times smaller negative charge (blue) at the opposite edge; in the plot the sign of $g^{\pm}$ is chosen to satisfy Eq. (\ref{eq:termination_NW}).
}
\end{figure}

We now analyze the response of a QH nanowire of width $W\gg 2 l$, such as the one  shown in Fig. \ref{fig:TL-NW} and we focus on the effect of the Coulomb coupling between different edges.
The $y$-direction is again divided into the three regions $\mathcal{R}_i$ shown in Fig. \ref{fig:3gates-device}, that are characterized by a different electrostatic configuration. 
%However, in contrast to the analysis in Sec. \ref{sec:EMP-gated}, we consider here a closed device.

%
%
%In Sec. \ref{sec:EMP-gated}, we discuss a model of the motion of the EMP in such a device can be modeled by studying separately the propagation in each region and 
%For simplicity, we assume that the width $W\gg 2 l$ of the QH nanowire  is constant and it varies , and so we can use the conductivity profile of a nanowire, i.e. $\sigma_{xy}(\textbf{r})=\sigma_{xy}\Theta(x)\Theta(W-x)$ to model the EMP propagation there.

According to the discussion in Sec. \ref{sec:EMP-gated}, the response in this setup can be modeled by studying separately the EMP dynamics in $\mathcal{R}_i$ and by appropriately matching the solutions at the boundaries to account for the applied voltage $V_{1,2}$, see Eqs. (\ref{eq:eq-motion-linearized-gate}) and (\ref{eq:matching-cond}).

For this reason, we begin our analysis by looking for self-consistent EMP excitations in a nanowire with the conductivity profile
\begin{equation}
\sigma_{xy}(\textbf{r})=\sigma_{xy}\Theta(x)\Theta(W-x).
\end{equation}
%
%In $\mathcal{R}_B$, neglecting
%To gain insight into the problem, let us begin by studying the EMP propagation in an infinite, undriven nanowire case and so we neglect for the moment the effect of the voltage drive.
In this case, the equation of motion (\ref{eq:FT-self-consistent-charge-straight-line}) in the momentum space, straightforwardly modifies as
\begin{multline}
\label{eq:FT-self-consistent-charge-NW}
\omega\rho_i (x,q,\omega )=2\pi q\sigma_{xy} (\delta(x)-\delta(x-W))\\
  \int d x' G_i\left(x-x',q,0\right)\rho_i \left(x',q,\omega \right).
\end{multline}
Here, the index $i=(1,2,B)$ labels the electrostatic configuration of $\mathcal{R}_i$; the corresponding Green's function  $G_i$ is given by Eq. (\ref{eq:G-function-app}).
%
% in particular 
%The Green's function $G$ depends on the electrostatic configuration. 
%In particular to model the EMP propagation in the region $\mathcal{R}_i$ of the device in Fig. \ref{fig:TL-NW}, we consider $G\rightarrow G_i$, with $G_i$ given in  Eq. (\ref{eq:approximate-GF}).
It is convenient to introduce the vector decomposition of the excess charge density
\begin{equation}
\label{eq:rho-electron-hole}
\rho_i(x,q,\omega)=\rho_0 \textbf{g}_i(q,\omega)^T\left(
\begin{array}{c}
\delta(x) \\
-\delta(x-W)
\end{array}
\right),
\end{equation} 
which leads to the matrix eigenvalue equation
\begin{equation}
\label{eq:eigenvalue-eqNW}
\omega \textbf{g}_i(q,\omega)=q \underline{\mu}_i(q) \textbf{g}_i(q,\omega),
\end{equation}
with an antisymmetric velocity matrix
\begin{equation}
\label{eq:mu-mat}
\underline{\mu}_i=\left(
\begin{array}{cc}
v^F_i & -v_i^I \\
v_i^I & - v^F_i
\end{array}
\right).
\end{equation}
The intra- and inter-edge velocities are defined respectively as 
\begin{subequations}
\label{eq:v-NW}
\begin{flalign}
v_i^F&= 2\pi\sigma_{xy}G_i(l,q,0),\\
v_i^I&= 2\pi\sigma_{xy}G_i(W,q,0)<v_i^F.
\end{flalign}
\end{subequations}

Because we are interested here in understanding the effects of the inter-edge Coulomb coupling, we restrict  our analysis to nanowires that are thin on the scale of the wavelength, i.e. $q W\ll 1$: when this condition is not met, the inter-edge coupling is negligible. 
%Note also that at this point the velocities $v^{F,I}_i$ depend on the momentum $q$. To use the local approximation of $G_i$ introduced in Sec. \ref{sec:EMP-gated}, see  Eq. (\ref{eq:estimation-local-G}), we take the $q l\rightarrow 0$ and $qW \rightarrow 0$ limits in Eq. (\ref{eq:v-NW}).
%The existence of this limit is guaranteed by the presence of the back gate.
%
%$v_B$ is defined in Eq. (\ref{eq:inter_edge_v}), while the inter-edge velocity is $v_1= 2\pi\sigma_{xy}G(W,q,0)<v_0$. To keep the notation simple, we neglect here the explicit dependence of the velocities on momentum $q$.
%

The antisymmetry of $\underline{\mu}_i$ is a general consequence of the Green's reciprocity theorem (i.e. $G(x,x')=G(x',x)$); also, the tracelessness  of $\underline{\mu}_i$ guarantees that the eigenvalues of the matrix come in pairs with the same absolute value and opposite sign.

The matrix $\underline{\mu}_i$ is easily diagonalized: it has eigenvalues $\pm v^{C}_i$ with $v^{C}_i=\sqrt{(v_i^F)^2-(v_i^I)^2}$ and corresponding normalized eigenvectors
\begin{equation}
\textbf{g}^{\pm}_i =\frac{1}{\sqrt{2}}\left(
\begin{array}{c}
 \sqrt{1\pm\frac{v^{C}_i}{v^F_i}} \\
 \sqrt{1\mp\frac{v^{C}_i}{v^F_i}} \\
\end{array}
\right).
\end{equation}

The excess charge density of each eigenvector is mostly localized at one edge of the nanowire, but  because of the inter-edge Coulomb interactions, it also drags a fraction 
\begin{equation}
r_i=\frac{(\textbf{g}^+_i)_R}{(\textbf{g}^+_i)_L}=\sqrt{\frac{v^F_i-v^{C}_i}{v^F_i+v^{C}_i}}=\frac{v^F_i-v^{C}_i}{v_i^I}
\end{equation}
of charge with opposite sign at the other edge. The edges of the nanowire at position $x=0$ and $x=W$ are labeled by $L$ and $R$, respectively, see Fig. \ref{fig:TL-NW}. \\

%Voltage and current density are straightforwardly modified from the case above, and in particular the voltage is given by
%\begin{multline}
% V_{\pm}(\textbf{r},z)=\frac{\rho_0 e^{-i q y}}{\sqrt{2}} \\
% \left(\sqrt{1\pm\frac{v}{v_0}}G\left(x,q,z\right)-\sqrt{1\mp\frac{v}{v_0}}G\left(W-x,q,z\right)\right),
%\end{multline}
%while, integrating over the nanowire, the conduction current flowing in the $y$-direction is 
%\begin{equation}
%I_{\pm}(y)=\frac{\rho_0 e^{-i q y}}{\sqrt{2}}(v_0-v_1)
%\left(\sqrt{1+\frac{v}{v_0}}+\sqrt{1-\frac{v}{v_0}}\right).
%\end{equation}
%Note that the current carried by the two eigenmodes is flowing in the same direction, in agreement with conventional Hall analysis.
%(??Voltage and current explicit expressions maybe not required??)

%We can now analyze the effect of metal electrodes and of an external drive in this setup. For convenience, we focus only on symmetric devices where the gates are placed at the same distance to the left and right edges of the nanowire, i.e. the capacitances per unit length $\tilde{c}_i$ are equal for the two edges. In this case, the matrix $\underline{\mu}$ has the same structure as in Eq. (\ref{eq:mu-mat}), but the explicit expression for the velocities $v_{0,1}$ is modified by the presence of the electrodes; for example if we consider a top gate the velocities are obtained by using the screened electrostatic Green's function $G$ defined in Eq. (\ref{eq:approximate-GF}). 

Let us now focus on the effect of the driving voltages. 
The main difference with Sec. \ref{sec:EMP-gated} is that the EMP equation of motion (\ref{eq:eq-motion-parts}) becomes here a system of coupled equations for $\textbf{g}$. Using the local approximation for $G$ and $V_a$ discussed there, we obtain
\begin{equation}
\label{eq:system_NW}
i \omega \textbf{g}(y,\omega)=\partial_y\left(\underline{\mu}(y) \textbf{g}(y,\omega)\right) +\sigma_{xy}\partial_y \textbf{V}_a(y,\omega),
\end{equation}
where the velocity matrix $\underline{\mu}(y)$  and the driving term 
\begin{equation}
\textbf{V}_a(y,\omega)= \left(
\begin{array}{c}
 V_a(0,y,0,\omega) \\
 V_a(W,y,0,\omega) \\
\end{array}
\right),
\end{equation}
are both piecewise functions with a constant value in $\mathcal{R}_i$.
In particular, $\underline{\mu}(y)$ in $\mathcal{R}_i$ reduces to the $q\rightarrow 0$ limit of the matrix $\underline{\mu}_i$ in   Eq. (\ref{eq:mu-mat}). Also, for the device in Fig. \ref{fig:TL-NW}, the driving voltage is equal at the two edges, and so 
\begin{equation}
\label{eq:simplified-Va_NW}
\textbf{V}_{a,(1,2)} (\omega) =V_{1,2}(\omega) \frac{d_B}{d_B+d_{1,2}} 
\left(
\begin{array}{c}
1\\
1 \\
\end{array}
\right),
\end{equation}
and $\textbf{V}_{a,B}=0$.
%  for simplicity, from now we consider only the limit $d_B\gg d_{1,2}$.

The partial differential equations (\ref{eq:system_NW}) can be easily decoupled by introducing the matrices of column eigenvectors $M_i=(\textbf{g}^{+}_i,\textbf{g}^{-}_i)$, that diagonalize $\underline{\mu}_i$. Defining the EMP eigenmodes $\textbf{u}_i\equiv M^{-1}_i\textbf{g}_i$ and $\underline{v}_{i}^C=\mathrm{diag}(+v^C_i,-v^C_i)$, one obtains the equations of motion
\begin{equation}
i\omega \textbf{u}_i(y,\omega)=\underline{v}_{i}^C\partial_y\textbf{u}_i(y,\omega),
\end{equation}
and the matching conditions
\begin{equation}
\label{eq:matching-NW}
\underline{v}_{B}^C\textbf{u}_B-\underline{v}_{1,2}^C\textbf{u}_{1,2}= \sigma_{xy} M_{1,2}^{-1} \textbf{V}_{a,(1,2)} (\omega).
\end{equation}
Note that using Eq. (\ref{eq:simplified-Va_NW}), the right hand side of (\ref{eq:matching-NW}) simplifies to $\sigma_{xy}\Gamma_{1,2} V_{1,2} \frac{ d_B}{d_B+d_{1,2}}  (1,1)^T$, with
\begin{equation}
\Gamma_{1,2}=  \frac{v^F_{1,2}}{\sqrt{2}v^C_{1,2}}\left(\sqrt{1+\frac{v^C_{1,2}}{v^F_{1,2}}}-\sqrt{1-\frac{v^C_{1,2}}{v^F_{1,2}}}\right).
\end{equation}\\

We can now find the total current $I_{1,2}$ flowing into the top gates in $\mathcal{R}_{1,2}$, which  is obtained by integrating the excess charge density $\rho_{1,2}$.
In a nanowire, $I_{1,2}$ is  composed of the sum of the contributions of the two counter-propagating EMP eigenmodes $u^{\pm}_{1,2}$ ($\pm$ indicates the sign of the velocity), i.e.
\begin{equation}
\label{eq:decomposition-charge-NW}
I_{1,2}= \frac{1}{\sqrt{2}} \left(\sqrt{1+\frac{v^C_{1,2}}{v^F_{1,2}}} I^{+}_{1,2}  +\sqrt{1-\frac{v^C_{1,2}}{v^F_{1,2}}} I^{-}_{1,2} \right),
\end{equation}
with 
\begin{equation}
I^{\pm}_{1,2}=- i\omega \frac{d_B}{d_B+d_{1,2}} \int_{\mathcal{R}_{1,2}} dy u^{\pm}_{1,2}(y,\omega).
\end{equation}

Note that the setup in Fig. \ref{fig:TL-NW} is a closed device, and so to proceed further in our analysis, we need to model the termination of the nanowire.
For simplicity, we assume, that the nanowire is thin, i.e. $W\ll L_i$; in this limit, one can neglect the dynamics of the EMP in the $x$-direction, and require that at the end points $y^t_i$ of the wire, the normal component of the current density vanishes. 
In terms of the EMP eigenmodes $u^{\pm}_{1,2}$, this boundary condition reduces to 
\begin{equation}
\label{eq:termination_NW}
u^+_{1,2}(y= y^t_{1,2})=-u^-_{1,2}(y= y^t_{1,2}).
\end{equation}

Using this condition and Eq. (\ref{eq:decomposition-charge-NW}), we find that $I^{+}_{1,2}=I^-_{1,2}$ for the device in Fig. \ref{fig:TL-NW}, and so, the total current flowing in the top gates reduces to $I_{1,2}=I^+_{1,2}/\Gamma_{1,2}$.
%The two currents have the same sign in our convention because of the boundary condition $u_+(y= y_t)=-u_-(y= y_t)$ and because the velocity of the excitations has the same absolute value but opposite sign.
Note now that because of the prefactor of the inhomogeneous term in Eq. (\ref{eq:matching-NW}), one obtains that $I^+_{1,2}\propto \Gamma_{1,2}$, and so the total current $I_{1,2}$ is independent of $\Gamma_{1,2}$.
%, when computing the total terminal-wise admittance matrix of the device, $\Gamma_{1,2}$ simplifies with the prefactor $1/\Gamma_{1,2}$ of the total current. 
Consequently, the terminal-wise admittance matrix of this QH nanowire   has the same matrix elements shown in Eq. (4) of  \cite{HITLpt1} (obtained without inter-edge interactions) but with EMP velocities renormalized by the Coulomb drag.

Also, Eqs. (\ref{eq:decomposition-charge-NW}) and (\ref{eq:termination_NW}) justify the equivalent circuit model for these devices proposed in \cite{HITLpt1}, where there are two circuits  characterized by two charge densities with opposite sign and moving in opposite direction connected in parallel. 
These charge densities are sketched in Fig. \ref{fig:TL-NW} as the blue and red components of $g^{\pm}$; in the plot their sign is chosen to satisfy Eq. (\ref{eq:termination_NW}).

\subsection{Near-field analysis\label{sec:field_distribution}}

In this section, we provide a more detailed discussion of the electromagnetic field at the edge of a QH droplet, which accounts also for near-field corrections. 
For simplicity, we now neglect the driving voltage and study only  self-consistent plasmonic excitations in a half-plane.

As discussed in Sec. \ref{sec:EMP-half-plane}, in the QH limit ($\sigma_{xx}\rightarrow 0$) and for sharp edges, a purely classical model of the EMPs has a pathology due to the electrostatic instability of a 1-dimensional line of charge \cite{Macdonald-Rice}.
This issue can be resolved by considering a conductivity tensor with a small but finite broadening $l'$ into the bulk of the material. 
For example, we consider a conductivity profile of the form 
\begin{equation}
\label{eq:conductivity-erf}
\sigma_{xy} (\textbf{r})= \frac{\sigma_{xy}}{2} \left(1+\text{erf}\left(\frac{x}{l'}\right)\right).
\end{equation}
%\begin{figure}
%a)\includegraphics[width=0.45\textwidth]{WFapp}
%b)\includegraphics[width=0.45\textwidth]{lprimeandx0}
%\caption{\label{fig:wfapp} 
%EMP charge density of a QH droplets with atomically defined edges and filling factors $\nu=1$ expected from a RPA analysis \cite{QEMP}. 
%In the 
%a) Comparison between the EMP charge density computed in RPA and  exact wavefunctions in Eq. (\ref{eq:wf-e0}) (red lines) and the normalized and shifted gaussian in Eq. (\ref{eq:wf-appe0}) (blue lines). We used $\epsilon_F=1.4\hbar\omega_c$  for the solid lines and  $\epsilon_F=0.6\hbar\omega_c$ for the dashed lines. The thick gray line indicates the physical termination of the QH material. b) Fitting parameters $l'$ (blue line) and $x_0$ (red line) in units $l_B$ as a function of the Fermi energy $\epsilon_F$.}
%\end{figure} 
%
%A justification for this choice is presented  in Appendix  \ref{sec:quantum-field}. In fact, the quantum treatment discussed in Appendix  \ref{sec:quantum-field}, for which $l'$ is of the order of the magnetic length $l_B=\sqrt{\hbar/(eB)}$, see Fig. \ref{fig:wfapp}.
%In fact, because of the spatial dependence of the conductivity, the delta function in Eq. (\ref{eq:FT-self-consistent-charge-straight-line}) becomes a normalized gaussian and the excess charge density has the form $\rho(\textbf{r})\propto \partial_x\sigma_{xy}(\textbf{r})\propto e^{-(x/l')^2}$ expected from a quantum mechanical analysis restricted to the value of filling factor $\nu=1$ \cite{QEMP, MacDonald_plasmon, Chamon}.
%
Because of the spatial dependence of the conductivity, the delta function in Eq. (\ref{eq:FT-self-consistent-charge-straight-line}) becomes  a normalized gaussian and so the excess charge density takes  now the form $\rho(\textbf{r})\propto \partial_x\sigma_{xy}(\textbf{r})\propto e^{-(x/l')^2}$.
%From a quantum mechanical analysis, one expects a similar EMP charge density in QH droplets with atomically defined edges and filling factors $\nu=1$ \cite{QEMP, MacDonald_plasmon}.
%%
% expected from a quantum mechanical analysis restricted to the value of filling factor $\nu=1$ \cite{QEMP, MacDonald_plasmon, Chamon}.
%

Note that the length $l'$ is a phenomenological parameter, whose value has to be  extracted from experiments or computed a priori.
Here, to estimate $l'$, we fit the excess charge density $\rho$ and the eigenfrequency $\omega$ obtained in our semiclassical model against the results obtained from a quantum mechanical analysis \cite{QEMP, MacDonald_plasmon,Mikhailov}.
A more detailed explanation of the quantum mechanical treatment, including a discussion of the leading quantum corrections to the EMP dynamics, can be found in Appendix \ref{sec:quantum-field}.
In particular,  we find that in QH droplets with atomically defined edges and filling factors $\nu=1$, the EMP charge density is also approximately gaussian,  see Fig. \ref{fig:wfapp}a), and so the conductivity (\ref{eq:conductivity-erf}) is well-suited to model these systems.

In this case, a good agreement of the results is achieved when $l'\propto l_B$, with $l_B=\sqrt{\hbar/(eB)}$ being the magnetic length; the proportionality constant is of order one and its value depends on the Fermi energy of the QH material. For example, when the Fermi energy lies in the middle of the cyclotron gap between the lowest and first Landau level,
\begin{equation}
\label{eq:lp-est}
l'\approx 0.75 l_B.
\end{equation}
Also, from the quantum mechanical treatment presented in Appendix \ref{sec:quantum-field}, it follows that the conductivity profile in Eq. (\ref{eq:conductivity-erf}) is not centered at the physical edge of the QH material, but  is shifted into the bulk by a length $x_0\propto l_B$. In particular, at the same Fermi energy that gives (\ref{eq:lp-est}), we obtain
\begin{equation}
\label{eq:x0-est}
x_0\approx 1.15 l_B.
\end{equation}
The precise value of $l'$ and $x_0$ is relevant for the discussion of the resonator-qubit coupling in Sec. \ref{sec:coupling}.
The general dependence of $l'$ and $x_0$ on the Fermi energy is shown in Fig. \ref{fig:wfapp}b).

\subsubsection{EMP velocity \label{sec:classical-eigen}}

%We now provide a more detailed discussion of the electromagnetic field at the edge of a QH droplet. In this analysis, we neglect the driving voltage and study only  self-consistent excitations in the half-plane.
%As discussed in Sec. \ref{sec:EMP-half-plane}, in the QH limit ($\sigma_{xx}\rightarrow 0$) and for sharp edges, a purely classical model of the EMP has a pathology due to the electrostatic instability of a 1-dimensional line of charge \cite{Macdonald-Rice}.
%This issue can be resolved by considering a conductivity tensor with a small but finite broadening $l'$ into the bulk of the material: for example, we consider here a conductivity profile of the form 
%\begin{equation}
%\label{eq:conductivity-erf}
%\sigma_{xy} (\textbf{r})= \frac{\sigma_{xy}}{2} \left(1+\text{erf}\left(\frac{x}{l'}\right)\right).
%\end{equation}
%%
%This choice is consistent with the quantum treatment discussed in Sec. \ref{sec:quantum-field}, for which $l'$ is of the order of the magnetic length $l_B=\sqrt{\hbar/(eB)}$, see Fig. \ref{fig:wfapp}.
%In fact, because of the spatial dependence of the conductivity, the delta function in Eq. (\ref{eq:FT-self-consistent-charge-straight-line}) becomes a normalized gaussian and the excess charge density has the form $\rho(\textbf{r})\propto \partial_x\sigma_{xy}(\textbf{r})\propto e^{-(x/l')^2}$ expected from a quantum mechanical analysis restricted to the value of filling factor $\nu=1$ \cite{QEMP, MacDonald_plasmon}.

%We have not been able to solve the resulting equation analytically, and thus we proceed numerically. 

When no external voltage is applied,  Eq. (\ref{eq:classical-rho-V-relation}) with the  conductivity profile (\ref{eq:conductivity-erf}) reduces to a homogeneous Fredholm integral equation of the second kind, that has to be solved for the EMP charge density $\rho$ and for the eigenfrequency $\omega$. 
There are several possible ways to proceed: we choose here an approach that can be easily generalized to include a finite diagonal  conductivity, as described in Sec. \ref{sec:dissipation-field}.

We work in the momentum space ($y\rightarrow q$), and we introduce the auxiliary function $p$ defined by 
\begin{equation}
\label{eq:p-def}
\rho(x,q,\omega)=p(x,q,\omega)\frac{e^{-(x/l')^2}}{\sqrt{\pi}l'}.
\end{equation}
 For simplicity of notation, we suppress the explicit dependence of $p$ on $q$ and $\omega$.
Also, we neglect at first the effect of the electrodes, and so we use the free space Green's function (\ref{eq:G-nogates}). The EMP equation of motion reduces to
\begin{equation}
\label{eq:integraleq-p}
\frac{\omega}{q v_p}  p(x)=\frac{2}{\sqrt{\pi}l'}\int_\mathbb{R} dx' e^{-(x'/l')^2} K_0\left(|q||x-x'|\right)p(x').
\end{equation}

This integral equation can be converted into a matrix eigenvalue problem by using the decomposition
\begin{equation}
\label{eq:legendre-dec}
p(x)=\sum_{n=0}^{\infty}p_n  \sqrt{n+\frac{1}{2}} P_n\left(\text{erf}\left(\frac{x}{l'}\right)\right),
\end{equation}
where $P_n$ are the Legendre polynomials \cite{Stegun}. This leads to
\begin{equation}
\label{eq:eigeneq-classical-matrix}
\frac{\omega}{q v_p} p_n=\sum_{m=0}^{\infty} \Lambda_{nm}p_m,
\end{equation}
with 
\begin{multline}
\label{eq:lambda_kernel}
\Lambda_{nm}=\sqrt{n+\frac{1}{2}}\sqrt{m+\frac{1}{2}}\int_{-1}^{1}ds P_n(s)\int_{-1}^{1}ds' P_m(s')\\
K_0\left(|ql'|\left| \text{erf}^{-1}(s)-\text{erf}^{-1}(s')\right| \right),
\end{multline}
and with $\text{erf}^{-1}$ being the inverse error function.

Note that although the matrix problem in Eq. (\ref{eq:eigeneq-classical-matrix}) supports an infinite number of eigenvalues, in the sharp edge limit, the slower (acoustic) modes are strongly damped and weakly coupled to external voltage sources \cite{Glazman,QEMP}; for this reason, we neglect them here and focus only on the fastest (optical) mode.
Taking the long-wavelength limit $ql'\rightarrow 0$ in the kernel of the integral (\ref{eq:lambda_kernel}), we obtain 
\begin{equation}
\label{eq:app-Lambda}
\Lambda_{nm}\approx -\log\left(\frac{(ql')^2 e^{\gamma}}{8}\right)\delta_{n0}\delta_{m0}-\tilde{\Lambda}_{nm},
\end{equation}
with $\gamma$ being the Euler constant and with 
\begin{multline}
\label{eq:lambda_tilde_kernel}
\tilde{\Lambda}_{nm}=\log\left(2 e^{\gamma}\right)\delta_{n0}\delta_{m0}+\sqrt{n+\frac{1}{2}}\sqrt{m+\frac{1}{2}}\times\\
\int_{-1}^{1}ds P_n(s)\int_{-1}^{1}ds' P_m(s')
\log\left(\left| \text{erf}^{-1}(s)-\text{erf}^{-1}(s')\right| \right).
\end{multline}
The first term in Eq. (\ref{eq:lambda_tilde_kernel}) guarantees that $\tilde{\Lambda}_{00}=0$. This decomposition is useful for perturbative considerations: for long wavelengths, the dominant term in the eigenvalue equation (\ref{eq:eigeneq-classical-matrix}) is the divergent term at $n=m=0$. This implies that the EMP eigenfrequency in this limit can be approximated as
\begin{equation}
\label{eq:eigen-freq-classical-lw}
\omega\approx-qv_p\log\left(\frac{(ql')^2 e^{\gamma}}{8}\right),
\end{equation}
and the EMP charge density (\ref{eq:p-def})   preserves the gaussian form, i.e. $p_n\approx \delta_{n0}$ in Eq. (\ref{eq:legendre-dec}).
This solution is consistent with the quantum analysis in Appendix \ref{sec:quantum-field} and it  also qualitatively agrees with the results of the smooth edge model of Aleiner and Glazman \cite{Glazman,Glazman2}.\\

It is also straightforward to modify the interaction kernel in the definition of $\Lambda_{nm}$ (\ref{eq:lambda_kernel}) to include the effects of external electrodes. For example, we consider now a top gate at a distance $d$ from the EMP. As discussed in Sec. \ref{sec:EMP-gated}, the metal electrode regularizes the long wavelength behavior of the EMP because of the positive image charge at $z=2d$, see Eq. (\ref{eq:approximate-GF}). 
For this reason,  when the gate is very close to the EMP, the perturbative treatment just presented becomes questionable; we find however that it still gives an excellent approximation, with an error below $4\%$ for $d\gtrsim 2l'$ at $q=0$.

\begin{figure}
a)\includegraphics[width=0.22\textwidth]{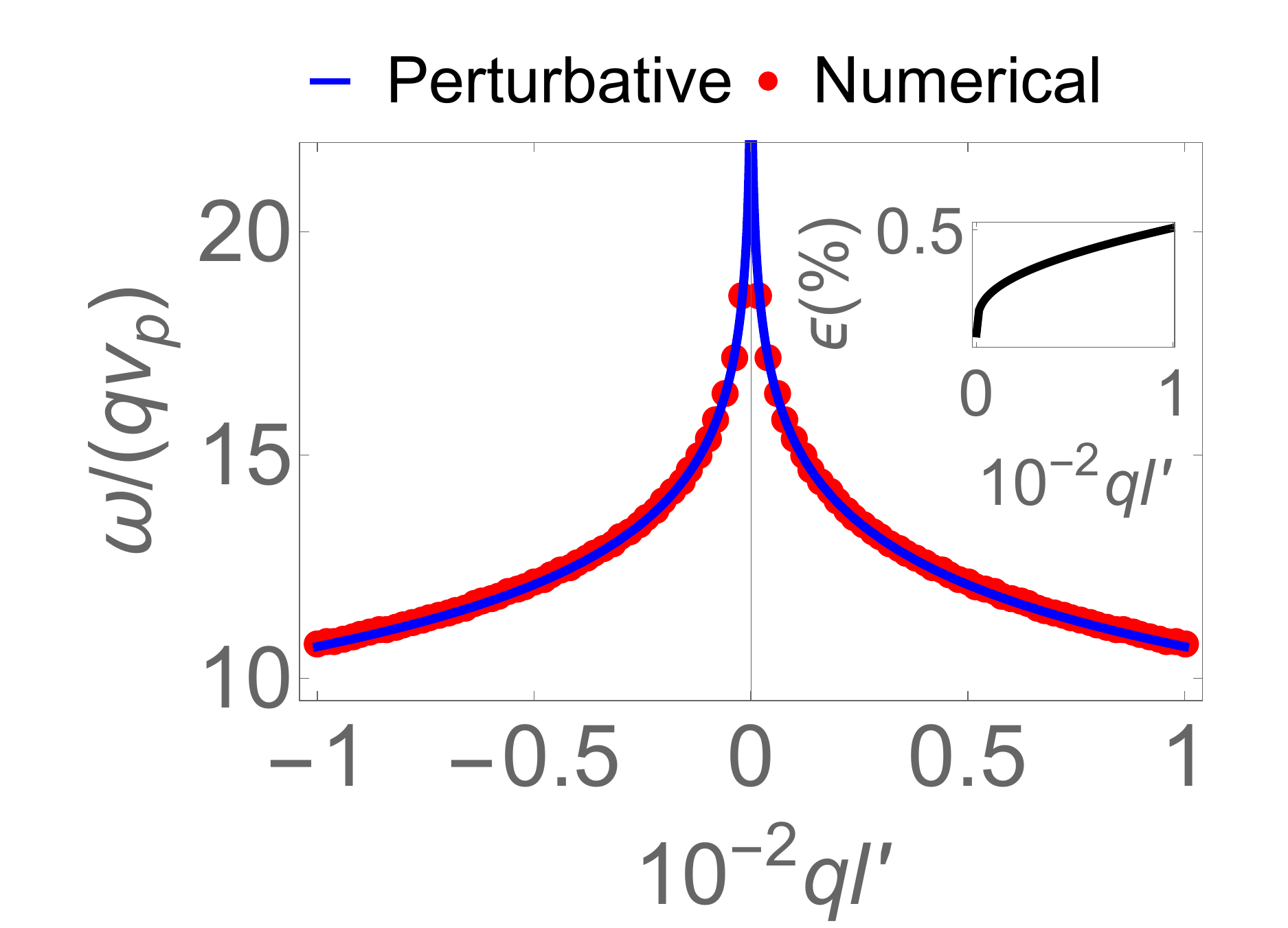}
b)\includegraphics[width=0.22\textwidth]{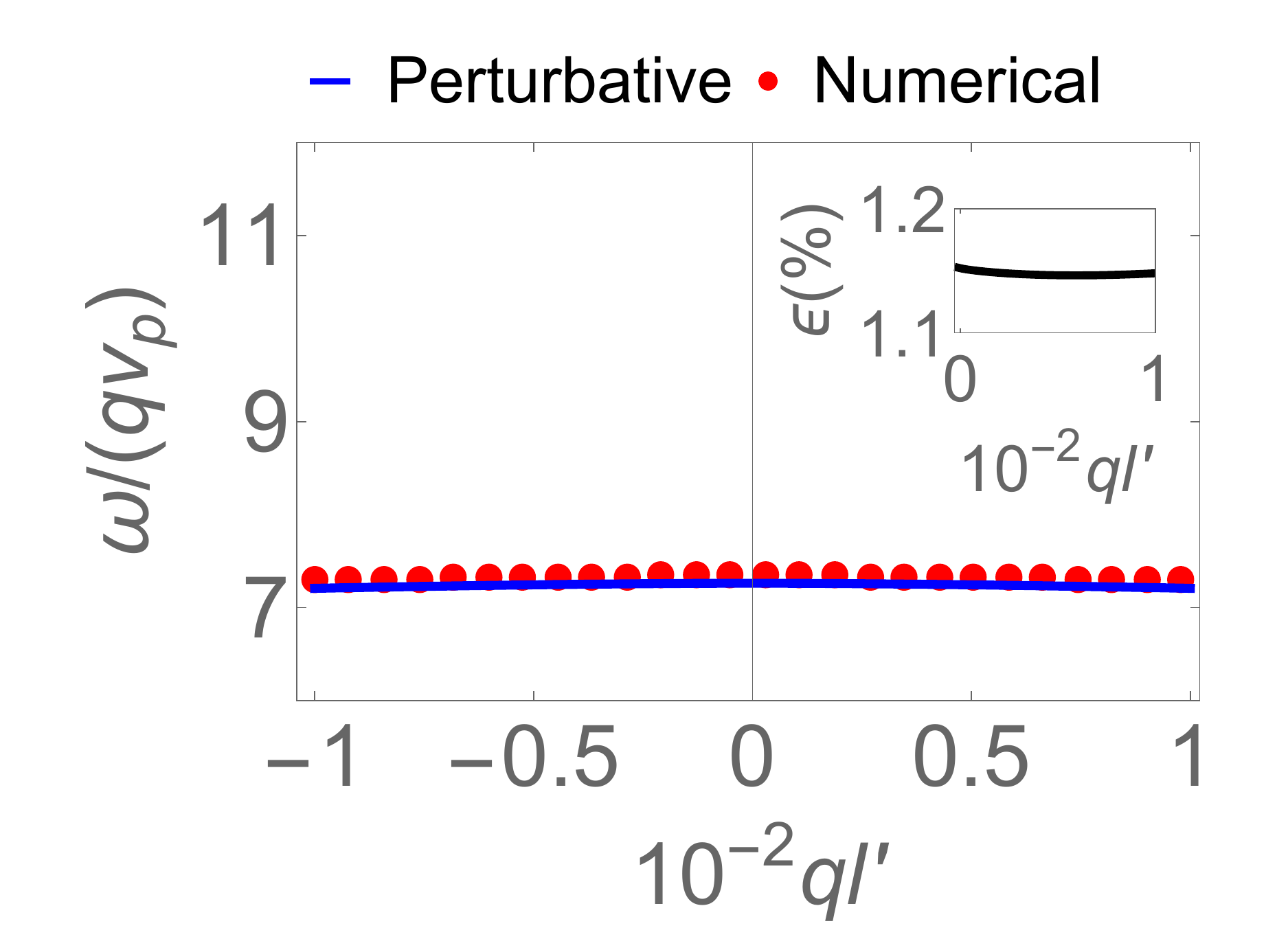}
\caption{\label{fig:pertvsnumericseigen} Eigenvalues of Eq. (\ref{eq:eigeneq-classical-matrix}): comparison between the numerical results and the perturbative solution  (\ref{eq:eigen-freq-classical-lw}). In the insets, we plot the percentage error in making the approximation. In a), we plot the eigenfrequency without considering screening gates, while in b) we include a top gate at  distance $d=10 l'$ from the EMP.}
\end{figure}

A comparison between numerics and the perturbative expansion for small wavevectors with and without electrodes is shown in Fig. \ref{fig:pertvsnumericseigen}: the correction due to $\tilde{\Lambda}$ is negligible in the parameter regime we are interested in.\\

At this point, we can also compare this result to the simple solution proposed in Sec. \ref{sec:EMP-half-plane}: we find that in the longwavelength limit the eigenfrequencies in Eqs. (\ref{eq:eigenfreq-classical-cutoff}) and (\ref{eq:eigen-freq-classical-lw}) coincide when 
\begin{flalign}
\label{eq:c0-erf}
&&
l=c_0l',
&\ \  \mathrm{with} \ \
c_0=\frac{e^{-\gamma/2}}{\sqrt{2}}\approx 0.53;
&&
\end{flalign}
the same value of $l$ is appropriate in the presence of electrodes.

For different conductivity profiles, one can proceed in a similar way. In the long-wavelength limit, we find that  the excess charge density takes again the form $\rho(\textbf{r})\propto \partial_x\sigma_{xy}(\textbf{r})$ and the eigenfrequency  can be estimated from the far-field  Eqs. (\ref{eq:eigenfreq-classical-cutoff}) and (\ref{eq:inter_edge_v}), but with a different value of $c_0$, which depends on the conductivity profile chosen. 
For example, using $\sigma_{xy}(\textbf{r})\propto \arctan(\sqrt{x/l'})$,  we obtain $c_0=4$, in  agreement with the solution of \cite{Glazman}.

\subsubsection{Electromagnetic field \label{sec:fields-classical}}
Here, we discuss the near-field behavior of the electromagnetic field and compare it to the results presented in Sec. \ref{sec:EMP-half-plane}. 
To do so, we use the perturbative solution to the eigenvalue problem (\ref{eq:eigeneq-classical-matrix}): we neglect the corrections caused by $\tilde{\Lambda}_{nm}$ and consider a gaussian charge density with broadening $l'$.
Without electrodes, the charge, potential and current are given by
\begin{subequations}
\label{eq:CVC-gaussian}
\begin{flalign}
\label{eq:charge-gaussian}
\rho(\textbf{r})&=\rho_0 e^{iqy} e^{-\left(\frac{x}{l'}\right)^2}/(2\pi\sqrt{\pi}l')\\
\label{eq:potential-gaussian}
V(\textbf{r},z) &=V_0 e^{iqy} \mathcal{G}_0(x,q,z)\\
\textbf{j}(\textbf{r})&=V_0 e^{iqy} \sigma_{xy}(\textbf{r})\left(\begin{array}{c}
-iq g_0(x,q) \\
\partial_x g_0(x,q)
\end{array}\right),
\end{flalign}
\end{subequations}
and they have to be compared to their far-field counterparts in Eq. (\ref{eq:CVC}).

Here, $\rho_0$ is a unspecified constant of units charge per meter, we defined $V_0=\rho_0/(8\pi^2\epsilon_S)$ and $\sigma_{xy}(\textbf{r})$ is given in Eq. (\ref{eq:conductivity-erf}).
The dimensionless function $\mathcal{G}$ depends on the specific electrostatic configuration and in free space is
\begin{equation}
\label{eq:G0-functnogate}
\mathcal{G}_0(x,q,z)=\frac{2}{\sqrt{\pi}l'}\int_{\mathbb{R}}dx' e^{-\left(\frac{x-x'}{l'}\right)^2} K_0\left(|q|\sqrt{x'^2+z^2}\right).
\end{equation}

\begin{figure}
\includegraphics[width=0.4\textwidth]{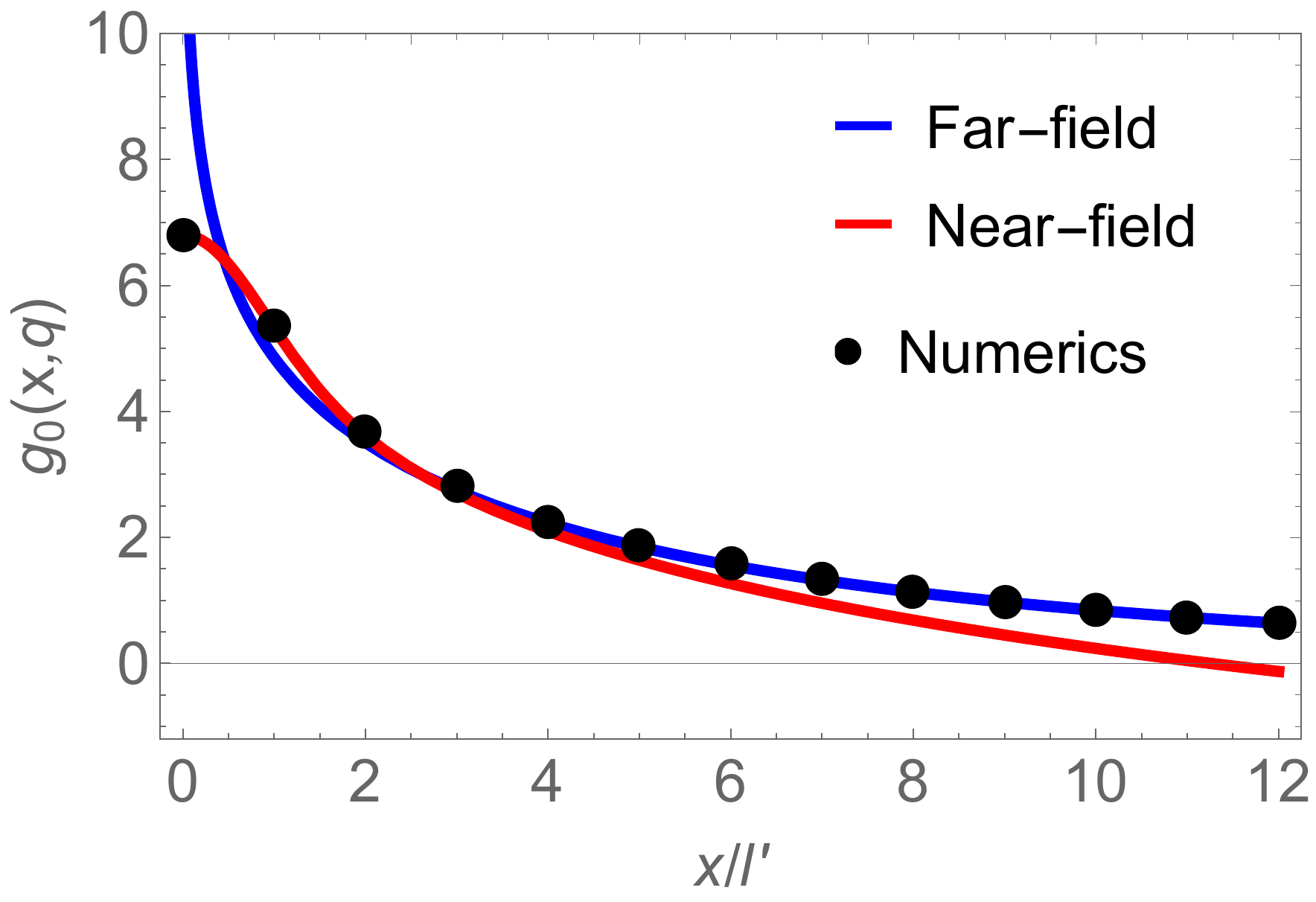}
\caption{\label{fig:approximation_g_funct} Near- and far-field approximation of the function $g_0(x,q)$, defined by the limit $z\rightarrow 0$ of Eq. (\ref{eq:G0-functnogate}). The black dots are obtained by solving the integral numerically, while the solid red and blue lines are the asymptotic solution in the near- and far-field limit, respectively. For illustrative purposes, we used a rather high value of $ql'=0.1$. }
\end{figure}

The value of $\mathcal{G}_0$ on the $z=0$ plane is particularly important for the qubit coupling described in Sec. \ref{sec:Hartree-text}, and so we define $\mathcal{G}_0(x,q,0)\equiv g_0(x,q)$. We now analyze two useful asymptotic limits of $g_0$, namely $|q|^{-1}\gg x,l'$ and $l'\ll |q|^{-1},x$. 
The first limit is useful to have a good estimation of $g_0$ in the near-field, i.e. when $x$ and $l'$ are of the same order of magnitude and both much smaller than $1/q$,  while the second limit gives a better estimation in the far-field, and, more generally, when the argument of the Bessel function is  not infinitesimal.
In the former case, we expand the Bessel function to the lowest order in $|q|$ and perform the integration, leading to 
\begin{equation}
\label{eq:g-appqseries}
g_0(x,q)\approx\, _1F_1^{(1,0,0)}\left(0;\frac{1}{2};-\frac{x^2}{l'^2}\right)-\log \left(\frac{(q l')^2e^{\gamma}}{16}\right),
\end{equation}
with $\, _1F_1^{(1,0,0)}$ being the derivative with respect to the first argument of the Kummer function of the first kind \citep{Stegun}.
In contrast, in the far-field limit, we approximate the gaussian by a delta function as in Sec. \ref{sec:EMP-half-plane}, and we obtain
\begin{equation}
\label{eq: g-app-delta}
g_0(x,q)\approx 2 K_0\left(\left| q x\right| \right).
\end{equation}
The two different approximations are shown in Fig. \ref{fig:approximation_g_funct}.

\begin{figure}[!ht]
a)\includegraphics[width=0.45\textwidth]{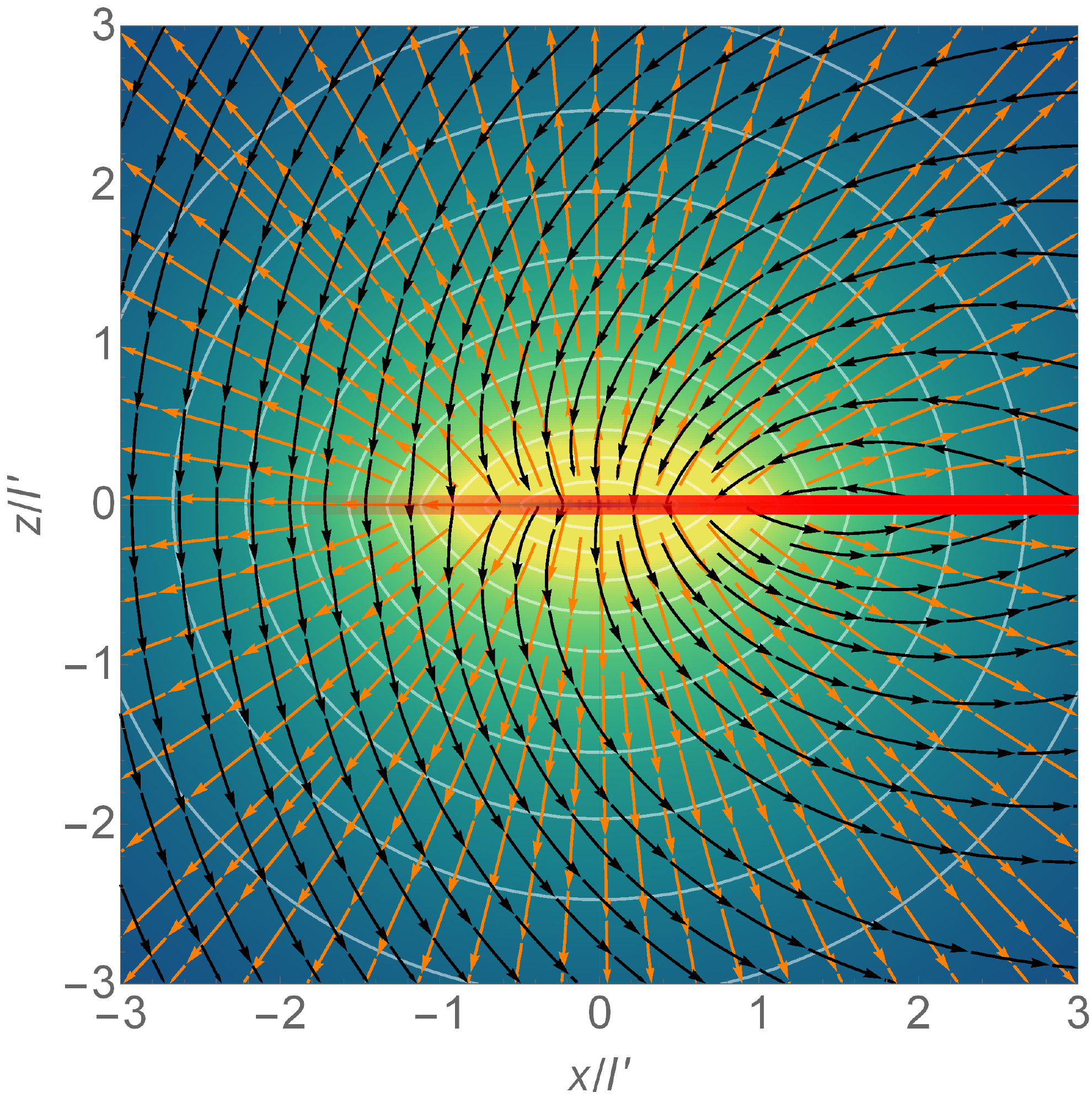}
b)\includegraphics[width=0.45\textwidth]{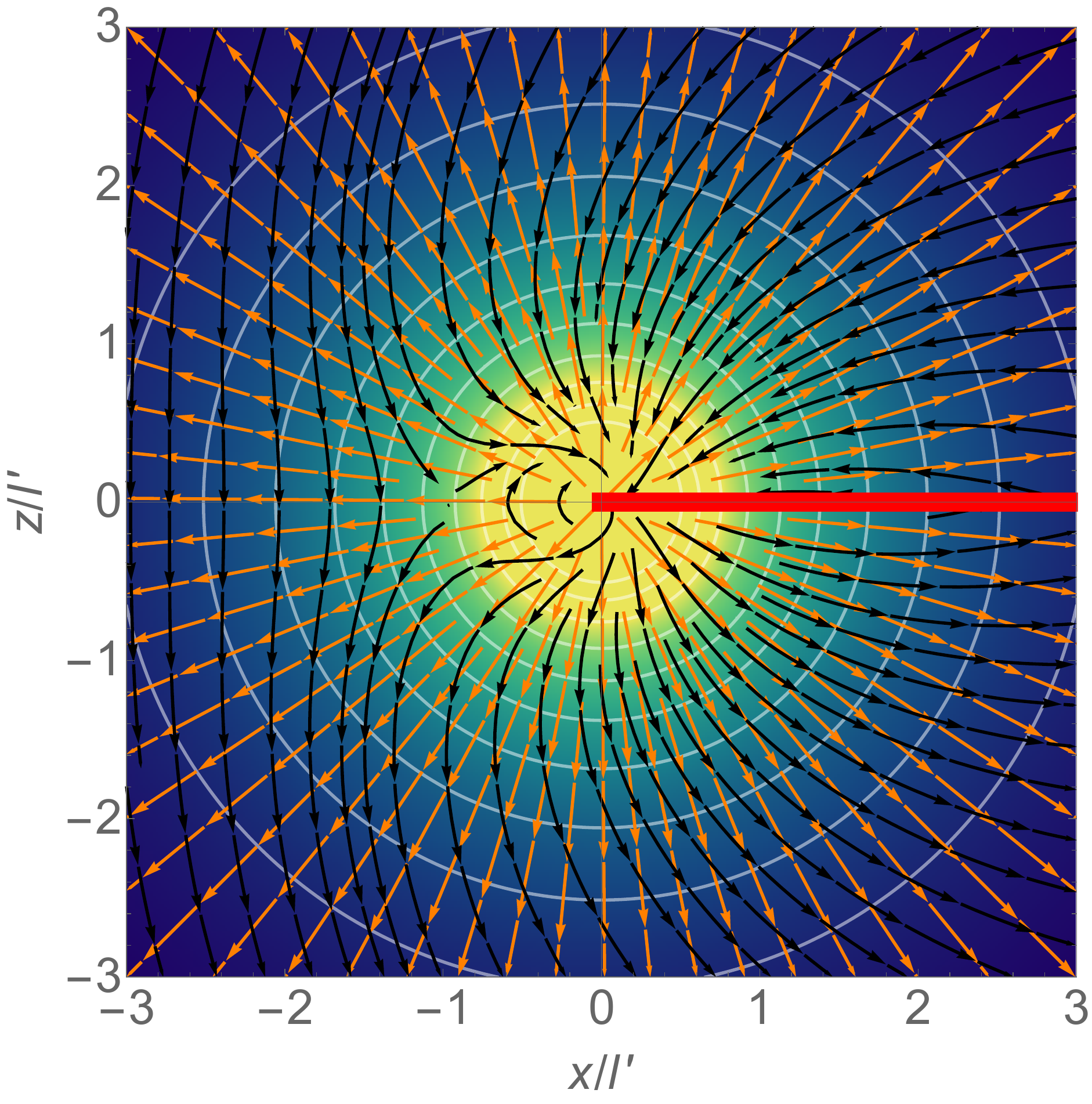}
\caption{\label{fig:em-fields} Electromagnetic field in the plane $y=0$ generated by a QH material in the $(x,y)$ plane and without electrodes. In a) we show the results obtained by  the potential and current density in Eq. (\ref{eq:CVC-gaussian}), and in b) we show the analogous far-field  obtained from Eq. (\ref{eq:CVC}). 
In the plots, we choose $ql'=0.01$ and the axis are in units $l'= l/c_0$ with $c_0=0.53$. The orange (black) stream lines represent the electric (magnetic) field, while in the background, we plotted the scalar potential $V$.
The thick red lines indicate the position of the QH material, and their opacity is weighted by the conductivity profile. }
\end{figure}

From Eq. (\ref{eq:CVC-gaussian}), it is straightforward to  compute the electric and magnetic fields
\begin{subequations}
\begin{flalign}
\textbf{E}(\textbf{r})&\approx-\nabla_{\textbf{r},z}V(\textbf{r},z),\\
\label{eq:B-field_VP}
\textbf{B}(\textbf{r})&=\frac{1}{c^2_S}\nabla_{\textbf{r},z}\times \int d\textbf{r}' G_0(\textbf{r},\textbf{r}',z)
\left( \begin{array}{c}
\textbf{j}(\textbf{r}')\\
0
\end{array}
\right),
\end{flalign}
\end{subequations}
where once again we use the electro-quasi static approximation for $\textbf{E}$ and neglected the small corrections due to the  time derivative of $\textbf{B}$ (i.e. $\nabla\times \textbf{E}\approx 0$). 
In Eq. (\ref{eq:B-field_VP}), we use $\textbf{B}=\nabla\times \textbf{A}$, where $\textbf{A}$ is the vector potential in the Lorenz gauge \cite{EQS_gauge}. 
Fig. \ref{fig:em-fields} shows a comparison of the electromagnetic fields in the cross section $y=0$ obtained from Eq. (\ref{eq:CVC-gaussian}) and from its far-field limit Eq. (\ref{eq:CVC}). In the plots, we consider two QH materials lying in the $(x,y)$ plane with a smooth and an abrupt conductivity profile and we neglect the effect of the external electrodes.\\

We can also verify the estimation of the impedance given in Sec. \ref{sec:EMP-half-plane}.
The conduction current $I_c$ can be computed without resorting to an artificial cut-off  length, and we find that Eq. (\ref{eq:cond-current}) is still applicable.
Also, to find the power flow, we start from the usual definition of the Poynting  vector $\textbf{S}=\frac{1}{2}\textbf{E}\times\textbf{H}^{*}$ and we resort to the electroquasi-static approximation $\nabla\times\textbf{E}\approx 0$, such that, up to a unimportant curl, $\textbf{S}\approx\frac{1}{2}V(\textbf{j}+i\omega\epsilon
_S\textbf{E})^{*}$, see e.g. Sec. 11 of \cite{Poynt_V}. Integrating $\textbf{S}$ along a circular cross-section with radius $R\rightarrow\infty$ in the $y=0$ plane,  the  average power flow reduces to
\begin{equation}
\label{eq:power-near}
P=\frac{I_c^2}{4\sigma_{xy}}(1-h(q)).
\end{equation}
The function $h(q)\propto e^{-(ql'/2)^2 }/K_0((ql'/2)^2)$ can be discarded in the long wavelength limit, and so we obtain the same result as Eq. (\ref{eq:power-flow}).\\

To conclude this section, we  comment on the effect of a top and of a side gate at distance $d$ from the edge of the QH material. The function $\mathcal{G}_0$ in (\ref{eq:G0-functnogate}) is modified as $\mathcal{G}_0\rightarrow\mathcal{G}_0+\mathcal{G}_{t,s}$, where the additive corrections are
\begin{subequations}
\label{eq:G-corr-exp-t-s}
\begin{flalign}
\mathcal{G}_t&=-\int_{\mathbb{R}}\frac{2d'}{\sqrt{\pi}l'} e^{-\left(\frac{x-x'}{l'}\right)^2} K_0\left(|q|\sqrt{x'^2+(z+2d)^2}\right),\\
\mathcal{G}_s&=-\int_{-d}^{\infty}\frac{2dx'}{\sqrt{\pi}l'} e^{-\left(\frac{x-x'}{l'}\right)^2} K_0\left(|q|\sqrt{(x+x'+2d)^2+z^2}\right),
\end{flalign}
\end{subequations}
for the top and side gate, respectively.

We are interested in understanding the effect of the electrodes on the electric field and on its gradient in the $z=0$ plane, where, in Sec. \ref{sec:coupling}, we place the qubit. To do so, the far-field asymptotic limit of the integrals $g_{t,s}\equiv\mathcal{G}_{t,s}(z\rightarrow 0)$ suffices, because for a finite value of $d$ the argument of the Bessel function does not diverge. 
By approximating the gaussian in the integrand (\ref{eq:G-corr-exp-t-s}) by a delta function, it is straightforward to verify that the electric field in $x$-direction decreases (increases) by introducing a top (side) gate. In contrast, the electric field gradient always decreases when we consider a side gate, while for a top gate the behavior depends on $d$:  the gradient increases if $d>x/2$ and it decreases otherwise.

%3d Plots of the fields in gaussian and long wavelength case and far-field limit coincided.
%Plots current density
%
%
%
%We can also compute the Poynting vector. Starting from the conventional definition and considering the EQS approximation, we subtract the divergence of the field and get to the Eq. in the text. Integrating, we get
%
%We also examine the electric field and its gradient in the case of gates, plot of field and gradient in x direction for top and side gate.
% 
%Brief Justification of EQS approximation and form of poyntin vector ecc... \\

%\end{equation}

\subsection{\label{sec:dissipation-field}Dissipation}
In this section, we extend the semiclassical model  presented of Sec. \ref{sec:classical-eigen} to include dissipation and we present a simple fitting of the results inspired by the analytic solution of \cite{Volkov}.

The decay of the EMPs is assumed to be caused by a finite and real diagonal part $\sigma_{xx}(\textbf{r})$ of the conductivity tensor in the system of equations (\ref{eq:classical-system-eq}). Imperfections in the dielectric or in the electrodes are neglected and can be  accounted for a posteriori \cite{Pozar}. 
Because of $\sigma_{xx}$, the relation in Eq. (\ref{eq:classical-rho-V-relation}) between the charge density and the screened potential is modified by the additional term in the right hand side
\begin{equation}
\sigma_{xx}(\textbf{r})\nabla^2_{\textbf{r}}V(\textbf{r},0,\omega)+\left(\nabla_{\textbf{r}}\sigma_{xx}(\textbf{r})\right) \cdot\nabla_{\textbf{r}} V(\textbf{r},0,\omega).
\end{equation}

%Here, we extend the calculations of Sec. \ref{sec:classical-eigen} to include the effect of dissipation and we discuss the numerical results that justify the simple solution presented in Sec. \ref{sec:dissipation}.
%
%We model dissipation by including a finite and real diagonal conductivity $\sigma_{xx}(\textbf{r})$ in the system (\ref{eq:classical-system-eq}).
%The relation in Eq. (\ref{eq:classical-rho-V-relation}) between the charge and the screened potential is modified by the additional term in the right hand side
%\begin{equation}
%\sigma_{xx}(\textbf{r})\nabla^2_{\textbf{r}}V(\textbf{r})+\left(\nabla_{\textbf{r}}\sigma_{xx}(\textbf{r})\right) .\nabla_{\textbf{r}} V(\textbf{r}).
%\end{equation}

%A similar model with sharp and frequency dependent conductivities  was solved analytically by Volkov and Mikhailov \cite{Volkov}. 
For simplicity, we now restrict our analysis to self-consistent excitations in a half-plane. We use the form of (\ref{eq:conductivity-erf}) for the two components of the conductivity tensor $\sigma_{xx} (\textbf{r})$ and $\sigma_{xy}(\textbf{r})$. We also neglect retardation effects and take the DC limit of the conductivity.

Fourier transforming the $y$-direction and using the free space Green's function (\ref{eq:G-nogates}), we obtain an integral equation for the auxiliary function $p(x)$ similar to Eq. (\ref{eq:integraleq-p}) with the additional integral in the right hand side
\begin{equation}
-i \frac{2 \sigma_{xx}}{\sqrt{\pi}\sigma_{xy} q l'} \hat{D}(x) \int_\mathbb{R} dx' e^{-(x'/l')^2} K_0\left(|q||x-x'|\right)p(x'),
\end{equation}
where we define the differential operator
\begin{equation}
\label{eq:D-dissip-op}
\hat{D}(x)=\partial_x+\frac{\sigma_{xx}(\textbf{r})}{\partial_x \sigma_{xx}(\textbf{r})} \left(\partial^2_{xx}-q^2\right).
\end{equation}
%\begin{equation}
%\hat{D}(x)=\frac{\partial}{\partial_x}+\frac{\sigma_{xx}(\textbf{r})}{\partial_x \sigma_{xx}(\textbf{r})}\left(\frac{\partial^2}{\partial_x^2}-q^2\right)
%\end{equation}

Following the procedure presented in Sec. \ref{sec:classical-eigen}, we discretize the integral equation by using the decomposition (\ref{eq:legendre-dec}) and we obtain the eigenvalue equation (\ref{eq:eigeneq-classical-matrix}) with the extra imaginary term
\begin{equation}
-i\frac{\sigma_{xx}}{\sigma_{xy}ql'}\sum_{m=0}^{\infty}\sum_{k=0}^{\infty} \mathcal{D}_{nk}\Lambda_{km}p_m,
\end{equation} 
with
\begin{multline}
\mathcal{D}_{nk}=\frac{2}{\sqrt{\pi}}\sqrt{n+\frac{1}{2}}\sqrt{k+\frac{1}{2}}\int_{\mathbb{R}}dxe^{-x^2/l'^2}\\
P_n\left(\mathrm{erf}\left(\frac{x}{l'}\right)\right)
\hat{D}\left(x\right)P_k\left(\mathrm{erf}\left(\frac{x}{l'}\right)\right).
\end{multline} 

Therefore, the problem including dissipation reduces to the diagonalization of the complex-valued matrix
\begin{equation}
\label{eq:lambdaDmat}
\Lambda_D=\left(\mathcal{I}-i\frac{\sigma_{xx}}{\sigma_{xy}ql'}\mathcal{D}\right)\Lambda,
\end{equation}
with $\mathcal{I}$ being the identity matrix.

For long wavelengths, we approximate $\Lambda$ by Eq. (\ref{eq:app-Lambda}) and neglect the $q^2$ correction in Eq. (\ref{eq:D-dissip-op}).
To find the complex EMP eigenfrequency $\omega$, we diagonalize $\Lambda_D$ numerically.\\

\begin{figure}
a) \includegraphics[width=0.45\textwidth]{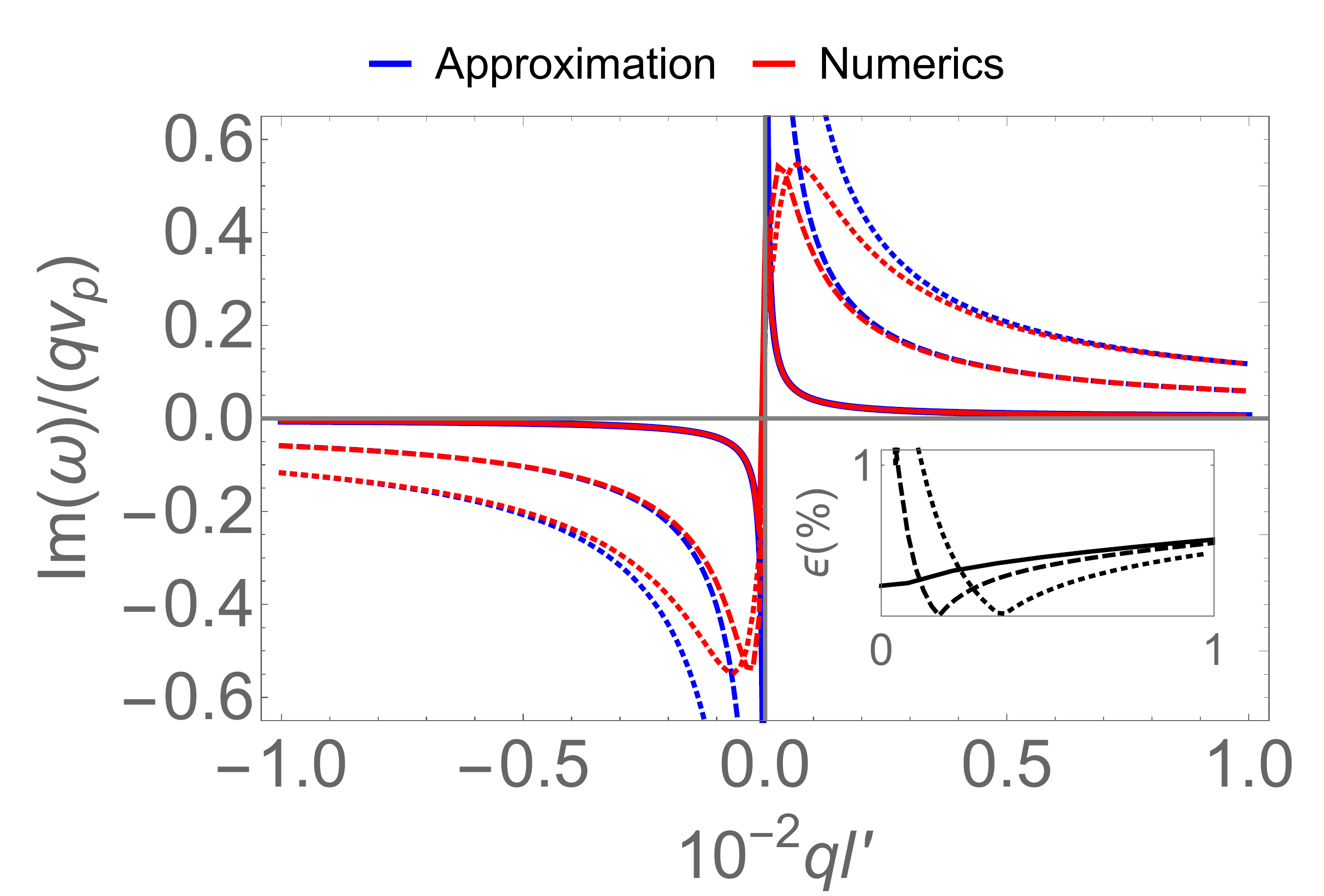}
b) \includegraphics[width=0.45\textwidth]{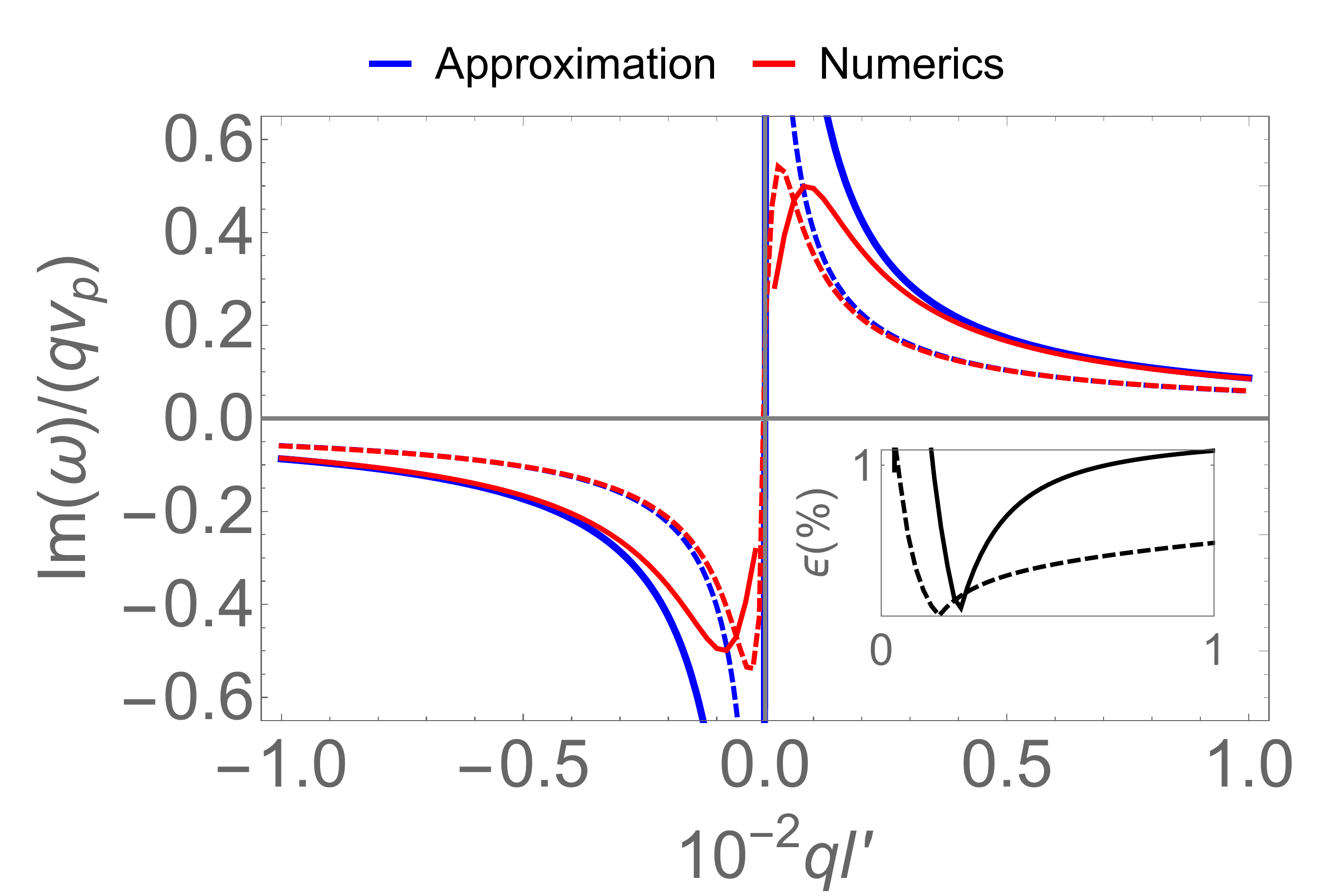}
c) \includegraphics[width=0.45\textwidth]{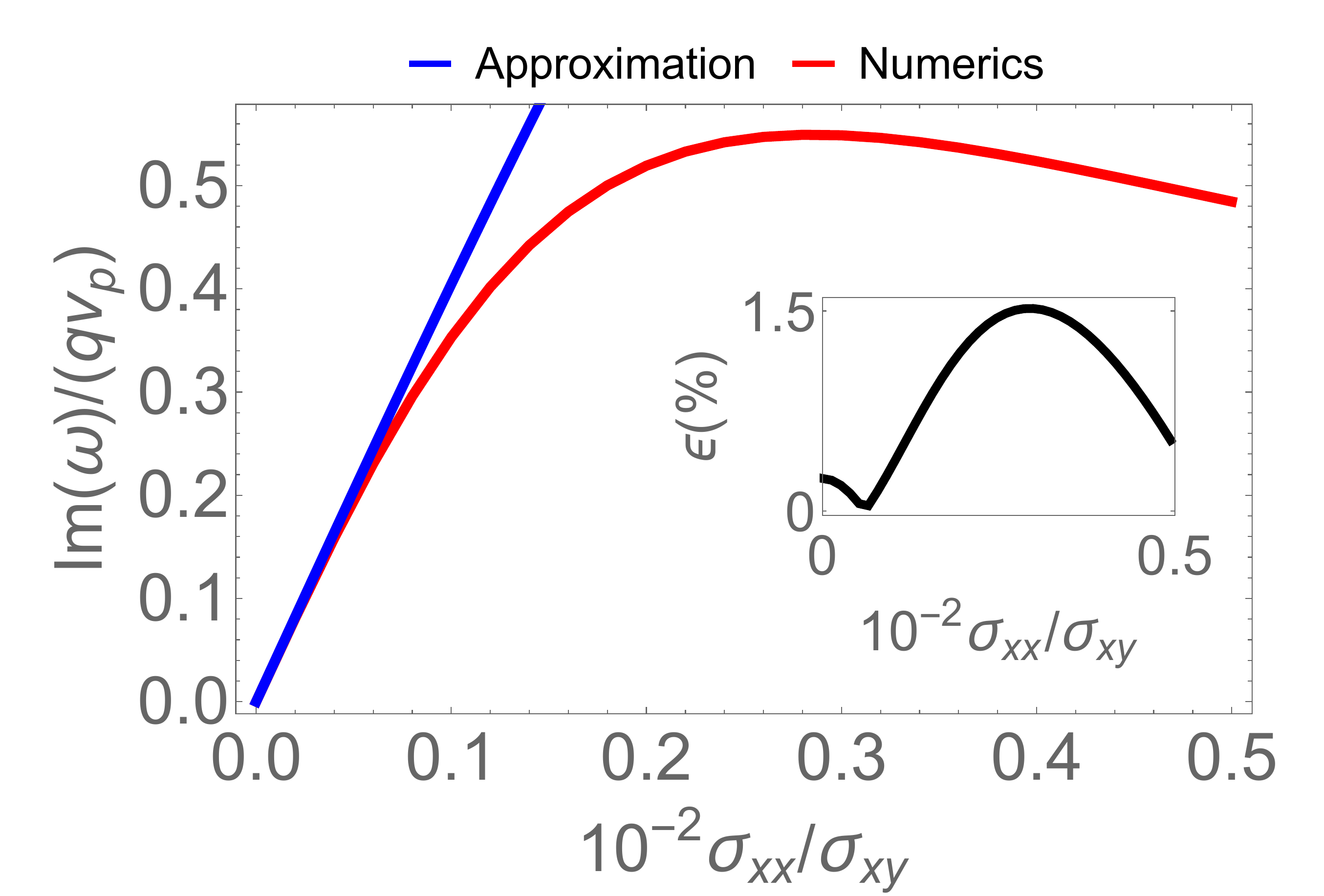}
\caption{\label{fig:dissipvsq} Attenuation of the EMPs: $\mathrm{Im}(\omega)$.
The red lines are computed numerically by diagonalizing (\ref{eq:lambdaDmat}), while the blue lines are obtained by using the complex-valued length $l$  (\ref{eq:l-appr}) in the far-field EMP eigenfrequency (\ref{eq:eigenfreq-classical-cutoff}), with the appropriate velocity dependent on the electrostatic configuration.
In the insets, we plot the percentage error made in $\mathrm{Re}(\omega)$ by this approximation.
In a) and b),  we plot $\mathrm{Im}(\omega)$ as a function of the wavevector $ql'$ at a fixed value of the diagonal conductivity.
In a), we do not include metal gates and use (\ref{eq:G-nogates}). The solid, dashed and dotted lines are obtained for $\sigma_{xx}/\sigma_{xy}=10^{-4}$, $\sigma_{xx}/\sigma_{xy}=10^{-3}$ and $\sigma_{xx}/\sigma_{xy}=2\times 10^{-3}$, respectively. 
In b), we show the effect of a top gate and use (\ref{eq:approximate-GF}). We consider $\sigma_{xx}/\sigma_{xy}=10^{-3}$ and plot with a solid line the value of $\mathrm{Im}(\omega)$ including a top gate at distance $d=10l'$; the dashed line is shown for comparison and is obtained without gates.
In c), we plot $\mathrm{Im}(\omega)$ as a function of the ratio of diagonal to off-diagonal conductivity $\sigma_{xx}/\sigma_{xy}$ and we use $ql'=10^{-3}$.
  }
\end{figure}

In Fig. \ref{fig:dissipvsq}a), we show how $\mathrm{Im}(\omega)$ varies as a function of the wavevector $q$ for different values of $\sigma_{xx}/\sigma_{xy}$.
Note that the dissipative term $\mathrm{Im}(\Lambda_D)$ is proportional to the ratio of two small parameters $\sigma_{xx}/(\sigma_{xy}ql')$, and so it is not necessarily small. Therefore, even for small values of the diagonal conductivity, one cannot generally neglect the redistribution of charges in the bulk due to $\sigma_{xx}$ \cite{Volkov}. However, the matrix element $(\mathcal{D}\Lambda)_{00}=0$ and so in the long wavelength and small dissipation limit (and  when $\sigma_{xx}/(\sigma_{xy}ql')\lesssim1$), Eq. (\ref{eq:eigen-freq-classical-lw}) still gives a good estimation of $\mathrm{Re}(\omega)$.
The dependence of $\mathrm{Im}(\omega)$ on the ratio of diagonal to off-diagonal conductivities $\sigma_{xx}/\sigma_{xy}$ obtained for $ql'=10^{-3}$ is shown in Fig. \ref{fig:dissipvsq}c).  \\

Our numerical solution  can be interpreted by considering the analytical solution of a closely related problem, provided by Volkov and Mikhailov \cite{Volkov}.
They consider a very sharp edge, modeled by $\underline{\sigma}(\textbf{r})\propto\Theta(x)$, and a frequency dependent diagonal conductivity of the form $ \sigma_{xx}(1+i\omega T)$, with $T$ being a characteristic scattering time. The imaginary part of $ \sigma_{xx}$ is related to the kinetic inductance of the QH material.

In free space and for long wavelengths, they calculated a EMP propagation velocity $\omega/q\approx-2v_p \log(|q| l)$, where, most importantly, $l$ is a complex number with the units of length: the lifetime of the EMP is then parametrized by the imaginary part of $l$.
In particular, they find 
\begin{equation}
l\propto-i\frac{v_p}{\omega}\frac{\sigma_{xx}}{\sigma_{xy}}(1+i\omega T).
\end{equation}

We remark that the presence of an imaginary part of $\sigma_{xx}$  in their treatment is required to avoid singularities of the Coulomb interaction kernel. In contrast, if one considers the smoother profile of the conductivity (\ref{eq:conductivity-erf}), varying from zero to the bulk value over a finite length $l'$, the problem is well-defined even for  $\mathrm{Im}(\sigma_{xx})=0$, as discussed in Sec. \ref{sec:field_distribution}. 
In this case, we find that $\mathrm{Im}(\omega)$ can be well approximated by using the far-field eigenfrequency $\omega$ in Eqs. (\ref{eq:eigenfreq-classical-cutoff}) and (\ref{eq:inter_edge_v}), with the complex-valued length
\begin{equation}
\label{eq:l-appr}
l\approx c_0 \left( l'- i\pi\frac{v_p}{\omega_0}\frac{\sigma_{xx}}{\sigma_{xy}} \right)\equiv l_0-i l_1,
\end{equation}
where $\omega_0\equiv\omega(\sigma_{xx}\rightarrow 0)\in\mathbb{R}$, and $c_0$ is the dimensionless constant defined in Eq. (\ref{eq:c0-erf}).

From Fig. \ref{fig:dissipvsq}a) and \ref{fig:dissipvsq}c), we observe that this estimation works reasonably well when $l_0\gtrsim l_1 $, i.e. $\sigma_{xx}/\sigma_{xy}\lesssim ql'$, and that it overestimates dissipation otherwise. 
In particular, when $\sigma_{xx}/\sigma_{xy}=10^{-4}$, the agreement is excellent in the range of wavelengths considered. 
%
%From the figure, we observe that in the range of wavelength considered the agreement is excellent when $\sigma_{xx}/\sigma_{xy}=10^{-4}$. 
For higher values of $\sigma_{xx}/\sigma_{xy}$, the approximation becomes worse at short wavevectors $ql'$. In fact, for small values of $ql'$, the numerical analysis suggests a finite value of attenuation, while the approximation scales as $\sim 1/q$ because of the divergence of $l_1=\mathrm{Im}(l)$ in Eq. (\ref{eq:l-appr}), and so the approximation overestimates the attenuation of the EMPs.
Also, in the range of parameters considered, the real part of the propagation frequency $\mathrm{Re}(\omega)$ does not change appreciably.\\

Note that the effect of metal electrodes can be straightforwardly included by appropriately modifying the integrand in the definition of $\Lambda$  (\ref{eq:lambda_kernel}).
When $\sigma_{xx}/(\sigma_{xy}ql')\lesssim 1$, and as long as the distance $d$ of the metal gate from the EMP satisfies $d\gg l'$, we find that one can well estimate the eigenfrequency $\omega$ by using the complex-valued length $l$ in (\ref{eq:l-appr}).
% also in the presence of metal gates at distance $d\gg l'$ from the EMP. 
In this case, the image charge at the electrodes needs to be included by appropriately adjusting $\omega_0$. For example,  for a top gate, we modify the EMP velocity (\ref{eq:inter_edge_v}) by using  the Green's function in (\ref{eq:approximate-GF}) instead of (\ref{eq:G-nogates}).

%The effect of metal electrodes can be straightforwardly included by appropriately modifying the matrix $\Lambda$ in Eq. (\ref{eq:lambdaDmat}). When $\sigma_{xx}/(\sigma_{xy}ql')\lesssim1$, we find that the simple model of Sec. \ref{sec:dissipation} applies also in this case, provided that $d\gg l'$ and that the EMP velocity is appropriately modified, see Sec. \ref{sec:EMP-gated}.  

In Fig. \ref{fig:dissipvsq}b), we show how the attenuation $\mathrm{Im}(\omega)$, obtained for $\sigma_{xx}/\sigma_{xy}=10^{-3}$ and $d=10l'$, varies as a function of the wavevector $q$  (solid lines). 
Comparing to the dissipation of the EMPs in free space (dashed lines), we observe that the  lifetime of the excitations is generally reduced by the interaction with the image charge, in agreement with the analysis of Volkov and Mikhailov \cite{Volkov}.\\

\subsubsection{\label{sec:dissipation}Quality factor of QH resonators}

To conclude this section about dissipation, we now discuss how a finite value of $\sigma_{xx}$ degrades the performance of the devices.
We restrict our analysis to QH materials with  abruptly terminated edges and a filling factor $\nu=1$, and so  the conductivity profile (\ref{eq:conductivity-erf}) is expected to be appropriate, see Appendix \ref{sec:quantum-field}. 
In particular, we focus on a QH droplet of perimeter $L_y$ and we consider for simplicity an electrostatic configuration that does not break the translational invariance in the $y$-direction, i.e. the direction of propagation of the excess charge density $\rho$.
In this case, the droplet supports plasmonic excitations that satisfy periodic boundary conditions for $\rho$ in $y$. The periodicity of $\rho$ restricts the allowed values of the wavevector to $q=2\pi n_q/L_y$, where $n_q\in \mathbb{N}$ is the wavenumber, and so the EMP eigenfrequency is quantized.

%Because at the resonant frequency obtained by evaluating  the dispersion relation (\ref{eq:eigenfreq-classical-cutoff}) at $q=2\pi n_q/L_y$, this circuit produces a large current for  a small voltages applied to 
In this paper, we refer to this device as a QH resonator, where the resonant frequency is obtained by evaluating  the dispersion relation (\ref{eq:eigenfreq-classical-cutoff}) at $q=2\pi n_q/L_y$.
We remark that a QH resonator differs from conventional microwave resonators, where the electromagnetic field propagates back and forth in the bulk of the material instead of chirally along the perimeter; a QH resonator can be designed to mimic a conventional one by appropriately breaking the translational invariance in the $y$-direction, see e.g. Fig. 3 of \cite{HITLpt1}.\\

In lossy microwave resonators, the resonator frequency $\omega_R\equiv\omega\left(q= 2\pi n_q/L_y \right)$ becomes complex-valued. The imaginary part of $\omega_R$ is related to the attenuation in the system and is often parametrized by the dimensionless quality factor $Q$, defined as \cite{Pozar}
\begin{equation}
\omega_R=\omega_0\left(1+\frac{i}{2Q}\right).
\end{equation}

To obtain an intuitive equation for $Q$, we further simplify the fitting formula of the complex-valued eigenfrequency $\omega$ discussed in Sec. \ref{sec:dissipation-field}  by expanding $\omega$  around the  real part $l_0$ of $l$ in (\ref{eq:l-appr}).
The result obtained with this expansion agrees reasonably well with the numerics in the same parameter region for which the use of the complex-valued length $l$ is appropriate, i.e. $\sigma_{xx}/\sigma_{xy}\lesssim ql'$.

With this simplification, the $Q$ factor reduces to
\begin{equation}
Q\approx-\frac{G\left(l_0,q,0\right)}{2l_1 \partial_{l_0}G\left(l_0,q,0\right)}=\frac{1}{2}\frac{\sigma_{xy}}{\sigma_{xx}}\left(\frac{\omega_0 \tau^*}{2\pi}\right)^2,
\end{equation}
where the timescale $\tau^*$ is defined by
\begin{equation}
\tau^*=\frac{1}{v_p}\sqrt{\frac{L_y}{n_q}}\left(-c_04\pi^2\epsilon_S\partial_{l_0}G\left(l_0,q,0\right)\right)^{-1/2}.
\end{equation} 
Here, $\tau^*$ represents the time required for an excitation with  velocity $v_p$  to travel for an effective relaxation length given by the geometric mean of the characteristic lengths (in the $x$- and $y$-direction) over which the electric field varies. For example, in the long wavelength limit and  without external electrodes, $\tau^*\approx \sqrt{(L_y/n_q) l'}/v_p$.

Note that $\omega_0\tau^*\propto (L_y)^{-1/2}$, and so longer resonators have lower $Q$. For this reason, the meta-material construction presented in \cite{HITLpt1} is particularly appealing to implement long and low-loss transmission lines.

Also, if we consider a metal gate placed at a distance $d$ from the edge of the QH material, such that $qd\ll 1$, we find that the attenuation of the EMP increases marginally, but the resonance frequency of the resonator decreases drastically because of the regularization of the $\log(q)$ singularity in the EMP velocity. This consideration implies that the $Q$ factor is lowered  by the electrodes; for example, for a top gate, $\omega_0\tau^*\propto\log(1+d^2/l_0^2)$, and so $Q$ scales logarithmically with $d$.\\

To give a quantitative example, we consider a realistic QH  resonator in GaAs ($\epsilon_S^*\approx 8.7$) of perimeter $L_y=35\mathrm{\mu m}$ and wavenumber $n_q=1$, with a top gate placed at $d=3\mathrm{\mu m}$ from the EMP and under the effect of a magnetic field $B=0.1\mathrm{T}$. The electron density is chosen to have a filling factor $\nu=1$. In this case, we expect a resonance frequency $\omega_0/(2\pi)\approx 10.5\mathrm{GHz}$ and $\tau^*\approx 42.4 \mathrm{ns}$. Considering a diagonal resistivity of a few Ohms per square \cite{Stromer_Res,Briggs_Res,QAH-exp,QAH-exp2}, experimentally achievable for a wide range of materials at the $\mathrm{mK}$ temperatures required for spin qubit operations, we can use  $\sigma_{xx}/\sigma_{xy}= 10^{-4}$, and we obtain $Q\approx 10^3$.

\section{\label{sec:coupling}Coupling to qubits}

\begin{figure}
\includegraphics[width=0.4\textwidth]{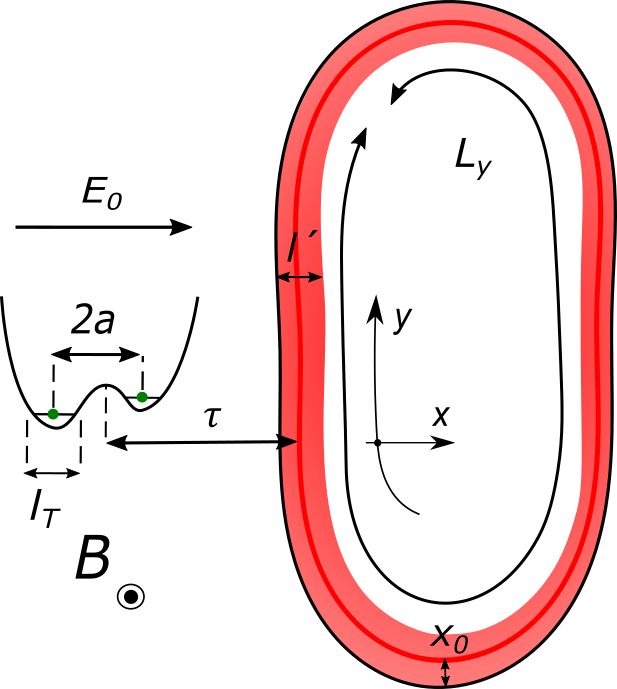}
\caption{\label{fig:qubit-resonator} QH resonator and ST qubit. To define the ST qubit, we consider two electrons in a double dot potential perpendicular to the edge of a QH resonator of perimeter $L_y$. The two dots are separated by a distance $2a$  and an external electric field $E_0$ is applied in the direction connecting the dots; the eigenstates of each dot are characterized by a confinement length $l_T$, see Eq.  (\ref{eq:lt-def}). The QH edge state  extends into the bulk for a length $l'=0.75 l_B$ (\ref{eq:lp-est}) and its center of mass (red solid line) is at a distance $\tau$ from the center of mass of the double dot; $\tau$ accounts for the shift $x_0=1.15l_B$ (\ref{eq:x0-est}) of the EMP charge density into the bulk of the QH material. An homogeneous magnetic field $B$ is applied in the direction perpendicular to the plane.}
\end{figure}

%\subsection{Qubit model \label{sec:qubit_model}}
There are several different proposals for coupling QH edge states and solid state qubits, e.g. via tunnel \cite{Loss} or Coulomb interactions \cite{Doherty2}.
In this section, we focus on the latter approach, and, in particular, we quantify the electrostatic coupling between the EMPs and  semiconductor qubits.
The coupling strongly depends on the type of qubit chosen and on its design, in particular qubits with higher susceptibility to the electric field have a larger coupling constant.

Here, for example, we analyze only singlet-triplet (ST) qubits and we consider the setup  shown in Fig. \ref{fig:qubit-resonator}. 
% to model them, we follow closely Ref. \cite{Burkard-Divincenzo}.

The model of the ST qubit is discussed in detail in Appendix \ref{sec:qubit_model}.
The qubit is defined by a double-well potential in the $(x,y)$ plane, where the centers of mass of the two dots are displaced by  $2a$ in the $x$-direction with respect to each other. In particular, we consider the quartic potential \cite{Burkard-Divincenzo}
\begin{equation}
\label{eq:quartic-potential}
V_C(x,y)=\frac{m\Omega^2}{2}\left(\left(\frac{x^2-a^2}{2a}\right)^2+y^2\right),
\end{equation}
where $m$ is the effective mass of the material and the frequency $\Omega$ quantifies the confinement strength.
A magnetic field $B$ is applied in the $z$-direction and an electric field $E_0$ is applied in the direction parallel to the two wells. 
$E_0$ creates a dipole moment between the two dots, which results into a finite detuning energy $\Delta$, i.e. a shift of the zero-point energy of the two dots. 
The field $B$ modifies the characteristic  length $l_T$ over which electrons are confined. This length is given by
\begin{equation}
\label{eq:lt-def}
l_T=\sqrt{\frac{\hbar}{m\Omega_T}},
\end{equation}
where we define the Fock-Darwin frequency
\begin{equation}
\label{eq:omega-T}
\Omega_T=\sqrt{\Omega^2+\frac{\omega_c^2}{4}},
\end{equation}
and the cyclotron frequency $\omega_c=eB/m$.
Also, we introduce the dimensionless constant
\begin{equation}
\label{eq:beta-def}
\beta=\left(\frac{\omega_c}{2\Omega_T}\right)^2=\left(1+\left(\frac{2\Omega}{\omega_c}\right)^2\right)^{-1}\in [0,1),
\end{equation}
which parametrizes the ratio of magnetic and harmonic confinement energy.

We consider a strongly depleted regime in which there are only two electrons  in the double dot. 
We then choose as computational states the usual antisymmetric (singlet) and symmetric (triplet) combinations of spins in the two dots; the energy gap between these states is given by the exchange interaction $J_{\text{ex}}$.
For simplicity,  we also neglect the effects of the spin-orbit coupling and of a magnetic field gradient, and so the singlet and triplet subspaces are decoupled.
In this case, $J_{\text{ex}}$ is obtained by combining Eqs. (\ref{eq:exchangeJ}), (\ref{eq:vint_NO}) and (\ref{eq:matrix elements})  and can vary over a few orders of magnitude for different qubit designs. 
In the following, we restrict the analysis to the value of the design parameters that guarantee an exchange energy in the microwave domain, i.e. $J_{\text{ex}}\lesssim 15 \mathrm{GHz}$.

The center of mass of the qubit is placed at a distance $\tau$ from the center of mass of the EMP of a QH resonator of perimeter $L_y$, see Fig. \ref{fig:qubit-resonator}.
The QH material is assumed to have a filling factor $\nu=1$ and we only examine the coupling to the lowest mode of the resonator, with wavenumber $n_q=qL_y/(2\pi)=1$. 
Also, we study the amplitude of the coupling between the qubit and the electric field evaluated in the cross section $y=0$, shown with orange lines in Fig. \ref{fig:em-fields}a).
%To maximize the coupling, we place the qubit in the antinode of the resonator.

Note that $\tau$ includes a shift $x_0$ (\ref{eq:x0-est}) of the EMP charge density into the bulk of the QH material which is caused by the  quantum corrections discussed in Sec. \ref{sec:field_distribution} and in Appendix \ref{sec:quantum-field}. The length $\tau$ is also assumed to be sufficiently large for the tunnel coupling to be unimportant and, for this reason, we only focus on the electrostatic coupling.
%The qubit is assumed to lie in the antinode of the electric field to maximize the coupling; we only examine the coupling to the lowest mode of the resonator, i.e. $n_q=1$.  Also,  we restrict our analysis to QH materials with filling factor $\nu=1$. 
%in the antinode of the 
%We consider a QH resonator of length $L_y$ and we place the qubit in the antinode of the electric field to maximize the coupling; we only examine the coupling to the lowest mode of the resonator, i.e. $n_q=1$.  Also,  we restrict our analysis to QH materials with filling factor $\nu=1$.
%The length $\tau$ characterizes the distance between the center of mass of the qubit and of the EMP, and it is assumed to be sufficiently large for the tunnel coupling to be unimportant; for this reason, we only focus on the electrostatic coupling.
The inductive coupling between the qubit and the magnetic field arising from the current flowing in the transmission line is also neglected. This is justified because the magnetic field generated by the EMP is of the order of few nano Tesla, and this results in a coupling strength of  a few $\mathrm{kHz}$. \\

We begin this section by examining the coupling between the ST qubit and the EMP in a lossless QH resonator. We provide a perturbative analysis of the interactions in Sec. \ref{sec:PT-coup} and a more detailed calculation in Sec. \ref{sec:Hartree-text}.  
We then consider the lossy resonators described in Sec. \ref{sec:dissipation-field},  and in Sec. \ref{sec:strong-coupl} we discuss the possibility of attaining strong resonator-photon coupling with these systems.

\subsection{Perturbation theory \label{sec:PT-coup}}

We now introduce an intuitive model useful to understand qualitatively the Coulomb coupling between EMPs and ST qubits.

First, we remark that in the absence of spin-orbit coupling and magnetic field gradient, an electric field does not couple the singlet and triplet subspaces. For this reason, the qubit dipole moment must be longitudinal,  i.e. $\propto \sigma_z$, ($\sigma_z$ is the Pauli matrix acting on the qubit), and so the resonator-qubit coupling has a form that is desirable for qubit read-out and scalability  \cite{Susanne1, Susanne2, Blais_longit}.

Also, the  qubit dipole moment depends on the externally applied electric field $E_0$. In particular, from the analysis presented in Appendix \ref{sec:qubit_model} (see Eqs. (\ref{eq:exchange-efield}), (\ref{eq:delta1}) and (\ref{eq:t2})), it follows that the non-detuned configuration ($E_0=0$) has no dipole moment, and so to the first order of perturbation theory, the qubit is not altered  by a homogeneous electric field $E$.
%
%that in the absence  the ST triplet quibit have a very low non-detuned configuration ($E_0=0$) is not altered by a homogeneous external electric field $E$, at least to the first order, because the qubit has no dipole moment.
%
%First, we stress that from Eqs. (\ref{eq:exchange-efield}), (\ref{eq:delta1}) and (\ref{eq:t2}), it follows that the non-detuned configuration ($\Delta=0$) is not altered by a homogeneous external electric field $E$, at least to the first order, because the qubit has no dipole moment.
 This property can be advantageous because it suppresses charge noise, however it also drastically reduces the electrostatic coupling with  transmission lines and resonators. \\

We discuss here two different procedures that can be followed to circumvent this problem and to obtain a finite photon-qubit interaction.

The first possibility is to use a non-homogeneous electric field. 
For example, one can consider an asymmetric structure, where one of the dots experiences a higher electric field than the other \cite{Doherty2}. 
In QH resonators, the electric field decays as $\sim 1/x$ in the direction perpendicular to the edge, see Eq. (\ref{eq:E_field_operator_far_field}), and so this asymmetry can be obtained  by placing the two dots perpendicularly to the resonator edge, as shown in Fig. \ref{fig:qubit-resonator}.
In this case, we can obtain a finite coupling also with no external electric field  $E_0=0$.

To have a simple model that captures the main physics of this device, we  Taylor expand the  electric field of the EMP in the $x$-direction (perpendicular to the edge) around the center of mass of the qubit, i.e. $E(x)\approx E(0)+\delta E x$, with $\delta E\equiv\left.\partial_x E\right|_{x\rightarrow 0}$, and we study the response of the qubit to the constant electric field gradient $\delta E$. 
We neglect here the spatial variation of the field in the other directions, which in the long wavelength limit vanishes at the qubit position, see Eq. (\ref{eq:E_field_operator_far_field}). 
If we consider $E_0=0$, the homogeneous component of the resonator field has no effect to linear order.
In contrast, $\delta E$ changes the double dot Hamiltonian (\ref{eq:qubit Hamiltonian-general}) by the addition of the quadratic term $-e\delta E x^2/2$ and, to the lowest order in $\delta E$, the exchange energy modifies as $J_{\text{ex}}\rightarrow J_{\text{ex}}+\delta  J_{\text{ex}}$, with
\begin{equation}
\label{eq:exchange-gradientefield}
\delta J_{\text{ex}}=\chi_t\frac{s}{\sqrt{2}(1-s^2)}  a^2 e \delta E.
\end{equation}
Here, $s$ is the overlap between the ground state wavefunctions of the two dots, i.e.
\begin{equation}
\label{eq:s-def}
s=e^{-a^2(1+\beta)/l_T^2},
\end{equation}
with $l_T$ and $\beta$ defined in Eqs. (\ref{eq:lt-def}) and (\ref{eq:beta-def}), respectively.
%we also introduce the dimensionless constant
%\begin{equation}
%\label{eq:beta-def}
%\beta=\left(\frac{\omega_c}{2\Omega_T}\right)^2=\left(1+\left(\frac{2\Omega}{\omega_c}\right)^2\right)^{-1}\in [0,1),
%\end{equation}
%which quantifies the ratio of magnetic and harmonic confinement energy.
The prefactor $\chi_t$ quantifies the  susceptibility of the qubit to a change in the tunnel coupling $t$ between the two dots; $\chi_t$ strongly depends on the qubit design. An explicit expression for $\chi_t$ is given in Eq. (\ref{eq:susc-tun}).

To find the interaction Hamiltonian, we now  quantize the electric field of the QH resonator as described in Appendix  \ref{sec:quantization}.
Considering, for simplicity, the far-field (in the sense of Sec. \ref{sec:EMP}) and long-wavelength asymptotic expression of $\textbf{E}(\textbf{r})$ in Eq. (\ref{eq:E_field_operator_far_field}),  we obtain
\begin{equation}
\label{eq:exchange-operator-gradientefield}
H_{\text{int}}=\frac{\hbar\gamma_1}{2} \sigma_z (\hat{a}^{\dagger}+\hat{a}),
\end{equation}
with $\hat{a}$ being the annihilation operator for a single boson in the cavity.
The coupling constant $\gamma_1$ is given by
\begin{equation}
\label{eq:gamma-grad-app}
\frac{\gamma_1}{2\pi}=-\sqrt{2}\frac{v_p}{L_y}\chi_t\frac{s}{1-s^2}  \frac{a^2}{\tau^2};
\end{equation}
$v_p$ is the characteristic plasmon velocity in Eq. (\ref{vp-scale}) evaluated at the filling factor $\nu=1$.

To have an estimation of $\gamma_1$, let us consider two weakly coupled dots;
%Two dots are weakly coupled when the  
% where $t,U_{\text{Ha,Fo}}\ll U_{\text{Hu,Ad}}\approx U_{\text{Hu}}/2$; 
in this case, Eq. (\ref{eq:gamma-grad-app}) simplifies to
\begin{equation}
\label{eq:gamma-grad-app-weak}
\frac{\gamma_1}{2\pi}\approx 2\sqrt{2}\frac{v_p}{L_y}\frac{t}{|U_{\text{Hu}}|}\mathrm{csch}\left(\frac{a^2}{l_T^2}(1+\beta)\right)  \frac{a^2}{\tau^2}.
\end{equation}
The energy $U_{\text{Hu}}$ is the on-site Hubbard energy,  which quantifies the Coulomb interactions caused by the  double occupation of a single dot. 
We consider the coupling between dots to be weak when $U_{\text{Hu}}$ is much greater than all the other energy contributions.
As discussed in Appendix \ref{sec:qubit_model}, using the quartic confinement potential (\ref{eq:quartic-potential}), one can find explicit expressions of $U_{\text{Hu}}$ and of the tunnel energy $t$ as a function of the qubit design parameters (i.e. $a,\Omega,B$). The result is obtained by combining Eqs. (\ref{eq:vint_NO}), (\ref{eq:Pmat}) and (\ref{eq:matrix elements}).

%Here, we define weak coupling as the limit for which $U_{\text{Hu}}$ is much greater than all the other energies; $U_{\text{Hu}}$ is the on-site Hubbard energy,  which quantifies the Coulomb interactions due to double occupancy of a single dot.

For example, for a realistic GaAs ($\epsilon_S^*\approx 8.7$) resonator of perimeter $L_y=20\mathrm{\mu m}$, the prefactor is $v_p/L_y\approx 2\mathrm{GHz}$; using also $t/|U_{\text{Hu}}|=0.1$,  $a=l_T$, $\tau=3a$ and $\beta= 0.01$, we obtain $\gamma_1/(2\pi)\approx 53\mathrm{MHz}$, comparable with the coupling strength in a strongly coupled spin-photon system \cite{Landig,Scarlino,Benito-exp}.

%for GaAs ($\epsilon_S^*\approx 8.7$), and considering a resonator of $40\mathrm{\mu m}$, the prefactor is $v_p/L_y\approx 1\mathrm{GHz}$, and thus this coupling can be quite large.
%\begin{equation}
%\gamma_1=\frac{e^2}{2\sqrt{2}\pi\hbar\epsilon_S L_y}\chi_t\frac{s}{1-s^2}  \frac{a^2}{\tau^2},
%\end{equation}

%To have an idea of the order of magnitude of this term, note that for GaAs ($\epsilon_S^*\approx 8.7$), and a resonator of $40\mathrm{\mu m}$, the prefactor is $v_p/L_y\approx 1\mathrm{GHz}$.

%We remark here the qubit dipole moment is longitudinal, i.e. $\propto \sigma_z$, and so the interaction Hamiltonian (\ref{eq:exchange-operator-gradientefield}) has a form that is desirable for qubit read-out and scalability  \cite{Susanne1, Susanne2, Blais_longit}.

It is also interesting to observe that there is a finite coupling to the electric field gradient $\delta E$ even if the dots are rotated and aligned parallel to the resonator edge. This coupling originates from the different magnetic field dependent phases between the wavefunctions of the two dots, and its magnitude is reduced, compared to Eq. (\ref{eq:gamma-grad-app}), by the multiplicative factor $\beta$ in Eq. (\ref{eq:beta-def}).\\

The second procedure to obtain a finite coupling in this setup is to move away from the sweet spot that suppresses charge noise and to include a small homogeneous electric field $E_0$ in the $x$-direction, see Fig. \ref{fig:qubit-resonator}.
The qubit then acquires a finite dipole moment and it becomes susceptible in the first order to the homogeneous (averaged) component of the electric field of the QH resonator $E(0)\equiv E$ \cite{Doherty1}.
Note that in this approach, the qubit is more vulnerable to charge noise; however, since $E$ is quite high, one can achieve a finite coupling strength even for small values of $E_0$, for which the qubit susceptibility to noise is still low.

Combining Eqs. (\ref{eq:exchange-efield}),  (\ref{eq:delta1}) and (\ref{eq:t2}), we find that the correction to $J_{\text{ex}}$ linear in $E$ is
\begin{equation}
\label{eq:exchange-detunedefield}
\delta J_{\text{ex}}=2e^2\left(\chi_t t_2+\chi_{\Delta} \Delta_1^2\right)  E_0 E.
\end{equation}
Here, $\chi_{\Delta}$ is defined in Eq. (\ref{eq:susc-det}) and is the susceptibility of the qubit to the detuning $\Delta$.
The quantities $\Delta_1$  and $t_2$ are the coefficients that relate the detuning $\Delta$ and the tunnel energy $t$ to the total homogeneous electric field ($E_{\text{tot}}=E_0+E$), respectively; explicit equations for $\Delta_1$  and $t_2$ are given in Eq. (\ref{eq:delta1-t2_design-par}).

%Using Eqs. (\ref{eq:exchange-efield}), it is immediate to obtain the correction to the exchange energy  
%\begin{equation}
%\label{eq:exchange-detunedefield}
%\delta J_{\text{ex}}=2e^2\left(\chi_t t_2+\chi_{\Delta} \Delta_1^2\right)  E_0 E.
%\end{equation}
%The quant(\ref{eq:delta1}) and (\ref{eq:t2})

Using  Eq. (\ref{eq:E_field_operator_far_field}), one obtains a longitudinal interaction Hamiltonian as in Eq. (\ref{eq:exchange-operator-gradientefield}), with coupling strength, which we will now call $\gamma_2$, given by
\begin{equation}
\label{eq:gamma-electricfield-app}
\frac{\gamma_2}{2\pi}=-4\frac{v_p}{L_y}\left(\chi_t t_2+\chi_{\Delta} \Delta_1^2\right)\frac{e E_0}{\tau}.
\end{equation}
%\begin{equation}
%\gamma_2=\frac{e^3 E_0}{\pi\hbar\epsilon_S L_y \tau}\left(\chi_t t_2+\chi_{\Delta} \Delta_1^2\right).
%\end{equation}

Considering  again two weakly coupled dots, $\gamma_2$ simplifies to 
\begin{equation}
\label{eq:gamma-electricfield-app-weak}
\frac{\gamma_2}{2\pi}\approx -\mathrm{sign}(E_0)6\sqrt{2}\frac{v_p}{L_y}\frac{t}{|U_{\text{Hu}}|}\mathrm{csch}\left(\frac{a^2}{l_T^2}(1+\beta)\right) \frac{b_0}{\tau},
\end{equation}
where we introduce the length $b_0$ defined by
\begin{equation}
\label{eq:b0-length}
b_0=\frac{e|E_0|}{m\Omega^2},
\end{equation}
%\begin{equation}
%\label{eq:b0-length}
%b_0=\frac{e|E_0|}{m\Omega^2}=\frac{e|E_0|l_T}{\hbar\Omega_T (1-\beta)},
%\end{equation}
which characterizes the shift of the single dot wavefunctions due to the external field $E_0$, see Eq. (\ref{eq:translation-h0}).

If we use the same realistic parameters used to estimate $\gamma_1$ in Eq. (\ref{eq:gamma-grad-app-weak}), i.e. $L_y=20\mathrm{\mu m}$, $t/|U_{\text{Hu}}|=0.1$,  $a=l_T$, $\tau= 3a$ and $\beta= 0.01$, and we consider the value of  $E_0$ for which $b_0=0.1a$,  we obtain $\gamma_2/(2\pi)\approx -\mathrm{sign}(E_0)48\mathrm{MHz}$.

Note  that a homogeneous electric field in the $y$-direction only shifts the qubit center of mass and its zero-point energy, and so in the rotated configuration, where the qubit is parallel to the QH edge, we obtain $\gamma_2=0$.

Also, we remark  that the total coupling $\gamma_{\text{tot}}$ is given by the sum of two contributions of  the same order of magnitude, i.e. $\gamma_{\text{tot}}=\gamma_1+\gamma_2$, one of which is externally tunable because of $E_0$. 
For example, by aligning $E_0$ to the electric field of the resonator (i.e. $E_0<0$), the total coupling increases, and for the parameters used, it reaches the value $\gamma_{\text{tot}}/(2\pi)\approx 100 \mathrm{MHz}$, while in the opposite limit ($E_0>0$), the coupling is minimized.
In devices with more qubits coupled to the same resonator, this tunability can be exploited to control selectively the coupling of each individual qubit \cite{Nigg,Mariantoni}. 

It is important to notice that both coupling terms are inversely proportional to the perimeter of the resonator $L_y$, and therefore shorter QH droplets have higher coupling to the qubit.
Additionally, as explained in Sec. \ref{sec:dissipation}, the  EMPs in shorter droplets have a higher lifetime, and, consequently, a higher $Q$-factor.
We remark again that longer transmission lines can be manufactured from shorter resonators by using the meta-material construction described in \cite{HITLpt1}.
\\

To conclude this analysis, we now comment on the effect of  metal  electrodes on the coupling constant.
Because of the electrodes, the electric field of the resonator changes as described in Sec. \ref{sec:fields-classical}; the Coulomb interactions in the double dot are also modified (see Eq. (\ref{eq:vint_NO})), but these corrections are negligible.
In particular, we consider here only two different configurations: a top and a side gate placed at distance $d$ from the center of mass of the EMP. A top gate decreases the averaged resonator field $E$ in the $x$-direction, but, when $d>\tau/2$, it increases the electric field gradient $\delta E$; for this reason, a top gate is more convenient to increase $\gamma_1$. In contrast, a side gate has the opposite effect: it increases $E$ and decreases $\delta E$, and so a side gate is advantageous to attain a higher value of $\gamma_2$, for which we require a finite $E_0$.
When $d\gg L_y$, the corrections to the electric field caused by the metal are negligible and Eqs. (\ref{eq:gamma-grad-app}) and (\ref{eq:gamma-electricfield-app}) are appropriate.  

\subsection{Hartree integral\label{sec:Hartree-text}}
The perturbative solution presented in Sec. \ref{sec:PT-coup} is expected to give a good estimation of the coupling strength in the far-field limit, i.e. when $\tau\gg a,l,l_T$ and $|q|\tau\ll 1$.
To verify the validity of  Eqs. (\ref{eq:gamma-grad-app}) and (\ref{eq:gamma-electricfield-app}), we find an effective Hamiltonian capturing the EMP-qubit coupling by computing the Hartree integral
 \begin{equation}
 \label{eq:Hartree_int}
H_{\text{int}}=\int d\textbf{r}\int d\textbf{r}'\rho_{\text{R}}(\textbf{r})G(\textbf{r},\textbf{r}',0) \rho_{\text{D}}(\textbf{r}'),
\end{equation}
and  by projecting the result onto the qubit subspace.  
Here, $G$ is the  Green's function of the electrostatic configuration chosen (and evaluated in the $z=0$ plane), $\rho_{\text{R}}$ is the charge density operator of the EMP in the resonator, obtained by selecting the term with the appropriate wavevector $q=2\pi n_q/L_y$ in  Eq. (\ref{eq:charge_density_operator}), and $\rho_{\text{D}}$ is the charge density of the double dot. We consider again a QH resonator with filling factor $\nu=1$ and wavenumber $n_q=1$.
The detailed solution of (\ref{eq:Hartree_int}) is presented in Appendix \ref{app:sub:EMP-QubitCoupling}.

This procedure accounts for the precise spatial profile of the electric field (and of its gradient) and it captures also near-field corrections; the resulting couplings $\gamma_{1,2}$ are given by
\begin{equation}
\label{eq:gamma1-full}
\frac{\gamma_1}{2\pi}=\frac{v_p}{L_y}\frac{\chi_ts a^2}{\sqrt{2}(1-s^2)}\left(\frac{2g(\tau)-g(\tau-a)-g(\tau+a)}{a^2}\right),
\end{equation}
and
\begin{equation}
\label{eq:gamma2-full}
\frac{\gamma_2}{2\pi}=\frac{v_p}{L_y} \left(\chi_{\Delta}\Delta_1^2+\chi_t t_2\right)eE_0\left(\frac{g(\tau+a)-g(\tau-a)}{a}\right).
\end{equation}
%\begin{multline}
%\label{eq:gamma2-full}
%\frac{\gamma_2}{2\pi}=-4\frac{v_p \sqrt{n_q}}{L_y}eE_0\left(\chi_{\Delta}\Delta_1^2 c_{\Delta}\left(\frac{g(\tau-a)-g(\tau+a)}{4}\right) \right. \\
%\left. \chi_t t_2 c_t \left(1-\frac{\tau^2}{a^2}\right)\left(\frac{g'(\tau-a)+g'(\tau+a)}{4}-\frac{g'(\tau)}{2}\right) \right).
%\end{multline}
%Here, we introduce the dimensionless prefactors $c_t=\frac{2}{3}\left(\frac{\tau^2}{a^2}-1\right)^{-1}$ and $c_{\Delta}=\left(\frac{a}{l_T}-\frac{3 l_T}{4a}\right)^{-1}$ and the function $g'(x)=\partial_x g(x,q)$, which depends on the electrostatic configuration considered.
The dimensionless function $g$ depends on the electrostatic configuration considered. In particular, in free space, it is related to the function $g_0$ (defined in Sec. \ref{sec:fields-classical} as the projection onto the $(x,y)$ plane of the EMP  potential $\mathcal{G}_0$) by the substitution $l'\rightarrow\lambda=\sqrt{l'^2+l_T^2}$, see Eqs. (\ref{eq:g-funct}) and (\ref{eq:G0-functnogate}).
When a top (side) gate are included, we have $g=g_0+g_{t (s)}$. The functions $g_{t,s}$ are  given in (\ref{eq:gfunct-gate}) and are obtained by using the substitution $l'\rightarrow\lambda$ in the $z=0$ limit of (\ref{eq:G-corr-exp-t-s}).
For simplicity of notation, we have dropped the explicit $q$ dependence in $g$.\\

\begin{figure}
a)\includegraphics[width=0.4\textwidth]{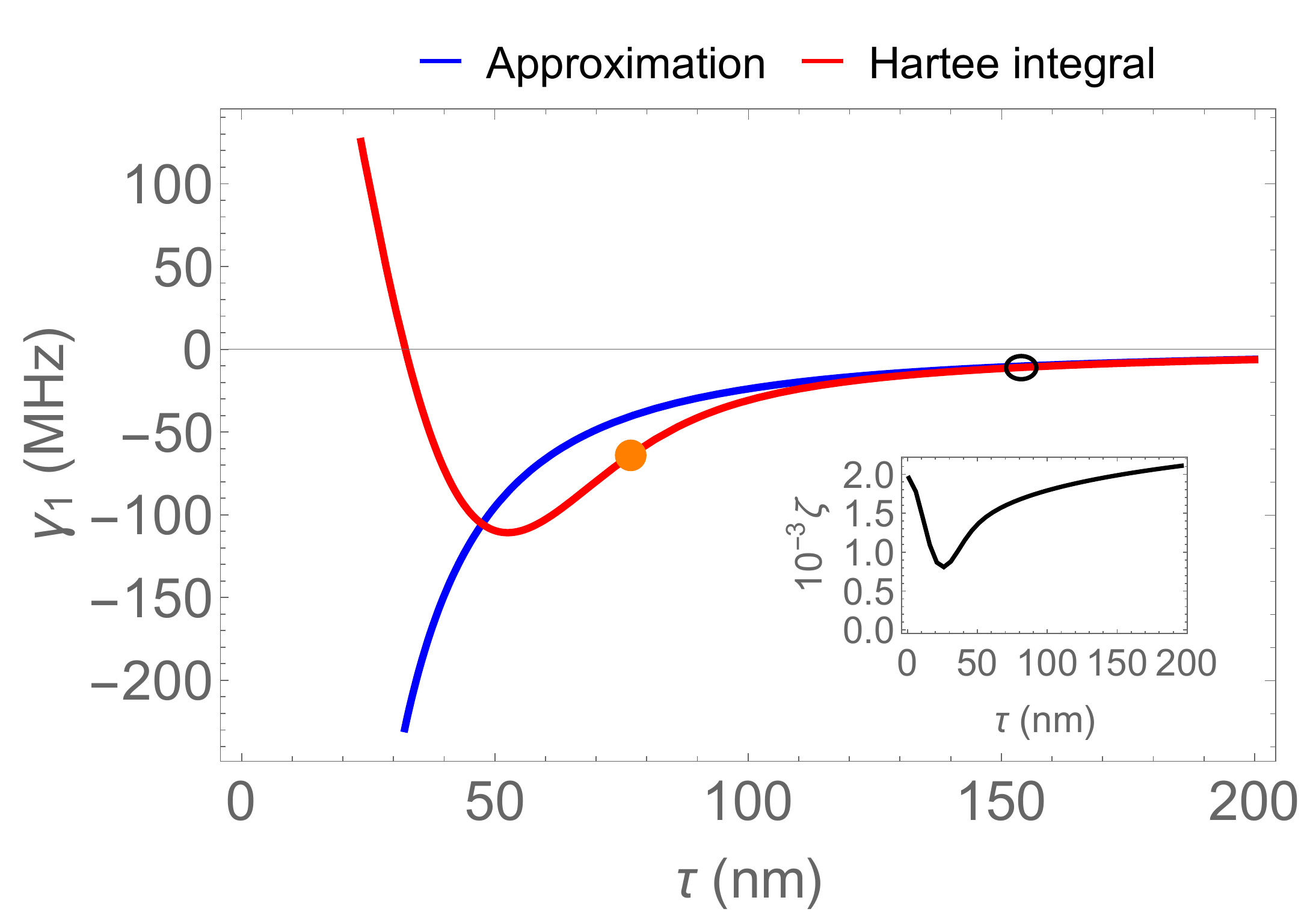}
b)\includegraphics[width=0.4\textwidth]{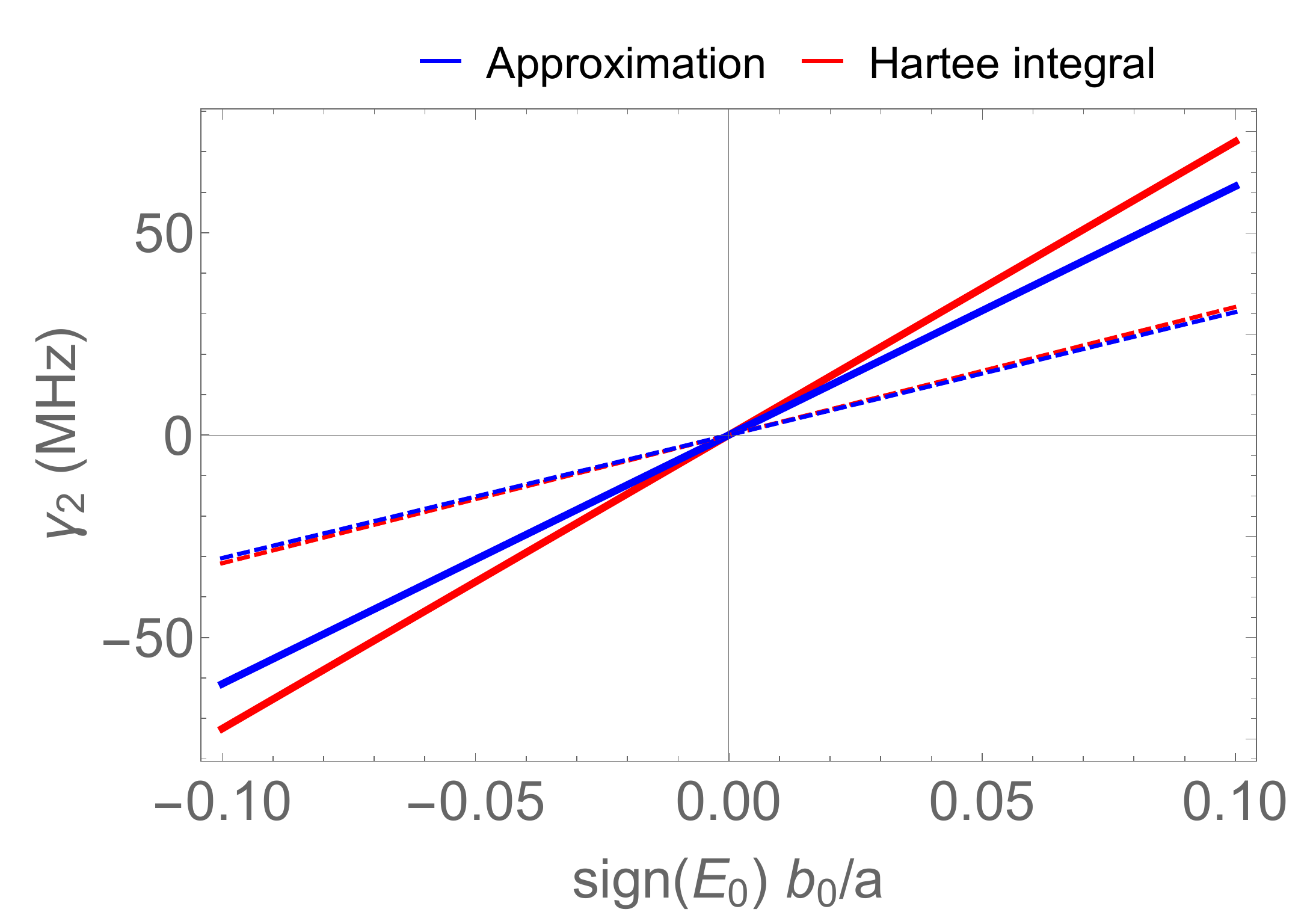}
\caption{\label{fig:coupling_grad_comparison} Coupling constants $\gamma_{1,2}$. For the plots, we use $2a=30\mathrm{nm}$, $\Omega=3.85 \mathrm{meV}$, $B=0.44\mathrm{T}$ and $L_y=27.7\mathrm{\mu m}$; we consider $l'=0.75l_B$ and $x_0=1.15 l_B$, as discussed in Sec. \ref{sec:field_distribution}. We also include the effect of a top gate placed at distance $d= 0.6\mathrm{\mu m}$ from the EMP. In a) we consider $E_0=0$. The blue line is obtained  using Eq. (\ref{eq:gamma-grad-app}), and the red line is obtained  using Eq. (\ref{eq:gamma1-full}). In the inset, we plot the dimensionless parameter $\zeta$, defined in Eq. (\ref{eq:zeta-leakage}), which quantifies the coupling between the computational and the non-computational subspace of the double dot.
In b) we show the dependence of the coupling $\gamma_2$ on a homogeneous electric field $E_0$. The length $b_0$ is defined in Eq. (\ref{eq:b0-length}) and it is proportional to  $|E_0|$; to include negative values of $E_0$, we rescale the horizontal axis by $\mathrm{sign}(E_0)$. The blue and red lines are obtained respectively by using Eqs. (\ref{eq:gamma-electricfield-app}) and (\ref{eq:gamma2-full}). 
The solid and dashed lines are obtained by using $\tau=\tau_{\text{s}}=x_0+a+l_T\approx 77\mathrm{nm}$, and $\tau=2\tau_{\text{s}}\approx 154\mathrm{nm}$. 
These values of $\tau$ are marked in a) by using an orange dot and a black circle, respectively. }
\end{figure}

In Fig. \ref{fig:coupling_grad_comparison}a), we show how the coupling $\gamma_1$ to the electric field gradient of the QH resonator  changes as a function of the distance $\tau$. 
For the plot, we consider a resonator with a perimeter $L_y= 27.7\mathrm{\mu m}$ and two dots very close to each other, with  $2a=30\mathrm{nm}$. We also consider an harmonic confinement potential  $\Omega=3.85 \mathrm{meV}$ and an external magnetic field $B=0.44\mathrm{T}$. 
For these parameters, the susceptibility to tunneling is $\chi_t\approx 0.8$, and  differs only slightly from the weak tunnel coupling expansion $\chi_t^{\text{weak}}\approx -4t/|U_{\text{Hu}}|\approx 0.6$.
We also include a top gate at a distance $d=0.6\mathrm{\mu m} $, which is required to slow the EMPs, but has no significant effect on the coupling. 
In this setup, the exchange energy and the resonance frequency of the resonator are both in the microwave domain: in particular, we find $J_{\text{ex}}\approx 12.2 \mathrm{GHz}$ and $\omega_R\approx 12.6 \mathrm{GHz}$, respectively.
The resonator frequency $\omega_R$ is calculated from the far-field result (\ref{eq:eigenfreq-classical-cutoff}) by using the Green's function (\ref{eq:approximate-GF}) in the EMP velocity (\ref{eq:inter_edge_v}).
The wavevector is $q=2\pi/L_y$ and the cut-off length is $l\approx 0.75 \times 0.53  l_B$; the two quantitative corrections to the magnetic length originate respectively from the quantum correction (\ref{eq:lp-est}) and from the spatial profile of the conductivity (\ref{eq:c0-erf}).

Also, the interaction with the resonator causes a finite coupling between the computational and the non-computational subspace of the double dot. This coupling is quantified by the dimensionless parameter $\zeta$, that is defined in Eq. (\ref{eq:zeta-leakage}) and is plotted in the inset of Fig. \ref{fig:coupling_grad_comparison}a). For the qubit designs considered here, we find that $\zeta$  is negligible and so the Hamiltonian (\ref{eq:exchange-operator-gradientefield}) provides a good description of the system.

From the figure, we observe  as expected that the two different approaches used for computing the coupling $\gamma_1$, i.e. Eqs. (\ref{eq:gamma-grad-app}) and (\ref{eq:gamma1-full}), differ when the qubit is close to the resonator edge, but they coincide in the far-field, when $\tau\gg a$. This limiting behavior can be easily understood considering that, except for the different length $l'\rightarrow \lambda$ in the definition of $g$, the combination of functions $g$ with different arguments in the parentheses in  Eq. (\ref{eq:gamma1-full}) is proportional to the discrete second derivative in $x$ of the EMP potential in the $(x,y)$ plane (\ref{eq:potential-gaussian}), and, consequently, to the discrete derivative of the electric field. This function reduces exactly to the continuum value of $\delta E$ when $a\rightarrow 0$.
In other words, from a detailed analysis, we find that the simple perturbative result for the exchange energy in Eq. (\ref{eq:exchange-gradientefield}) has the correct form, but the  continuous gradient $\delta E$ is replaced by its discrete analog.\\

More care is required when examining the coupling term  $\gamma_2$.
In this case, in fact, we find that the Hartree integral  and the perturbative treatment presented in Sec. \ref{sec:PT-coup} do not agree quantitatively.
In fact, a direct estimation of $\gamma_2$ from (\ref{eq:Hartree_int}) leads to Eq. (\ref{eq:gamma2-full-Hartree}). This equation differs from Eq. (\ref{eq:gamma-electricfield-app}) even in the far-field limit, where the two approaches should coincide.
The reason for this disagreement is discussed in detail in  Appendix \ref{app:sub:EMP-QubitCoupling}.
To summarize, this difference can be traced back to the explicit dependence on the averaged resonator field $E$ of the Fock-Darwin  wavefunctions in Eq. (\ref{eq:FD-wf-full}), which is neglected in the Hartree integral.
For this reason, the qubit susceptibility to $E$ obtained by this method differs from the one calculated in Sec. \ref{sec:PT-coup}, see Eqs. (\ref{eq:hartree-suscept}) and (\ref{eq:exchange-detunedefield}); the latter equation provides a more accurate estimation of the susceptibility.
Because the terms neglected in the Hartree integral are not expected to change the function in parentheses in (\ref{eq:gamma2-full-Hartree}), which is proportional to the discrete derivative of the EMP potential $V$, we adjust  the prefactor in Eq. (\ref{eq:gamma2-full-Hartree}) by using the ad-hoc substitution  shown in  (\ref{eq:subst-susc-E}); with this procedure, we obtain Eq. (\ref{eq:gamma2-full}).\\

In Fig. \ref{fig:coupling_grad_comparison}b), we show the correction to the coupling energy by including a small homogeneous electric field $E_0$. The parameters used in the plot are the same as in Fig. \ref{fig:coupling_grad_comparison}a) and we select the two different values of $\tau$ labeled by an orange and a hollow dot in the figure; both values of $\tau$ are large enough to guarantee a negligible overlap between the EMP and the qubit wavefunction, and so the tunnel coupling is unimportant. 
After the substitution (\ref{eq:subst-susc-E}), we observe that the results are in good agreement in the far-field (dashed lines), while they differ slightly in the near-field (solid lines).
We remark again that $\gamma_1$ and $\gamma_2$ can be in the same order of magnitude for small $E_0$, and so, depending on the sign of $E_0$, the total coupling $\gamma_{\text{tot}}=\gamma_1+\gamma_2$ can be significantly  increased or decreased.

%Comparing the two approaches, we find as expected  that Eqs. (\ref{eq:gamma-grad-app}) and (\ref{eq:gamma-electricfield-app}) both capture qualitatively the far-field behavior of the coupling.
%However, while Eqs. (\ref{eq:gamma-grad-app}) and (\ref{eq:gamma1-full}) are also in a excellent quantitative agreement, we find that Eq. (\ref{eq:gamma2-full}) differs from (\ref{eq:gamma-electricfield-app}) by the multiplicative factors $c_{t,\Delta}$.
%These prefactors are related to the different approximation scheme used; in particular, they originate from the projection of the orbital Hamiltonian $H_O$ (\ref{eq:orbital-hamilt}) onto a Fock-Darwin basis which depends on  the electric field of the resonator, as explained in Appendix \ref{app:qubit-Hamiltonian}. 
%These corrections are neglected in the Hartree integral approach, but they are expected to be relevant to quantitatively characterize $\gamma_2$, and thus we include them a posteriori by setting $c_{t,\Delta}\rightarrow 1$ in Eq. (\ref{eq:gamma2-full}).

\subsection{Strong EMP-qubit coupling \label{sec:strong-coupl}}
We can now discuss the possibility of attaining strong  coupling between the EMP and the qubit.  
The coupling is strong when $\gamma$ is larger than all the losses in the system.
The lifetime of ST  qubits (in GaAs) is often limited by the dephasing time, which is of the order of $100\mathrm{ns}$ \cite{Bluhm1}, but this lifetime can  be increased up to $200\mathrm{\mu s}$ by spin echo  \cite{Bluhm2}.
For the relevant frequency range, the estimated EMP lifetime is  of the order of $Q/\omega_R\sim 10^3/(1\mathrm{GHz})\sim 1\mathrm{\mu s}$, and so we consider this factor as the limiting timescale and we define the dimensionless ratio $\Gamma$ of coupling strength and attenuation in the resonator 
\begin{equation}
\Gamma\equiv \frac{\gamma}{\mathrm{Im}(\omega_R)}.
\end{equation}
When $|\Gamma|>1$, the resonator and the qubit are strongly coupled.

We now restrict our analysis to the case $E_0=0$, and we consider the near-field expression of $\gamma_1$ in Eq. (\ref{eq:gamma1-full}); the values of $\Gamma$ that we find here can be approximately doubled by a finite $E_0$.
Also, we include a top gate at distance $d$ and we obtain the complex resonator frequency $\omega_R$ by combining Eqs. (\ref{eq:eigenfreq-classical-cutoff}), (\ref{eq:approximate-GF}) and (\ref{eq:l-appr}).\\

\begin{figure}
a)\includegraphics[width=0.4\textwidth]{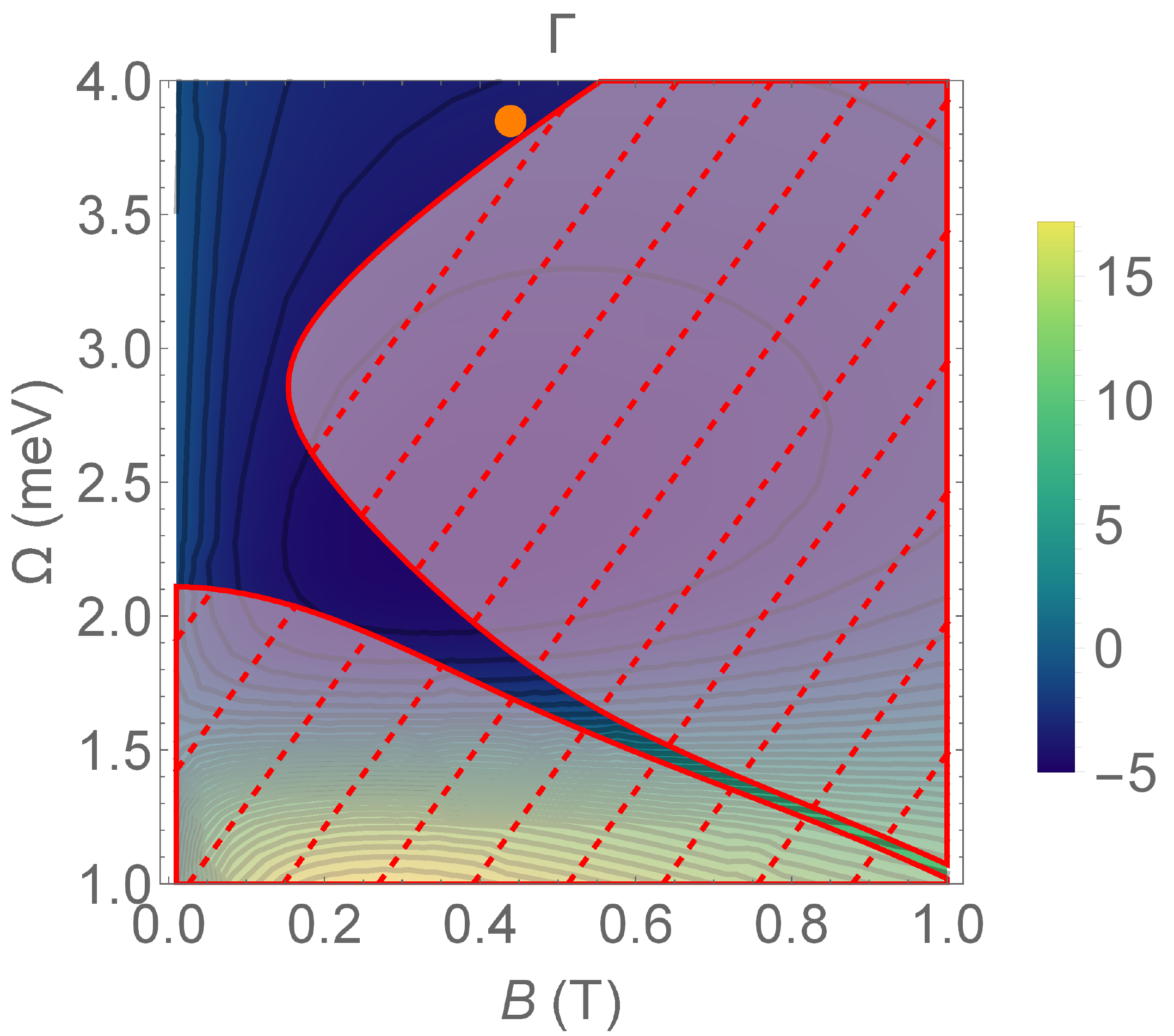}
b)\includegraphics[width=0.4\textwidth]{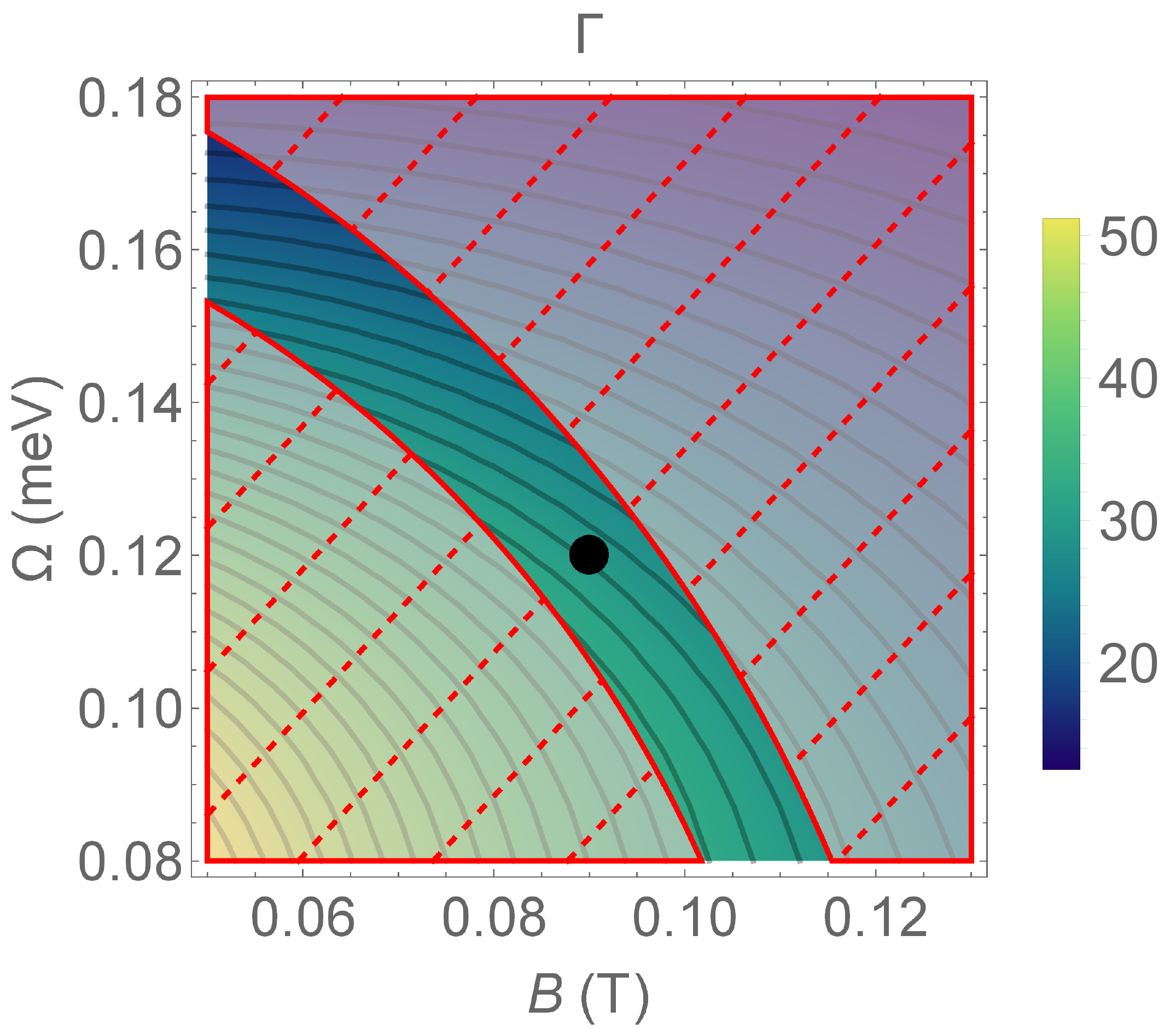}
\caption{\label{fig:coupling_gamma} Ratio $\Gamma$ of the coupling constant $\gamma_1$ and the attenuation of the QH resonator as a function of the magnetic field $B$ and of the harmonic confinement strength $\Omega$. 
In a) we consider two dots very close to each other, with  $2a=30\mathrm{nm}$ and we include a top gate at distance $d=0.6\mathrm{\mu m}$ from the system.
In b) we consider two dots  $2a=100\mathrm{nm}$ apart and we included a top gate at distance $d=2.2\mathrm{\mu m}$.
The orange and black dots in a) and b) mark the parameters used in Fig. \ref{fig:coupling_grad_comparison} ($B=0.44\mathrm{T}$ and $\Omega=3.85 \mathrm{meV}$) and in Fig. \ref{fig:coupling_gamma_Ly_dep} ($B=90\mathrm{mT}$ and $\Omega=0.12 \mathrm{meV}$), respectively.
In both figures, we consider a resonator of length $L_y=27.7\mathrm{\mu m}$ and we place the qubit at a distance $\tau=\tau_s=x_0+a+l_T$ from the QH edge ($x_0=1.15 l_B$). 
The value of $\tau_s$ depends on $B$ and $\Omega$; for the parameters corresponding to the orange (black) point, it is $\tau_s=77\mathrm{nm}$ ($\tau_s=242\mathrm{nm}$).
The regions of parameters marked by red dashed lines are characterized by an exchange coupling outside the microwave domain, i.e. $J_{\text{ex}}>15\mathrm{GHz}$.  }
\end{figure}

In Fig. \ref{fig:coupling_gamma}a), we show how $\Gamma$ changes as a function of the perpendicular magnetic field $B$ and of the harmonic confinement strength $\Omega$ for two quantum dots that are placed $2a=30\mathrm{nm}$ apart. 
For the plot, the distance between the qubit and the resonator edge is kept fixed to a minimal value $\tau=\tau_s=x_0+a+l_T$; also, the resonator has a perimeter $L_y=27.7\mathrm{\mu m}$ and we choose $d=0.6\mathrm{\mu m}$.
Note that the resonance frequency of the resonator depends on the magnetic field via the magnetic length; however, this choice of $L_y$ and $d$  guarantees that,  for all the values of $B$ considered,  the resonator frequency remains in the microwave domain, i.e. $\mathrm{Re}(\omega_R)\lesssim 15\mathrm{GHz}$.
Also, in the plot, we highlight the regions of parameters for which the exchange energy $J_{\text{ex}}$ takes values outside the  microwave domain and we exclude them from the discussion.
We observe that there is a large range of values of $\Omega$ and $B$ in the allowed region, for which $|\Gamma|$ is greater than one and so strong coupling is indeed possible.
As an example, in the figure, we marked with an orange dot the point corresponding to the orange dot in Fig. \ref{fig:coupling_grad_comparison}. 
For this choice of parameters and using the realistic value of diagonal conductivity $\sigma_{xx}/ \sigma_{xy}=10^{-4}$, we obtain $ \Gamma\approx -4$.\\

\begin{figure}
\includegraphics[width=0.4\textwidth]{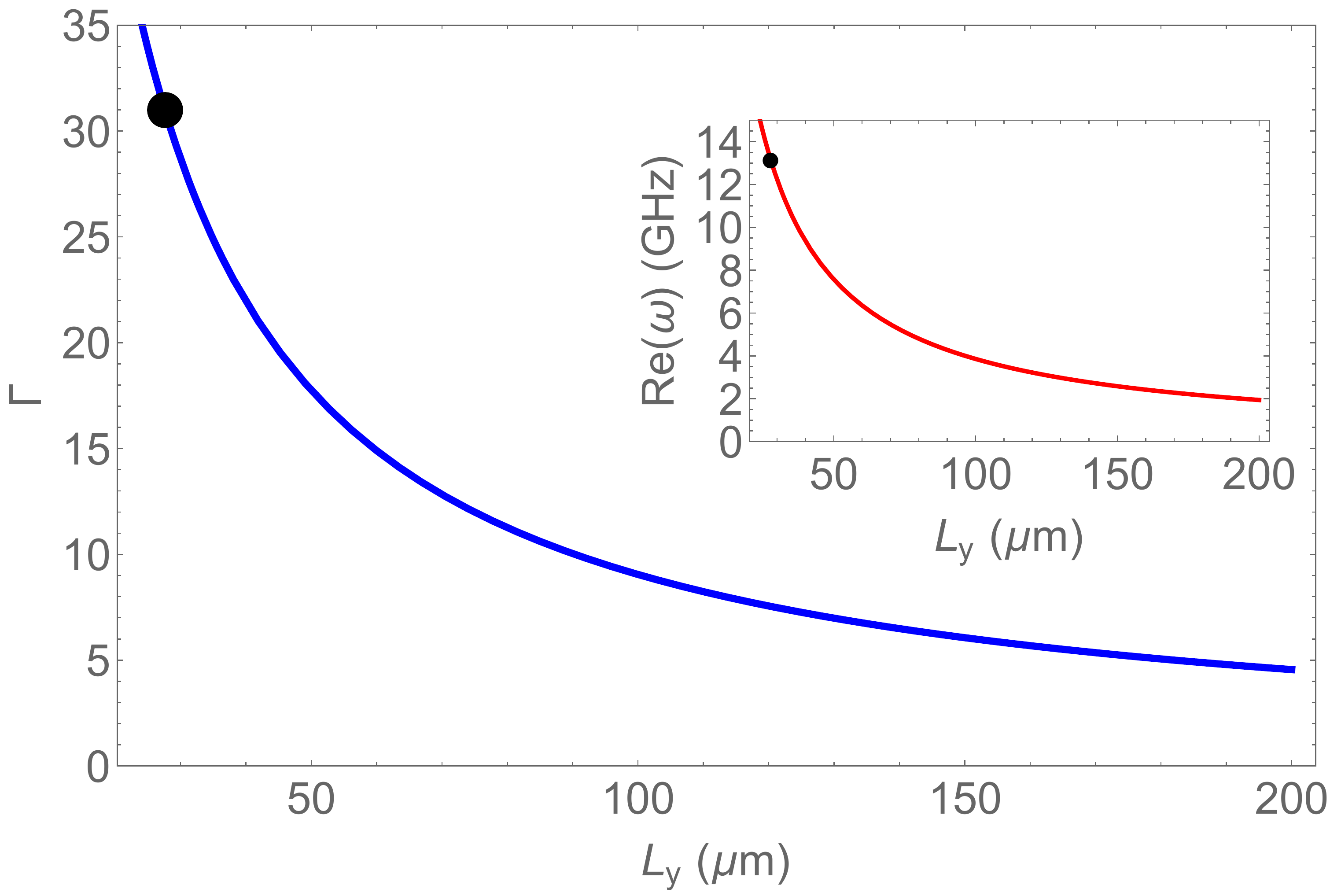}
\caption{\label{fig:coupling_gamma_Ly_dep} Ratio $\Gamma$ of the coupling constant $\gamma_1$ to the attenuation of the QH resonator as a function of perimeter of the QH resonator $L_y$. For the plot, we use $2a=100\mathrm{nm}$, $d=2.2\mathrm{\mu m}$, $\tau=242\mathrm{nm}$, $B=90\mathrm{mT}$ and $\Omega=0.12 \mathrm{meV}$.
In the inset, we show the dependence of the resonance frequency on $L_y$.
The black dots mark the special value $L_y=27.7\mathrm{\mu m}$: for this value of $L_y$, the value of $\Gamma$ here is equal to the value marked by the black dot  in Fig.\ref{fig:coupling_gamma}b).
}
\end{figure}

The ratio $\Gamma$ can be increased in different ways.
For example, one can use high quality QH materials with a lower value of $\sigma_{xx}$ or one can optimize the configuration of the electrodes  to improve the electrostatic coupling \cite{Benito-exp,Benito3,Petta}. 

Another possibility that can greatly enhance the interaction strength is to modify the qubit design, for example by lowering the harmonic confinement potential $\Omega$. In Fig. \ref{fig:coupling_gamma}a), we observe that by reducing $\Omega$, one can achieve higher values of $|\Gamma|$. To remain in the microwave domain, however, a finer tuning of $\Omega$ and $B$ is required.
This enhancement is related to the fact that when the value of $\Omega$ decreases (and $B$ is low enough to guarantee $\beta\ll 1$),  the confinement length $l_T$ increases. Because the susceptibility of the qubit to the electric field (and to its gradient) varies exponentially with $(a/l_T)^2$, the coupling $\gamma$ can be made significantly larger.
In this way, one can achieve strong coupling even in state-of-the-art qubit designs where the two dots are hundreds of nanometers apart.

As an example, in Fig. \ref{fig:coupling_gamma}b), we show how $\Gamma$ changes as a function of $\Omega$ and $B$ when the dots are placed at a distance of $2a=100\mathrm{nm}$. 
Note that the scale of  both $\Omega$ and $B$ are reduced by approximately  an order of magnitude compared to Fig. \ref{fig:coupling_gamma}a), and, for this reason, in this design a more careful tuning of the parameters is required to remain in the microwave domain.
For this plot, we used a resonator of perimeter $L_y=27.7\mathrm{\mu m}$ and a top gate placed $d=2.2\mathrm{\mu m}$ apart, for which we obtain $\mathrm{Re}(\omega_R)\lesssim 15\mathrm{GHz}$.
We observe now values of $\Gamma$ that are approximately an order of magnitude higher than in Fig. \ref{fig:coupling_gamma}a).
In particular, at the point marked by the black dot, i.e. $\Omega=0.12 \mathrm{meV}$ and $B=90\mathrm{mT}$, one obtains $J_{\text{ex}}=-6\mathrm{GHz}$, $\mathrm{Re}(\omega_R)=13.1\mathrm{GHz}$. When $\tau=\tau_s\approx 240\mathrm{nm}$ and $\sigma_{xx}/ \sigma_{xy}=10^{-4}$, the
coupling constant is $\gamma_1\approx 211 \mathrm{MHz}$ and the quality factor of the resonator is $Q\approx 965$: using  these values, one obtains the dimensionless coupling ratio $\Gamma\approx 31$. \\

We conclude this analysis by briefly discussing the dependence of $\Gamma$ on the perimeter $L_y$ of the resonator. This dependence  is shown in Fig. \ref{fig:coupling_gamma_Ly_dep}; for the plot we used the parameters marked by the black dot in Fig. \ref{fig:coupling_gamma}b).
As expected, $\Gamma$ decreases approximately as $\sim 1/L_y$ and it has the same scaling as $\mathrm{Re}(\omega_R)$. 
We observe that, despite this decrease, the coupling remains strong for resonators with a perimeter up to $100\mathrm{\mu m}$ long: this property can be exploited to entangle spin qubits over large distances.

\section{Conclusions and outlook}

We analyze the physics of QH devices and their coupling to qubits.
The electric current in these devices is carried by magnetoplasmonic excitations localized  at the edge of the QH material that propagate along the perimeter.
By using a semiclassical model capturing the main features of these excitations, we compute the spatial profile of the electromagnetic field in a variety of relevant cases and we justify the phenomenological model used in  \cite{HITLpt1}.
%Unlike previous treatments of sharp edges \cite{Volkov}, we do not require the presence of a finite complex-valued diagonal component of the conductivity tensor to avoid the divergence of the interaction kernel, but 
We consider a conductivity tensor which varies from zero to the bulk value over a length $l'$ of the order of the magnetic length. 
%For this reason our solution is qualitatively similar to the solution of the smooth edge model in  \cite{Glazman}; 
This  approach is justified by a quantum mechanical treatment of the EMPs, and, by comparing the results of the two calculations, we extract the value of $l'$.
We also account numerically for the dissipation due to a finite real-valued $\sigma_{xx}$ and we find a simple fitting formula to estimate the $Q$ factor of the QH resonator.

Using these results, we quantify the Coulomb coupling between EMPs and singlet-triplet qubits. In particular, we discuss two coupling schemes: the coupling to the gradient of the  electric field of the resonator and the coupling mediated by an externally applied electric field.
For both cases,  we find a simple analytic expression that can be used to estimate the coupling strength and we compare it to a more detailed solution based on the computation of the Hartree interaction integral.
We find that the coupling strength obtained for the two mechanisms is comparable and so the photon-qubit coupling can be tuned over a wide range of values.
Finally, we  discuss the possibility of achieving the strong photon-qubit coupling regime by comparing the strength of photon-qubit interaction to the estimated attenuation of the resonator; we conclude that strong photon-qubit coupling is indeed achievable with state-of-the-art qubit designs.\\

Some effects have been neglected in this analysis that might have an effect on the quantitative value of the interaction strength. 
For example, the ground state projection used to derive the double dot Hamiltonian is questionable for certain qubit designs. The neglected terms which mix high orbital states in the two dots, become important when the qubit confinement energy is lowered and thus we expect a quantitative change in the estimated value of the exchange energy  and of the susceptibility of the qubit. 
We believe that the higher orbital states do not change the qualitative picture discussed here, and that including their effect, interaction strengths of $100\mathrm{MHz}$ and higher can still be reached.
Also, we did not include other effects that are relevant for the qubit design such as spin-orbit coupling and magnetic field gradient, which are required to implement qubit gates. These additional terms in the Hamiltonian mix the singlet and triplet sectors, and thus they can potentially lead to some qualitative difference in the interaction strength, and to additional transversal coupling terms $\propto \sigma_{x,y} (\hat{a}^{\dagger}+\hat{a})$.
We believe that these extra terms can be made small by a careful qubit design, but we did not analyze them quantitatively.

\section{Acknowledgements}
The authors would like to thank D. Reilly, 
A.C. Doherty, A. Ciani, V. Langrock and F. Hassler for useful discussions.
This work was supported by the Alexander von Humboldt foundation.

\begin{appendix}

\section{\label{sec:quantum-field}Quantum corrections}

We discuss here the effect of quantum corrections in a QH material with filling factor $\nu=1$. We assume  that the boundary of the QH droplet is atomically defined, and so we neglect the edge reconstruction  caused by the static Coulomb interactions \cite{Chamon}.
In the long wavelength limit, the total EMP propagation velocity can be decomposed into a sum of two velocites: an electrostatic term and a quantum correction \cite{QEMP,Thouless_plasmon,MacDonald_plasmon,Mikhailov}. In a RPA analysis \cite{QEMP}, one finds that the former term is proportional to the matrix element of the Coulomb interactions 
\begin{equation}
\label{eq:classical-vrpa}
v_c(q)=\frac{e^2}{\hbar}\int dxdx'\left|\psi_0(x)\right|^2 G(x-x',q,0)\left|\psi_0(x')\right|^2,
\end{equation}
while the latter is given by the group velocity of a single electron wavepacket
\begin{equation}
v_q=\frac{1}{\hbar}\frac{\partial \epsilon_0(k)}{\partial k}.
\end{equation}
Here, $G$ is the electrostatic Green's function of the electrostatic configuration chosen, e.g. without gate it is given in  Eq. (\ref{eq:G-nogates}); $\epsilon_0(k)$ is the band dispersion of the lowest Landau level caused by presence of an edge and $\psi_0$ is the corresponding single-electron wavefunction. Both quantities need to be evaluated at the Fermi momentum $k_F$.
Also, in the RPA analysis \cite{QEMP}, the EMP charge density is proportional to the absolute value squared of the single electron wavefunction, i.e. $\rho\propto \left|\psi_0(x)\right|^2$.

To estimate these velocites, we can use the model Hamiltonian $H=\pmb{\pi}^2/(2m)$ and impose the boundary condition of vanishing wavefunction at $x=0$. Here, $m$ is the effective mass of the QH material and $\pmb{\pi}=\textbf{p}+e\textbf{A}$ is the dynamical momentum ($\textbf{p}=-i\hbar \nabla$ is the canonical momentum and $\textbf{A}$ is the vector potential). The eigensystem in the Landau gauge is \cite{QEMP} 
\begin{subequations}
\begin{flalign}
\Psi_0(\textbf{r})&=C e^{iky}\psi_0(x),\\
\label{eq:wf-e0}
\psi_0(x)&= e^{-\frac{1}{2}\left(\frac{x}{l_B}+k l_B\right)^2}H_{\xi_0(k)}\left(\frac{x}{l_B}+k l_B\right),\\
\label{eq:dispersion-e0}
\epsilon_0(k)&=\hbar \omega_c  \left(\xi_0(k)+\frac{1}{2}\right),
\end{flalign}
\end{subequations}
where $H_{\xi_0(k)}$ are the Hermite functions, $C$ is a normalization constant, $\omega_c=eB/m$ is  the cyclotron frequency and $l_B=\sqrt{\hbar/(eB)}$ is the magnetic length. We defined the monotonic function $\xi_0(k)$, which is the lowest solution of the dispersion relation
\begin{equation}
H_{\xi_n(k)}(k l_B)=0.
\end{equation}
Note that by imposing $\nu=1$, we are restricting the possible values of the Fermi energy to a fixed interval, i.e. $\epsilon_F/(\hbar\omega_c)\in(1/2,3/2)$.

\begin{figure}
\includegraphics[scale=0.5]{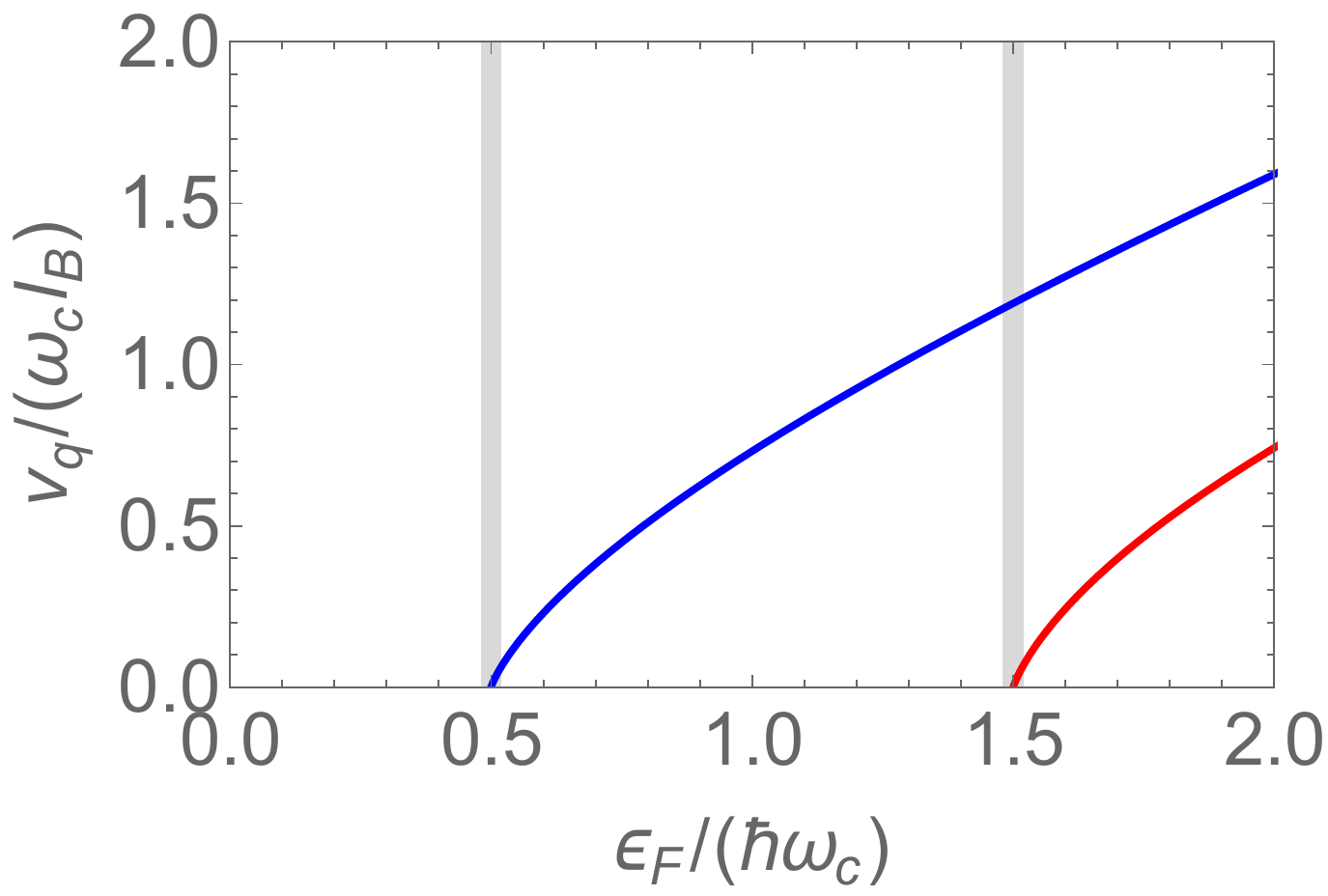}
\caption{\label{fig:quantum-v}Quantum velocities as a function of the Fermi energy $\epsilon_F$. The blue and red lines are the quantum velocities associated to the first and second Landau level, respectively.}
\end{figure}

The quantum contribution to the velocity can be now estimated from Eq. (\ref{eq:dispersion-e0}) and it can be expressed in terms of the Fermi energy $\epsilon_F$ by inverting the function $\xi_0$; the results are shown in Fig. \ref{fig:quantum-v}. 
The characteristic scales of quantum and electrostatic velocities in GaAs are comparable, $\omega_c l_B\approx 7.2\times 10^4\sqrt{B/\mathrm{T}}\mathrm{m/s}$, $v_p\approx 4\times 10^4\mathrm{m/s}$, however their prefactors usually differ by an order of magnitude because of the $\log(q)$ divergence of $v_c$. The presence of a metal electrode reduces $v_c$ leaving $v_q$ approximately unchanged and so the two velocities become comparable. However, even in this case, we do not expect the quantum corrections to modify the qualitative picture as long as $d\gg l_B$ and so, for simplicity, we neglect them in the text.  

The eigenfunctions in Eq. (\ref{eq:wf-e0}) are shifted gaussians weighted by Hermite functions. In the regime considered, where $\epsilon_F$ is relatively close to the lowest Landau level, the corrections due to the Hermite functions are small and we approximate $|\psi_0(x)|^2$ by a normalized gaussian, i.e. 
\begin{equation}
\label{eq:wf-appe0}
|\psi_0(x)|^2\approx \frac{e^{-\left(x-x_0\right)^2/l'^2}}{\sqrt{\pi}l'},
\end{equation}
where the shift $x_0$ and the broadening $l'$ are both of the order of the magnetic length.
Note that in RPA, the EMP charge density is $\rho \propto|\psi_0(x)|^2 $ \cite{QEMP,Mikhailov}, so this form of $\rho$ is consistent with the results obtained with the semiclassical analysis presented in Sec. \ref{sec:field_distribution}. 

\begin{figure}
a)\includegraphics[width=0.45\textwidth]{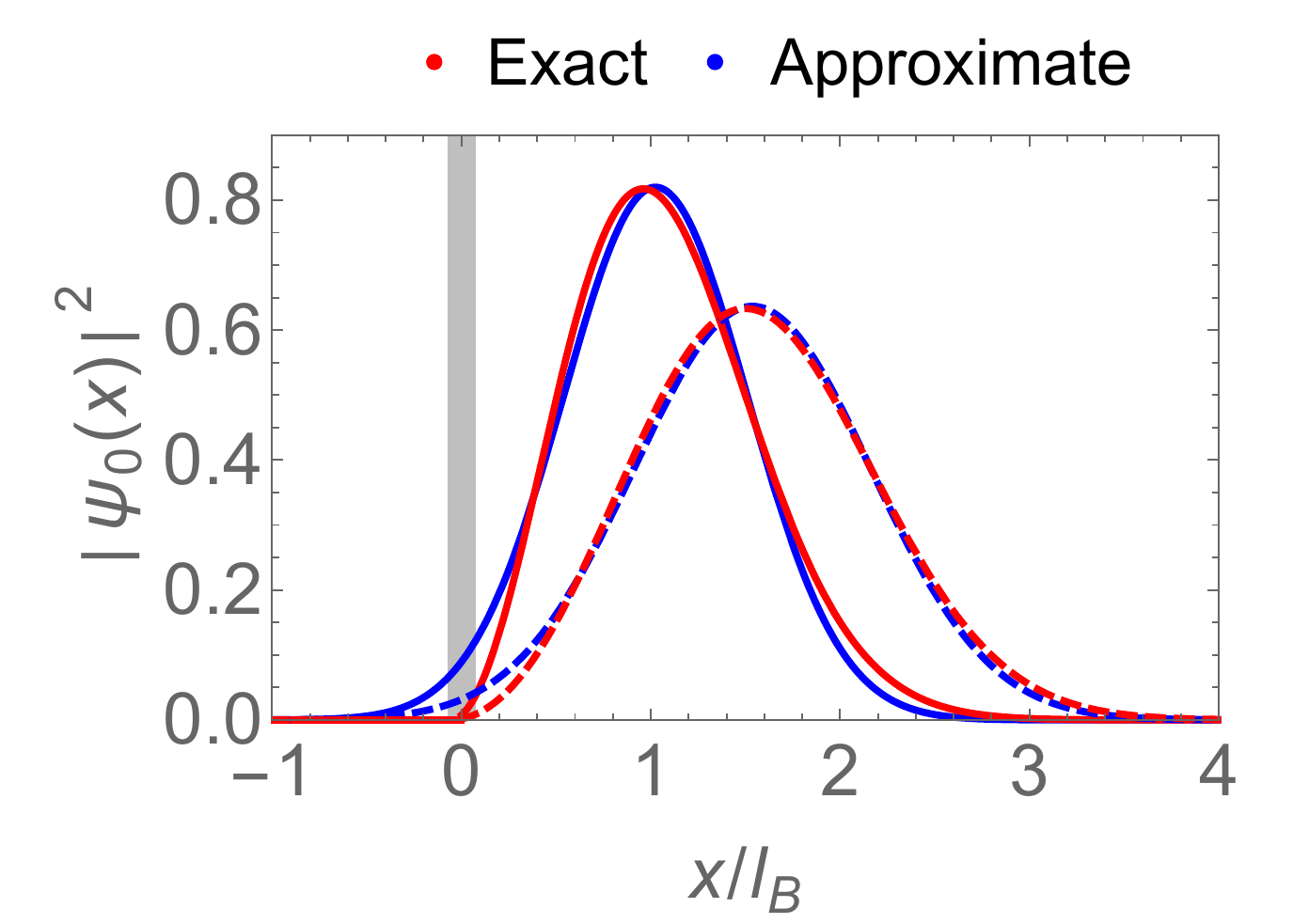}
b)\includegraphics[width=0.45\textwidth]{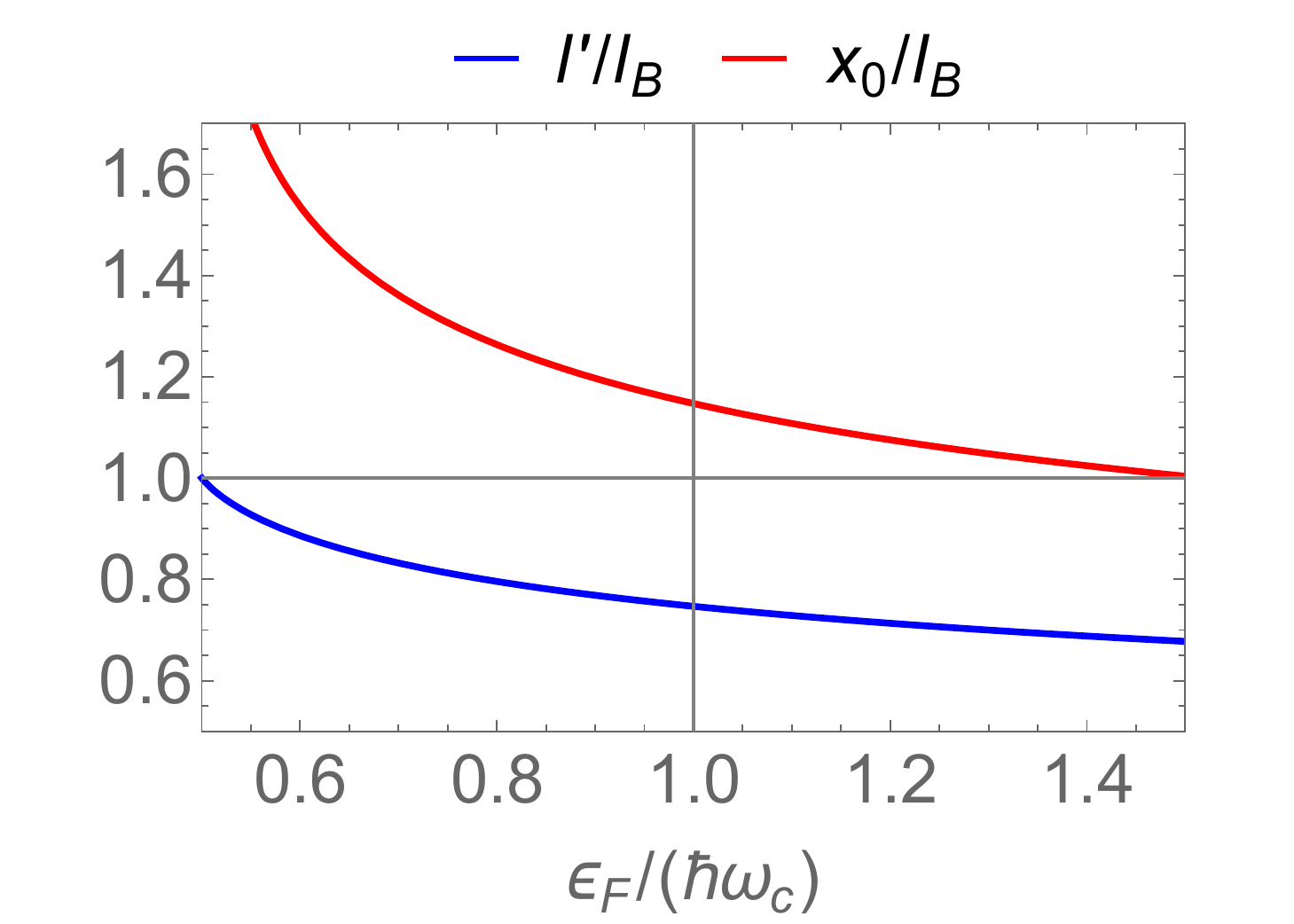}
\caption{\label{fig:wfapp} a) Comparison between the exact wavefunctions in Eq. (\ref{eq:wf-e0}) (red lines) and the normalized and shifted gaussian in Eq. (\ref{eq:wf-appe0}) (blue lines). We used $\epsilon_F=1.4\hbar\omega_c$  for the solid lines and  $\epsilon_F=0.6\hbar\omega_c$ for the dashed lines. The thick gray line indicates the physical edge of the QH material. b) Fitting parameters $l'$ (blue line) and $x_0$ (red line) in units $l_B$ as a function of the Fermi energy $\epsilon_F$.}
\end{figure}

Fig. \ref{fig:wfapp}a) shows a comparison between the gaussian approximation and the exact wavefunction in Eq. (\ref{eq:wf-e0}); in  Fig. \ref{fig:wfapp}b) we also show  the dependence of the fitting parameters $l'$ and $x_0$ on the Fermi energy.
In our analysis, we fix the Fermi energy to the middle of the  Landau level gap, and so we use $l'\approx 0.75 l_B$ and $x_0\approx 1.15 l_B$.

Also, note that using the charge density in Eq. (\ref{eq:wf-appe0}), the integral in Eq. (\ref{eq:classical-vrpa}), which defines the electrostatic velocity, can be computed analytically, leading to
\begin{equation}
\frac{v_c}{v_p}= e^{-\frac{(ql')^2}{4}}K_0\left(\frac{(ql')^2}{4}\right)=-\log\left(\frac{(ql')^2e^\gamma}{8}\right)+\mathcal{O}\left((ql')^2\right),
\end{equation}
in  agreement with Eq. (\ref{eq:eigen-freq-classical-lw}).\\

\subsection{Quantization of the EMPs \label{sec:quantization}}

Once the EMP eigenfrequency and spatial profile are known, one can use standard bosonization procedure to obtain the quantum mechanical operator  \cite{Wen_ChiralLL,Thouless_plasmon,MacDonald_plasmon}.
In particular, considering only the fastest EMP mode, the Hamiltonian of the QH system reduces to the usual sum of harmonic  oscillators
\begin{equation}
H_{R}=\sum _{q>0} \hbar \omega(q)\left(\hat{a}_q^{\dagger}\hat{a}_q+\frac{1}{2}\right),
\end{equation}
where the resonance frequencies $\omega(q)$ at different wavevectors are approximately given by Eq. (\ref{eq:eigen-freq-classical-lw}), and the bosonic ladder operators satisfy the canonical commutation relations $[\hat{a}_{q'},\hat{a}_q^{\dagger}]=\delta_{q,q'}$. 

The EMP charge density can also be expressed in terms of these bosonic operators \cite{Vignale}. Neglecting quantum corrections that mix monopole and acoustic modes \cite{Thouless_plasmon} and using the RPA solution for the EMP charge density (consistent with the analysis in Sec. \ref{sec:field_distribution}), we obtain
\begin{equation}
\label{eq:charge_density_operator}
\rho(\textbf{r})=-e\left| \psi_0(x) \right|^2 \sum _{q>0} \sqrt{\frac{q}{2 \pi  L_y}} \left(e^{i q y} \hat{a}_q^{\dagger }+e^{-i q y}\hat{a}_q\right),
\end{equation}
with $\left| \psi_0(x) \right|^2$  defined in Eq. (\ref{eq:wf-appe0}).

Comparing Eqs. (\ref{eq:charge-gaussian}) and (\ref{eq:charge_density_operator}), we find that for a fixed resonator perimeter $L_y=2\pi n_q/q$ (with $n_q$ being the wavenumber), the two equations coincide if $\rho_0e^{iqy}\rightarrow -e q \left(e^{i q y} \hat{a}_q^{\dagger }+e^{-i q y}\hat{a}_q\right)/\sqrt{n_q}$.
Using this result, one can quickly derive the  quantum mechanical operators from the classical analysis in Sec. \ref{sec:classical-eigen}. For example, the electric field operator of the resonator in the far-field limit and without electrodes is 
\begin{equation}
\label{eq:E_field_operator_far_field}
\textbf{E}(\textbf{r})=\frac{e\sqrt{n_q}}{2\pi \epsilon_S L_y}
\left(\begin{array}{c}
\frac{-x}{x^2+z^2}\\
0\\
\frac{-z}{x^2+z^2}
\end{array}\right)\left(e^{i q y} \hat{a}_q^{\dagger }+e^{-i q y}\hat{a}_q\right).
\end{equation}
Note that the characteristic scale of the electric field can be rewritten in terms of $v_p$ as $\frac{e}{2\pi \epsilon_S L_y}=\frac{h v_p}{e\nu L_y} $, with $\nu$ being the filling factor.\\

% Hamiltonian of the system can be written in terms of bosonized modes 
%
%
%LL and quantization of charge density and long wavelength expansion of e field
%
%
%
%Justification of l via Quantum calculations (RPA) in sharp edge.
%Discuss quantitative difference due to quantum velocity. This does not change the qualitative picture given above, but it simply renormalizes the velocity of propagation.
%Plot of RPA response 
%\\
%Plot of eigenfunctions and fitting of lengthscales.
%\\
%Explanation of second quantized Hamiltonian with Luttinger liquid (and connection to RPA) and plot of quantum velocities.\\
%The excess charge density operator is given by
%
%with $\psi_0(x)$ being the single-particle wavefunction in the lowest LL.\\
%The long-wavelength and far field electric field operator for a single photon in the cavity is
%\begin{equation}
%\label{eq:E_field_operator_far_field}
%\textbf{E}(\textbf{r})=\frac{e}{2\pi \epsilon_S L_y}e^{-iq y}
%\left(\begin{array}{c}
%\frac{x}{x^2+z^2}\\
%0\\
%\frac{z}{x^2+z^2}
%\end{array}\right)(a_q+a_q^{\dagger})
%\end{equation}

%
%
%(???Remark: I am thinking if I should put all this near-field analysis in the Appendix, and maybe just shift some of the plots (maybe Fig. \ref{fig:em-fields} and \ref{fig:dissipvsq} in the far-field analysis). However, I still find it quite interesting and at a least part of the analysis seems worth to be in the main text. Then I would not know how split what goes to appendix and what doesn't. Maybe sec. \ref{sec:quantum-field} can be put in the appendix as it is. Ask David's opinion about this ???)

\section{Qubit model \label{sec:qubit_model}}
\subsection{Double dot Hamiltonian \label{sec:double-dot-Ham}}
We now present the model adopted to describe the singlet-triplet qubit shown in Fig. \ref{fig:qubit-resonator}.
Our derivation follows closely Ref. \cite{Burkard-Divincenzo}.

We begin by considering the Hamiltonian $H_D$ of a collection of $N_e$ electrons with charge $-e$ (with $e>0$) and effective mass $m$ confined in the double-dot potential $V_C$ defined in Eq. (\ref{eq:quartic-potential}), and under the effect of a constant magnetic and electric field ($\textbf{B}$ and $\textbf{E}$ respectively). 
$H_D$ can be decomposed into a sum of three terms
\begin{equation}
\label{eq:qubit Hamiltonian-general}
H_{D}=\sum_{i=1}^{N_e}\left(H_Z(\textbf{r}_i)+H_O(\textbf{r}_i)+\sum_{j\neq i} U_{\text{int}}(\textbf{r}_i ,\textbf{r}_j)\right).
\end{equation}

The Zeeman Hamiltonian $H_Z$ splits the energy of spin up and down electrons and it is given by
\begin{equation}
H_{Z}(\textbf{r}_i)=-\frac{\hbar \mu g}{2} \textbf{B}\cdot\pmb{\sigma}_i,
\end{equation}
where $\mu$ is the Bohr magneton and $g$ is the $g$-factor; $\pmb{\sigma}_i$ is the vector of Pauli matrices acting on the spin of the $i$th electron.

The orbital component $H_O$ can be written in the following way
\begin{equation}
\label{eq:orbital-hamilt}
H_{O}(\textbf{r}_i)=\frac{\pmb{\pi}^2_i}{2 m}+\phi(\textbf{r}_i)+V_C(\textbf{r}_i).
\end{equation}
Here, $\pmb{\pi}_i$ is the dynamical momentum of the $i$th electron $\pmb{\pi}_i=\textbf{p}_i+e\textbf{A}(\textbf{r}_i)$, and $\textbf{A}$ is the vector potential; $\phi$ is the scalar potential and for constant $\textbf{E}$ field it reduces to $\phi(\textbf{r}_i)=-e\textbf{E}\cdot\textbf{r}_i$.  
%Here, following \cite{Burkard-Divincenzo}, we use a confinement potential $V_C$ with a simple analytic form, i.e. a double harmonic well displaced with respect to each other in the $x$ direction by an amount $2a$
%\begin{equation}
%V_C(\textbf{r}_i)=\frac{m \Omega^2}{2} \left(\left(\frac{x_i^2-a^2}{2a}\right)^2+y_i^2\right).
%\end{equation}
%The frequency $\Omega$ quantifies the strength of the confinement.
Also, we consider $\textbf{E}=E \hat{e}_x$ and $\textbf{B}=B \hat{e}_z$; $E$ is the total (homogeneous) electric field in the direction connecting the dots, and so it is the sum of the resonator field and the externally applied field.

The Coulomb interactions between the electrons are included in the term $U_{\text{int}}(\textbf{r}_i,\textbf{r}_j)$, that can be estimated by
\begin{equation}
\label{eq:int-hamil}
U_{\text{int}}(\textbf{r}_i,\textbf{r}_j)=\frac{e^2}{2}G(\textbf{r}_i,\textbf{r}_j),
\end{equation}
with $G(\textbf{r}_i,\textbf{r}_j)$ being the electrostatic Green's function for the configuration considered, potentially including the screening effect due to the  image charge at the electrodes.
%; the relevance  of this contribution will be discussed in more detail later on.
%We do not include the screening due to the electrons in the semiconductor since in depleted regions this screening length is expected to be large compared to the bulk one.

\subsection{Single dot basis \label{sec:basis-states}}
A convenient basis for the problem is provided by the eigenstates of the  single-particle Hamiltonian
\begin{equation}
h_{\pm}=\frac{\pmb{\pi}^2}{2 m}-eEx+\frac{m \Omega^2}{2} \left(\left(x\pm a\right)^2+y^2\right).
\end{equation}

The Hamiltonians $h_{\pm}$ are related to the well-known Fock-Darwin Hamiltonian 
\begin{equation}
h_0=\frac{\pmb{\pi}^2}{2 m}+\frac{m \Omega^2}{2} \left(x^2+y^2\right)
\end{equation}
by a magnetic translation in the $x$-direction, i.e.
\begin{equation}
\label{eq:translation-h0}
h_{\pm}=T_B^x(\pm a-b)h_0 T^x_B(\mp a+b)+e E (\pm a-b/2),
\end{equation}
where
\begin{equation}
T_B^x(X)=e^{i X (y/l_B^2+\pi_x/\hbar)}
\end{equation}
is the magnetic translation operator that shifts only the scalar potential, leaving the kinetic energy invariant, see e.g. \cite{Barnes}.
Here, $l_B=\sqrt{\hbar/(eB)}$ is the magnetic length and we introduce the electric length $b= e E/(m \Omega^2)$, see Eq. (\ref{eq:b0-length}).

We now fix the vector potential and we work in the symmetric gauge $\textbf{A}=(-By/2,Bx/2,0)$.
The eigenstates of $h_0$ are  easily found by introducing the bosonic ladder operators 
\begin{equation}
\alpha_{\pm}=-i l_T\frac{\pm p_y+i p_x}{2\hbar}-\frac{\pm y+i x}{2 l_T}
\end{equation}
and realizing that in terms of these operators $h_0$ is simply a sum of independent harmonic oscillators
\begin{equation}
h_0=\hbar \omega_+\left(\alpha_{+}^{\dagger}\alpha_{+}+\frac{1}{2}\right)+\hbar \omega_-\left(\alpha_{-}^{\dagger}\alpha_{-}+\frac{1}{2}\right),
\end{equation}
with frequencies $\omega_{\pm}=\Omega_T(1\pm \sqrt{\beta})$.
The Fock-Darwin frequency $\Omega_T$, its related oscillator length $l_T$, and the ratio $\beta$ are defined by Eqs. (\ref{eq:omega-T}), (\ref{eq:lt-def}), and (\ref{eq:beta-def}), respectively.

The eigenfunctions of $h_{\pm}$ can then be easily constructed from
\begin{equation}
\label{eq:FD-wf-full}
\Psi_{n_+ n_-}^{\pm}=T_B^x(\pm a-b)\frac{(\alpha_+^{\dagger})^{n_+}(\alpha_-^{\dagger})^{n_-}}{\sqrt{n_+! n_-!}}\Psi_{00},
\end{equation}
where the ground state wavefunction of $h_0$ is the normalized gaussian
\begin{equation}
\Psi_{00}=\frac{e^{-(x^2+y^2)/(2l_T^2)}}{\sqrt{\pi} l_T}.
\end{equation}

Explicitly, the ground state of $h_{\pm}$ has energy
\begin{equation}
\epsilon_{00}^{\pm}=\hbar \Omega_T+ e E (\pm a-b/2)
\end{equation}
and its wavefunction is, up to an overall phase,
\begin{equation}
\label{eq:Psi_00_pm}
\Psi_{00}^{\pm}=\frac{e^{\pm i\sqrt{\beta} a y/l_T^2}}{\sqrt{\pi} l_T}e^{-(y^2+(x-b\pm a)^2)/(2l_T^2)}.
\end{equation}

The first excited states are the states with $n_+=n_--1=0$ and their energy gap with respect to the ground state is $\Delta\epsilon=\hbar \Omega_T(1-\sqrt{\beta})$.
In strongly confined double dots, where the confinement potential $\Omega$ is the dominant energy scale and $\beta\ll 1$, this energy gap is large, and so we can project the problem onto the ground state and neglect mixing to higher orbital states \cite{Burkard-Divincenzo}.
%For this approximation to be valid, we also require that the inter-dot distance $a\gtrsim l_T$, such that the mixing terms between the ground state of the one dot and the higher orbital momentum states of the other dot are small.
%We analyze this coupling in more detail in Sec.\ref{app:sub:higher_orbitals}.

%In this shifted Fock-Darwin basis, the orbital and interaction Hamiltonians in Eqs. (\ref{eq:orbital-hamilt}) and (\ref{eq:int-hamil}) are respectively given by
%\begin{subequations}
%\begin{flalign}
%H_O &=\left(
%\begin{array}{cc}
% \eta_{-} & \eta_{-+} \\
% \eta_{-+} & \eta_{+} \\
%\end{array}
%\right), \\
%U_{\text{int}} &=\left(
%\begin{array}{cccc}
%V_{\text{Hu}} & V_{\text{M}} & V_{\text{M}}  & V_{\text{Fo}} \\
%V_{\text{M}} & V_{\text{Ad}} & V_{\text{Ha}} & V_{\text{M}} \\
%V_{\text{M}} & V_{\text{Ha}} & V_{\text{Ad}} & V_{\text{M}} \\
%V_{\text{Fo}} & V_{\text{M}} & V_{\text{M}}  & V_{\text{Hu}} \\
%\end{array}
%\right);
%\end{flalign}
%\end{subequations}

We use greek letters to indicate the matrix elements in this shifted Fock-Darwin basis, i.e. $\Psi_{00}^{\alpha}$, with $\alpha\in(-,+)$.
The matrix elements of the orbital Hamiltonian in Eq. (\ref{eq:orbital-hamilt}) are 
%\begin{equation}
%\label{eq:NO_HO}
%H_O=\left(
%\begin{array}{cc}
% \eta_{-} & \eta_{-+} \\
% \eta_{-+} & \eta_{+} \\
%\end{array}
%\right), 
%\end{equation}
\begin{equation}
\label{eq:NO_HO}
H_O^{\alpha\beta}=\left(
\begin{array}{cc}
 \eta_{-} & \eta_{-+} \\
 \eta_{-+} & \eta_{+} \\
\end{array}
\right), 
\end{equation}
where we define
\begin{subequations}
\begin{flalign}
\eta_{\mp}&=\hbar  \Omega_T f(a\pm b)-e E l_T \left(\frac{b}{2}\pm a\right),\\
\eta_{-+}&= e^{-a^2(1+\beta)/l_T^2} \left(\hbar \Omega _T  f(b) -e E l_T \left(a+\frac{b}{2}\right)\right),
\end{flalign}
\end{subequations}
and the function
\begin{multline}
f(x) = 1+\frac{3}{8}(1-\beta ) \left(\frac{l_T^2}{4 a^2}-\frac{a^2}{l_T^2}-1\right)+\\
(1-\beta ) \frac{x}{l_T} \left(\frac{a}{l_T}+\frac{3}{8} \frac{x}{l_T} \left(\frac{l_T^2}{a^2}+\frac{x^2}{3 a^2}-2\right)\right).
\end{multline}
Note that in this section, we use an unfortunate notation because the letter $\beta$ is used to label the dots and to parametrize the ratio of harmonic and magnetic confinement strength, defined in Eq. (\ref{eq:beta-def}); however, the meaning of $\beta$   is apparent from the context.

In the Fock-Darwin  basis the matrix elements of the interaction Hamiltonian in Eq. (\ref{eq:int-hamil}), have some symmetry, in particular, for any $\alpha,\beta\in(-,+)$, such that $\alpha\neq \beta$, the following relations are true
\begin{subequations}
\label{eq:symmetries}
\begin{flalign}
\langle \alpha,\alpha|U_{\text{int}}|\alpha,\alpha\rangle&\equiv V_{\text{Hu}},\\
\langle \alpha,\alpha|U_{\text{int}}|\beta,\beta\rangle&\equiv V_{\text{Ad}},\\
\langle \alpha,\beta|U_{\text{int}}|\beta,\alpha\rangle&\equiv V_{\text{Ha}},\\
\langle \alpha,\beta|U_{\text{int}}|\alpha,\beta\rangle&\equiv V_{\text{Fo}},\\
\langle \alpha,\alpha|U_{\text{int}}|\alpha,\beta\rangle&=\langle \alpha,\alpha|U_{\text{int}}|\beta,\alpha\rangle\equiv V_{\text{M}}.
\end{flalign}
\end{subequations}
Explicitly, the interaction elements are
\begin{subequations}
\label{eq:vint_NO}
\begin{flalign}
\frac{V_{\text{Hu}}}{\hbar v_p/ l_T}&=\sqrt{2\pi^3}-\frac{\pi}{d},\\
\frac{V_{\text{Ad}}}{\hbar v_p/ l_T}&=e^{-2 a^2 (1+\beta)} \left(\sqrt{2 \pi ^3}-\frac{\pi}{d}\right),\\
\frac{V_{\text{Ha}}}{\hbar v_p/ l_T}&=\sqrt{2 \pi ^3} e^{-a^2} I_0\left(a^2\right)-\frac{\pi}{d},\\
\frac{V_{\text{Fo}}}{\hbar v_p/ l_T}&=\sqrt{2 \pi ^3} e^{-a^2 (2+\beta)} I_0\left(\beta  a^2\right)- e^{-2 a^2 (1+\beta)}\frac{\pi}{d},\\
\frac{V_{\text{M}}}{\hbar v_p/ l_T}&=\sqrt{2 \pi ^3} e^{-a^2 (5+3 \beta)/4} I_0\left(\frac{1-\beta}{4} a^2\right)-e^{-a^2(1+\beta)}\frac{\pi}{d}.
\end{flalign}
\end{subequations}
Here, the lengths $a$ and $d$ are in units $l_T$, $I_0$ is the modified Bessel function of the first kind, $v_p$ is defined in Eq. (\ref{vp-scale}) and it has to be evaluated at the filling factor $\nu=1$.  We include the lowest order correction in $l_T/d$ due to the presence of an ideal metal gate at a distance $d$ from the double dot; also we consider $a,b\ll d$ so that the exact position of the gate (i.e. if it is on the side or at the top of the qubit) does not matter.\\

\subsection{Orthonormal basis and second quantization}

The basis states $\Psi_{00}^{\pm}$ are not orthonormal, and thus the overlap matrix $S_{\alpha\beta}=\langle\Psi_{00}^{\alpha}|\Psi_{00}^{\beta}\rangle$ is not the identity.
In general, an orthonormal basis $|O \rangle$ can be constructed by applying a linear transformation $P$ to the non-orthonormal states $|NO \rangle$, i.e. 
\begin{equation}
|O \rangle=|NO \rangle P.
\end{equation}
Any single-particle operator $A$ transforms from one basis to another according to
\begin{equation}
A_O=P^{-1}S^{-1} A_{NO} S^{-1}(P^{-1})^{\dagger}.
\end{equation}

The transformation $P$ can be found by requiring that the identity operator in the orthonormal basis transforms into the overlap matrix in the non-orthonormal basis, and so we obtain
\begin{equation}
\label{eq:L-trnasf}
S=(P^{-1})^{\dagger}P^{-1},
\end{equation}
and the transformation rule modifies as 
\begin{equation}
\label{eq:transf-singlepart}
A_O=P^{\dagger} A_{NO} P=\sum_{\alpha\beta}P^{\dagger}_{i\alpha}A_{NO}^{\alpha\beta}  P_{\beta j} .
\end{equation}
For two-particle operators $B$, the transformation straightforwardly generalizes as
\begin{equation}
\label{eq:transf-twopart}
B_O^{ijkl}=\sum_{\alpha\beta\gamma\delta}P^{\dagger}_{i\alpha} P^{\dagger}_{j\beta} B_{NO}^{\alpha\beta\gamma\delta} P_{\gamma k} P_{\delta l}.
\end{equation}
In the following, we use roman letters to indicates matrix elements in the orthonormal basis.

If we also require $P$ to be an Hermitian matrix, i.e. $P=P^{\dagger}$, we obtain from  Eq. (\ref{eq:L-trnasf}) that $P=S^{-1/2}$. Then, in our case, $P$ reduces to the symmetric real matrix
\begin{equation}
\label{eq:Pmat}
P= \frac{1}{2} \left(
\begin{array}{cc}
\frac{1}{\sqrt{1+s}}+\frac{1}{\sqrt{1-s}} & \frac{1}{\sqrt{1+s}}- \frac{1}{\sqrt{1-s}} \\
\frac{1}{\sqrt{1+s}}-\frac{1}{\sqrt{1-s}} & \frac{1}{\sqrt{1+s}}+ \frac{1}{\sqrt{1-s}} \\
\end{array}
\right),
\end{equation}
with $s$ being the overlap of the ground state wavefunctions of the two dots defined in Eq. (\ref{eq:s-def}).\\

Now that we have an orthonormal basis, we can introduce the associated fermionic operators $c_{i\sigma}$, where $i,j,k,l=(-,+)$ labels the position of the dot $x=\mp a$ and $\sigma,\sigma'=(\uparrow,\downarrow)$ labels the spin.
We then rewrite the double dot Hamiltonian in Eq. (\ref{eq:qubit Hamiltonian-general}) in the second quantized form
\begin{multline}
\label{eq:second-quantized-Hd}
H_D=\sum_{i,\sigma \sigma'}H_Z^{\sigma \sigma'} c^{\dagger}_{i\sigma} c_{i\sigma'} +\sum _{ij,\sigma}H_O^{ij} c^{\dagger}_{i\sigma} c_{j\sigma}+\\
\frac{1}{2}\sum _{ijlm,\sigma \sigma'}V_{\text{int}}^{ijkl}c^{\dagger}_{i\sigma}c^{\dagger}_{j\sigma'} c_{k\sigma'}c_{l\sigma}.
\end{multline}

Here, $H_Z=-\frac{\hbar\mu g}{2} B \tau_z$ ($\tau_i$ is the $i$th Pauli matrix acting on the spin degree of freedom) and the matrix elements of $H_O$ ($V_{\text{int}}^{ijkl}$) are found by applying the transformation rules in Eq. (\ref{eq:transf-singlepart}) (Eq. (\ref{eq:transf-twopart})) to the non-orthogonal matrix elements in Eq. (\ref{eq:NO_HO}) (Eq.  (\ref{eq:vint_NO})) with $P$ given in Eq. (\ref{eq:Pmat}).
In particular, we define, with $i\neq j$
\begin{subequations}
\label{eq:matrix elements}
\begin{flalign}
U_{\text{Hu}}&=\sum_{\alpha\beta\gamma\delta}P_{i\alpha} P_{i\beta} V_{\text{int}}^{\alpha\beta\gamma\delta} P_{\gamma i} P_{\delta i},\\
U_{\text{Ad}}&=\sum_{\alpha\beta\gamma\delta}P_{i\alpha} P_{i\beta} V_{\text{int}}^{\alpha\beta\gamma\delta} P_{\gamma j} P_{\delta j},\\
U_{\text{Ha}}&=\sum_{\alpha\beta\gamma\delta}P_{i\alpha} P_{j\beta} V_{\text{int}}^{\alpha\beta\gamma\delta} P_{\gamma j} P_{\delta i},\\
U_{\text{Fo}}&=\sum_{\alpha\beta\gamma\delta}P_{i\alpha} P_{j\beta} V_{\text{int}}^{\alpha\beta\gamma\delta} P_{\gamma i} P_{\delta j},\\
t &=\sqrt{2}\sum_{\alpha\beta}P_{i\alpha} H_O^{\alpha\beta} P_{\beta j}+\sqrt{2} \sum_{\alpha\beta\gamma\delta}P_{i\alpha} P_{i\beta} V_{\text{int}}^{\alpha\beta\gamma\delta} P_{\gamma i} P_{\delta j},\\
\epsilon_I &=\frac{1}{2}\sum_{\alpha\beta}\left(P_{-\alpha} H_O^{\alpha\beta} P_{\beta -}+P_{+\alpha} H_O^{\alpha\beta} P_{\beta +} \right),\\
\Delta &=\frac{1}{2}\sum_{\alpha\beta}\left(P_{-\alpha} H_O^{\alpha\beta} P_{\beta -}-P_{+\alpha} H_O^{\alpha\beta} P_{\beta +} \right).
\end{flalign}
\end{subequations}

Each energy contribution has a clear physical meaning.
$U_{\text{Hu}}$ and $U_{\text{Ad}}$ are respectively the on-site and off-site Hubbard terms that quantify the on-site and off-site Coulomb interaction energy due to a double occupation of the dots. The energies $U_{\text{Ha}}$ and $U_{\text{Fo}}$ are respectively the Hartree and Fock (exchange) contributions of the Coulomb interactions.
Furthermore, $t$ is a tunneling energy between the two dots and it  includes a small renormalization due to the Coulomb interactions. $\Delta$ is the detuning energy and is related to the dipole moment of the homogeneous electric field in the direction connecting the dots, which raises the ground state energy of one of the dots compared to the other.

It is important to remark that an external electric field $E$ only influences the orbital matrix elements $\epsilon_I,t$ and $\Delta$. The dipole energy $\Delta$ vanishes when no electric field is applied, and explicitly it has the form
\begin{equation}
\label{eq:delta1}
\Delta=\Delta_1  e E +\mathcal{O}(e E)^3,
\end{equation}
while the tunneling energy can be written as
\begin{equation}
\label{eq:t2}
t=t_0+t_2 (e E)^2+\mathcal{O}(e E)^4.
\end{equation}
The corrections to the lowest order in the electric field are given by
\begin{subequations}
\label{eq:delta1-t2_design-par}
\begin{flalign}
\Delta_1 &=\frac{1}{\sqrt{1-s^2}}\frac{l_T^2}{4 a}\left(3-4 \frac{a^2}{l^2_T}\right),\\
t_2&=-\frac{3 l_T^2}{4\hbar \Omega_T \sqrt{2} (1-\beta)}\text{csch}\left( \frac{a^2}{l_T^2}(1+\beta) \right).
\end{flalign}
\end{subequations}

\subsection{Singlet-Triplet Hamiltonian}

We now restrict the analysis to the two-electron sector.
A convenient basis to write the Hamiltonian is then the singlet and triplet basis, defined as
\begin{subequations}
\label{eq:ST-basis-ladder}
\begin{flalign}
|S,\mp\rangle &=c_{\mp\uparrow}^{\dagger }c_{\mp\downarrow}^{\dagger }|0\rangle, \\
|T,\uparrow\downarrow \rangle &=c_{-\uparrow\downarrow }^{\dagger }c_{+\uparrow\downarrow }^{\dagger }|0\rangle,\\
|(S,T),0\rangle &=\frac{c_{-\uparrow}^{\dagger }c_{+\downarrow}^{\dagger }\mp c_{-\downarrow}^{\dagger }c_{+\uparrow}^{\dagger }}{\sqrt{2}}|0\rangle.
\end{flalign}
\end{subequations}
The singlet states $|S,\mp\rangle$ are the states where two electrons with opposite spin occupy the same dot centered at position $x=\mp a$,  the triplet states $|T,\uparrow\downarrow\rangle$ are the states where electrons in different dots have their spin aligned in the direction of the arrow, and the states $|S,0\rangle$  and $|T,0\rangle$ are the antisymmetric and symmetric combination of spins in the two dots, respectively.
We use the following ordering of the states 
\begin{equation}
(|S,-\rangle ,|S,+\rangle ,|S,0\rangle ,|T,\uparrow\rangle ,|T,\downarrow\rangle ,|T,0\rangle )^T.
\end{equation}

When the effects of spin-orbit coupling and of a  magnetic field gradient are neglected, the singlet-triplet decomposition guarantees that the two-electron Hamiltonian is block diagonal,
\begin{equation}
\label{eq:full_H_Q}
H_D=H_S\oplus H_T.
\end{equation}

Using the Wick theorem and  subtracting $2\epsilon_I$ to all the diagonal terms, the two blocks are given by
\begin{equation}
\label{eq:HS-1state}
H_S=\left(
\begin{array}{ccc}
 U_{\text{Hu}}+2 \Delta & U_{\text{Ad}} & t \\
 U_{\text{Ad}} & U_{\text{Hu}}-2 \Delta & t \\
 t & t & U_{\text{Ha}}+ U_{\text{Fo}} \\
\end{array}
\right)
\end{equation}
and
\begin{equation}
\label{eq:HT-1state}
H_T=\left(
\begin{array}{ccc}
 U_{\text{Ha}}-U_{\text{Fo}}+U_Z & 0 & 0 \\
 0 & U_{\text{Ha}}-U_{\text{Fo}}-U_Z & 0 \\
 0 & 0 & U_{\text{Ha}}-U_{\text{Fo}} \\
\end{array}
\right).
\end{equation}

The energy contributions in the singlet sector $H_S$ are given in Eq. (\ref{eq:matrix elements}).
In the triplet sector $H_S$, we also define the Zeeman energy $U_Z=-\hbar\mu g B$, which separates in energy the  $|T,\uparrow\downarrow\rangle$  states because of the applied $B$ field. $U_{\text{Fo}}$ is particularly important for ST qubits because it splits the energy of the singlet and triplet states $|(S,T),0\rangle$ with vanishing angular momentum and it allows to use these states as a qubit basis.

It is important to remark here that the block Hamiltonians (\ref{eq:HS-1state}) and (\ref{eq:HT-1state}) are obtained by restricting the total Hilbert space to the subsector spanned by the ground state of each dot \cite{Burkard-Divincenzo}. 
We expect this approximation to hold quantitatively in  strongly confined quantum dots, where the energy splitting between the ground and first excited state is higher than the on-site Hubbard energy $U_{\text{Hu}}$;  even if this condition is not fulfilled, we  believe this model still provides a   qualitative understanding of the system. 
%
%the ground state projection works reasonably well as long as $a\gtrsim l_T$, while in the opposite limit, the mixing with higher orbital states becomes more relevant.
%To keep the description simple, for now we neglect these states;  their effect on the exchange energy and on the coupling strength is discussed in more detail in Appendix \ref{app:sub:higher_orbitals} and in Sec.\ref{sec:Coulomb-intqubit-res}, respectively.\\

%  the mixing of higher   we neglect the  is that the low energy physics of the double dot is fully  described by a projection onto the ground state  the and (\ref{eq:HT-1state}) is that the two dots are  Hilbert space of the double dot is spanned only the ground state energy of each dot
%the o  analysis, we include for simplicity only the ground state energy of each dot and we neglect the coupling to higher orbital states. These additional terms become relevant when the dots are very close to each other and their effect is discussed in more detail in Appendix \ref{app:sub:higher_orbitals}.
%(?Perhaps after splitting the paper I will put a bit more discussion about this aspect here and move some of the appendixes in the main text?)

The triplet sector is already diagonal in this basis, while the singlet eigenstates $(|\tilde{S},-\rangle,|\tilde{S},+\rangle,|\tilde{S},0\rangle)^T$ are obtained by a unitary rotation $M_S^{\dagger}$ of the singlet basis $(|S,-\rangle,|S,+\rangle,|S,0\rangle)^T$.
$M_S$ is the matrix of normalized column eigenvectors of $H_S$.\\

When no in-plane electric field is applied, i.e. $E=0$, there is no dipole moment between the two dots and $\Delta=0$. In this situation, $M_S$ has the simple form 
\begin{equation}
\label{eq:MS_mat}
M_S=\left(
\begin{array}{ccc}
 -\frac{1}{\sqrt{2}} & A(\epsilon_2) & A(\epsilon_0) \\
  \frac{1}{\sqrt{2}} & A(\epsilon_2) & A(\epsilon_0) \\
   0 &  B(\epsilon_2) & B(\epsilon_0) \\
\end{array}
\right),
\end{equation}
where we introduced the functions
\begin{subequations}
\begin{flalign}
A(x)&=\frac{2U_{\text{Ha,Fo}}-x}{\sqrt{4 t^2+2 \left(2U_{\text{Ha,Fo}}-x\right)^2}}\\
B(x)&=-\mathrm{sign}(t)\sqrt{1-2A(x)^2}.
\end{flalign}
\end{subequations}
In this case, the eigenenergies $\epsilon_i$ are given by
%\begin{equation}
%M_S=\left(
%\begin{array}{ccc}
% -\frac{1}{\sqrt{2}} & A(\epsilon_2) & A(\epsilon_0) \\
%  \frac{1}{\sqrt{2}} & A(\epsilon_2) & A(\epsilon_0) \\
%   0 &  B(\epsilon_2) & B(\epsilon_0) \\
%\end{array}
%\right).
%\end{equation}
%
%Here, we defined the functions
%\begin{subequations}
%\begin{flalign}
%A(x)&=\frac{2U_{\text{Ha,Fo}}-x}{\sqrt{4 t^2+2 \left(2U_{\text{Ha,Fo}}-x\right)^2}}\\
%B(x)&=-\mathrm{sign}(t)\sqrt{1-2A(x)^2},
%\end{flalign}
%\end{subequations}
%the eigenenergies
\begin{subequations}
\label{eq:zero-det-eigenenergy}
\begin{flalign}
\epsilon_0&=U_{\text{Hu,Ad}}+U_{\text{Ha,Fo}}-\sqrt{2 t^2+\left(U_{\text{Hu,Ad}}-U_{\text{Ha,Fo}}\right)^2}, \\
\epsilon_1&=U_{\text{Hu}}-U_{\text{Ad}}, \\
\epsilon_2&=U_{\text{Hu,Ad}}+U_{\text{Ha,Fo}}+\sqrt{2 t^2+\left(U_{\text{Hu,Ad}}-U_{\text{Ha,Fo}}\right)^2},
\end{flalign}
\end{subequations}
where we defined the combination of the Coulomb interaction energies
\begin{subequations}
\begin{flalign}
U_{\text{Hu,Ad}}&=\frac{U_{\text{Hu}}+U_{\text{Ad}}}{2}, \\
U_{\text{Ha,Fo}}&=\frac{U_{\text{Ha}}+U_{\text{Fo}}}{2} .
\end{flalign}
\end{subequations}

The computational basis is usually defined by the singlet-triplet states $(|\tilde{S},0\rangle,|T,0\rangle)^T $; the energy gap between these states is
\begin{equation}
\label{eq:exchangeJ}
J_{\text{ex}}=2 U_{\text{Fo}}+U_{\text{Hu},\text{Ad}}-U_{\text{Ha},\text{Fo}}-\sqrt{2 t^2+\left(U_{\text{Hu},\text{Ad}}-U_{\text{Ha},\text{Fo}}\right)^2}.
\end{equation}
%Since the main contribution to $J_{\text{ex}}$ is generally the Fock energy $U_{\text{Fo}}$, we refer to $J_{\text{ex}}$ as the exchange energy.
This energy gap is often called exchange energy, because in the limit of weakly coupled dots, where $t,U_{\text{Ha,Fo}}\ll U_{\text{Hu,Ad}}$, it reduces to the Fock interaction energy $J_{\text{ex}}\approx 2U_{\text{Fo}}$.\\

%(???Plot of exchange splitting against $\Omega, B$ for the two chosen $a$???)\\

%For example, by considering two dots at a distance of $2a \approx 30\mathrm{nm}$ in a magnetic field $B\approx0.2\mathrm{T}$ and in a harmonic confinement potential $\Omega\approx4\mathrm{meV}$, we obtain  $J_{\text{ex}}\approx -3.5\mathrm{GHz}$.\\

%(?Plot of exchange splitting against $\Omega, B$?)\\

It is now informative to verify what happens when a small, homogeneous electric field $E$ is applied in the direction connecting  to the two dots.
This term has two effects: it detunes the two dots leading to a finite dipole moment $\Delta$ between the dots and  it modifies the tunnel energy $t\rightarrow t+\delta t$ due to the change in the potential  landscape.

Both effects can be straightforwardly accounted for by conventional perturbation theory. The lowest non-trivial correction in $\delta t$ and $\Delta$ to the exchange energy is given by
\begin{equation}
\label{eq:exchange-efield}
\delta J_{\text{ex}}=\chi_t \delta t +\chi_{\Delta}\Delta^2.
\end{equation}
The susceptibilities to tunneling and detuning can be found explicitly and they are given by
\begin{subequations}
\begin{flalign}
\label{eq:susc-tun}
\chi_{t}&=-\frac{2t}{\sqrt{2t^2+(U_{\text{Hu,Ad}}-U_{\text{Ha,Fo}})^2}},\\
\label{eq:susc-det}
\chi_{\Delta}&=\frac{2}{\epsilon _1-\epsilon_0}\left(1-\frac{U_{\text{Hu,Ad}}-U_{\text{Ha,Fo}}}{\sqrt{2t^2+(U_{\text{Hu,Ad}}-U_{\text{Ha,Fo}})^2}}\right).
\end{flalign}
\end{subequations}
%(Plot susceptibilities in the regime with highest coupling?)
These susceptibilities to tunneling and detuning are the ones used in Eqs. (\ref{eq:exchange-gradientefield}) and (\ref{eq:exchange-detunedefield}).

Note that the lowest order correction is linear in the tunnel energy and quadratic in the detuning. In our model,   both these terms lead to a quadratic correction in the electric field, i.e. $\delta J_{\text{ex}}\propto E^2$, because $\Delta\propto E$ and  $\delta t\propto E^2$, see Eqs. (\ref{eq:delta1}) and (\ref{eq:t2}).

%This means that the non-detuned configuration is not altered by an homogeneous electric field at least to the first order. This property is advantageous as it suppresses charge noise, however the absence of a qubit dipole moment decreases also the electrostatic coupling with a resonator.

%The effect of a finite detuning $\Delta$ to lowest order correction to the exchange energy in $\Delta$ is given by
%\begin{equation}
%J_{\text{ex}}(\Delta)-J_{\text{ex}}=8\Delta^2 \frac{ A_0\left(\epsilon_3\right)^2}{\epsilon _1-\epsilon_3}.
%\end{equation}
%
%We remark that a finite homogeneous electric field not only detunes the two dots, but it also influences the tunneling energy $t$. Therefore, $J_{\text{ex}}$,  $A_0$ and $\epsilon_3$ all  depend on the electric field through $t$ and the correction to the  exchange energy in the lowest order of electric field is
%\begin{equation}
%\label{eq:exchange-efield}
%J_{\text{ex}}(E)=\left(J_{\text{ex}}+8(eEl_T)^2\Delta_1^2 \frac{ A_0\left(\epsilon_3\right)^2}{\epsilon _1-\epsilon_3}\right)_{t=t_0+t_2 (eEl_T)^2}
%\end{equation}
%where $\Delta_1$ and $t_2$ are defined respectively from Eqs. (\ref{eq:delta1}) and (\ref{eq:t2}).

\section{Solution of the Hartree integral \label{app:sub:EMP-QubitCoupling}}

Here, we derive the  results presented in Sec. \ref{sec:Hartree-text}. The coupling between two charge densities is captured by the Hartree integral in Eq. (\ref{eq:Hartree_int}). We neglect exchange interactions because of the negligible tunnel coupling between the EMP excess charge density and the double dot.
% \begin{equation}
% \label{eq:Hartree_int}
%H_{\text{int}}=\int d\textbf{r}\int d\textbf{r}'\rho_{\text{TL}}(\textbf{r})G(\textbf{r},\textbf{r}') \rho_{\text{Q}}(\textbf{r}'),
%\end{equation}
%where $G$ is the electrostatic Green's function of the configuration chosen.
%
%The excess charge density operator $\rho_{\text{TL}}$ of a QH  transmission line is defined by Eq. (\ref{eq:charge_density_operator}). 
We consider   a resonator of perimeter $L_y$, with $L_y$ being much longer than the other lengths in the problem.
To define the charge density operator of the double dot system $\rho_{\text{D}}$, we first introduce the charge density operator in the singlet-triplet basis
\begin{equation}
\rho_{\text{ST}}=\rho_S \oplus \rho_T,
\end{equation}
with 
\begin{subequations}
\begin{flalign}
\rho_S&=\left(
\begin{array}{ccc}
2 \rho_{11} & 0 & \sqrt{2}\rho_{12} \\
0 & 2 \rho_{22} & \sqrt{2}\rho_{12}^* \\
 \sqrt{2}\rho_{12}^* & \sqrt{2}\rho_{12} & \rho_{11}+\rho_{22} \\
\end{array}
\right),\\
\rho_T&=\left(\rho_{11}+\rho_{22}\right) \mathcal{I}_3.
\end{flalign}
\end{subequations}
Here, $\mathcal{I}_3$ is the $3\times 3$ identity matrix and $\rho_{ij}$ is the matrix element of the charge density operator in the orthonormal  basis
\begin{equation}
\rho=-e P
\left(
\begin{array}{cc}
\left| \Psi_{00}^{-} \right|^2 & (\Psi_{00}^{-})^{*} \Psi_{00}^{+} \\(\Psi_{00}^{+})^{*}\Psi_{00}^{-} & \left| \Psi_{00}^{+}\right|^2 \\
\end{array}
\right) P.
\end{equation}
$\Psi_{00}^{\pm}$ and $P$ are defined in Eqs. (\ref{eq:Psi_00_pm}) and (\ref{eq:Pmat}), respectively.

The charge density $\rho_{\text{D}}$ of the eigenstates of the Hamiltonian (\ref{eq:full_H_Q}) is related to $\rho_{\text{ST}}$ by a rotation $M_S^{\dagger}$ acting on the singlet subspace, i.e. $\rho_{\text{D}}=M_S^{\dagger}\rho_{\text{S}}M_S \oplus \rho_{\text{T}}$.
Since we are mainly interested in the situation where the  homogeneous electric field $E_0$ in the $x$-direction  is small, we compute the $M_S$ to the first order in perturbation theory and we get
\begin{equation}
\label{eq:MSdet}
M_S=M_S^0\left(1-2 \sqrt{2}\Delta_1 e E_0 \left(
\begin{array}{ccc}
 0 & \frac{A\left(\epsilon _2\right)}{\epsilon _1-\epsilon _2} & \frac{ A\left(\epsilon _0\right)}{\epsilon _1-\epsilon _0} \\
 \frac{ A\left(\epsilon _2\right)}{\epsilon _2-\epsilon _1} & 0 & 0 \\
 \frac{A\left(\epsilon _0\right)}{\epsilon _0-\epsilon _1} & 0 & 0 \\
\end{array}
\right)\right) C.
\end{equation}
Here, $M_S^0$ is the matrix of normalized column eigenvectors  obtained when $E_0=0$ and it is given by Eq. (\ref{eq:MS_mat}). 
To linear order in $E_0$, the eigenenergies $\epsilon_i$  in Eq. (\ref{eq:zero-det-eigenenergy}) are unchanged because the first correction is $\propto E_0^2$ (see Eqs. (\ref{eq:delta1}), (\ref{eq:t2}) and (\ref{eq:exchange-efield})). 
However, the eigenstates of the singlet Hamiltonian are modified  by the finite detuning $\Delta_1 e E_0$, leading to the corrections to $M_S^0$ shown in Eq. (\ref{eq:MSdet}). The diagonal matrix $C=\text{diag}(C_1,C_2,C_0)$ is required to renormalize the eigenstates.
% in particular
%\begin{equation}
%C_{0}=\left(1+\left(\frac{2\sqrt{2} A(\epsilon_{0})\Delta_1 }{\epsilon_1-\epsilon_{0}} eE_0\right)^2\right)^{-1/2}.
%\end{equation}

We consider now the setup in Fig. \ref{fig:qubit-resonator} and we assume that the QH  material has a filling factor $\nu=1$.
In this case, the interaction Hamiltonian reduces to
\begin{equation}
\label{eq:Hintfullkappa}
H_{\text{int}}=\left(M_S^{\dagger}
\kappa_S(q)M_S\oplus \kappa_T(q)
\right) \otimes
\hat{a}_q+h.c.,
\end{equation}
where 
\begin{subequations}
\begin{flalign}
\kappa_T(q)&=\left(\kappa_{11}(q)+\kappa_{22}(q)\right)\mathcal{I}_3, \\
\kappa_S(q)&=\left(
\begin{array}{ccc}
2 \kappa_{11} (q) & 0 & \sqrt{2} \kappa_{12}(q)\\
0 & 2\kappa_{22}(q) & \sqrt{2} \kappa_{12}^*(-q)\\
\sqrt{2} \kappa_{12}^*(-q) & \sqrt{2} \kappa_{12}(q) &  \kappa_{11}(q)+\kappa_{22}(q)
\end{array}
\right).
\end{flalign}
\end{subequations}
%\begin{equation}
%\kappa_{\mp} (q)=\frac{\left(1+\sqrt{1-s^2}\right)\kappa^{N}_{\mp}(q)-\left(1-\sqrt{1-s^2}\right)\kappa^{N}_{\pm}(q) -2s \mathrm{Re}(\kappa^{N}_{-+}(q))}{1-s^2}
%\end{equation}

We define the $2\times 2$ matrix $\kappa(q)=\kappa^{\dagger}(-q)$
\begin{multline}
\frac{\kappa(q)}{2\pi}=\frac{\hbar v_p}{L_y}\sqrt{n_q}e^{-(\frac{q l_T}{2})^2}\times\\
P\left(
\begin{array}{cc}
g(\tau+b_0+a,q) & s e^{q a \sqrt{\beta}} g(\tau+b_0,q)\\
s e^{q a \sqrt{\beta}} g(\tau+b_0,q) & g(\tau+b_0-a,q)
\end{array}\right) P,
\end{multline}
where $b_0$ is the shift of the double dot eigenfunctions due to the externally applied electric field $E_0$ and is defined in Eq. (\ref{eq:b0-length}).
The dimensionless function $g$ depends on the electrostatic configuration of the system; for example in free space $g=g_0$, and
\begin{equation}
\label{eq:g-funct}
g_0(X,q)=\frac{2}{\sqrt{\pi}}\int_{\mathbb{R}} \frac{dx}{\lambda} e^{-\frac{(x-X)^2}{\lambda^2}}  K_0(\left| q x\right| ),
\end{equation}
with  $\lambda=\sqrt{l'^2+l_T^2}$.

The presence of a top or side gate at distance $d$ from the EMP center of mass leads to an additive correction to $g_0$, respectively given by
\begin{subequations}
\label{eq:g-gates-coupling}
\begin{flalign}
g_t(X,q)&=-\frac{2}{\sqrt{\pi}}\int_{\mathbb{R}} \frac{dx}{\lambda} e^{-\frac{(x-X)^2}{\lambda^2}}  K_0\left(\left| q \right| \sqrt{x^2+4d^2}\right),\\
g_s(X,q)&\approx-g(X-2d,q).
\end{flalign}
\end{subequations}
The estimation of $g_s$ is accurate in the limit $d-(\tau+b+a)\gg l_T$ and $d\gg l'$, such that one can safely extend the domain of integration in $x$ and $x'$ from $[-d,\infty)$ to $\mathbb{R}$.

Note that $g_i(X,q)$ introduced here has a similar functional form of the function $g_i(x,q)$ defined in Sec. \ref{sec:fields-classical} as the projection onto the $z=0$ plane of the EMP potential. The two definitions coincide if  we substitute $l'\rightarrow \lambda$ in Eqs. (\ref{eq:G0-functnogate}) and (\ref{eq:G-corr-exp-t-s}).
In particular, we use  a small $q$ expansion as in Eq. (\ref{eq:g-appqseries}) to approximate $g_0$, while  we use a far-field approximation, analogous to Eq. (\ref{eq: g-app-delta}) to estimate $g_{t,s}$. We then obtain
\begin{subequations}
\label{eq:gfunct-gate}
\begin{flalign}
g_t(X,q)&\approx-2 K_0\left(\left| q \right| \sqrt{X^2+4d^2}\right),\\
g_s(X,q)&\approx-2 K_0\left(\left| q\right|\left|X-2d\right|\right).
\end{flalign}
\end{subequations}

Also, at small wavevectors ($qa\sqrt{\beta}\ll 1$), the matrix $\kappa$ is Hermitian and it can be factorized, leading to the conventional electrostatic interaction term $\propto \hat{a}^{\dagger}_q+\hat{a}_q$.\\

At this point, we proceed to compute the effective qubit-resonator Hamiltonian. To do so, we make the usual choice of computational basis states, i.e. we take $|\tilde{S},0\rangle$ and $|T,0\rangle$.  As long as the qubit energy splitting $J_\text{ex}$ is close to the resonance frequency of the resonator $\omega_{R}$, the effective coupling Hamiltonian is efficiently computed  by a Schrieffer-Wolff transformation \cite{Bravyi}.
The lowest order Schrieffer-Wolff Hamiltonian in the long wavelength and small detuning approximation is
\begin{equation}
\label{eq:effective-H-SW}
H^{(0)}_{\text{eff}}=\frac{J_{\text{ex}}}{2}\sigma_z+\hbar\omega_{R}\hat{a}^{\dagger}\hat{a}+\frac{\hbar\gamma}{2}\sigma_z \left(\hat{a}^{\dagger}+\hat{a}\right),
\end{equation}
with $\omega_R$ being the frequency of the resonator and with the coupling strength being
\begin{multline}
\label{eq:coupling-strength-full}
\frac{\gamma}{2\pi}=\frac{v_p \sqrt{n}}{L_y}\left(\frac{\chi_{\Delta}\Delta_1eE_0}{\sqrt{1-s^2}}\left(g(\tau+b_0-a)-g(\tau+b_0+a)\right)+ \right. \\
\left. \frac{\sqrt{2}\chi_t s}{1-s^2} \left(g(\tau+b_0)-\frac{g(\tau+b_0+a)+g(\tau+b_0-a)}{2}\right) \right).
\end{multline}
For simplicity of notation, we dropped the argument $q$ from the function $g$, the wavenumber $n_q$ and from the ladder operators $\hat{a}_q$; $b_0$ is defined in Eq. (\ref{eq:b0-length}).

Let us consider what happens when $E_0=0$ (and so $b_0=0$). In this case, we can easily find that $\gamma_1$ is given by Eq. (\ref{eq:gamma1-full}).
As explained in the text, this result is in quantitative agreement with the perturbative solution Eq. (\ref{eq:gamma-grad-app}). This agreement is understood by considering that the term in parentheses in (\ref{eq:gamma1-full}) is the discrete gradient of the electric field. This statement is valid if the length $l'$ in the definition of $g$ as the projection in the $z=0$ plane of Eq. (\ref{eq:G0-functnogate}) is substituted by $\lambda$.

We now attempt to quantify $\gamma_2$ by linearizing  Eq. (\ref{eq:coupling-strength-full}) in $E_0$. With this procedure, we obtain
\begin{equation}
\label{eq:gamma2-full-Hartree}
\frac{\gamma_2}{2\pi}\approx-\frac{ v_p \sqrt{n} }{2L_y}\frac{2\chi_{\Delta}\Delta_1 eE_0 a}{\sqrt{1-s^2}}\left(\frac{g(\tau+a)-g(\tau-a)}{a}\right);
\end{equation}
%\begin{equation}
%\label{eq:gamma2-full-Hartree}
%\frac{\gamma_2}{2\pi}\approx\frac{ v_p \sqrt{n} }{L_y}\frac{eE_0\chi_{\Delta}\Delta_1^2}{{1}-{3 l_T^2}/{(4a^2)}}\frac{a}{l_T}\left(\frac{g(\tau+a)-g(\tau-a)}{a}\right);
%\end{equation}
we neglected a small additional term of the form
\begin{equation}
\frac{4eE_0\chi_t t_2 v_p \sqrt{n}}{3L_y}  \left(\frac{g'(\tau-a)+g'(\tau+a)}{2}-{g'(\tau)}\right),
\end{equation}
where $g'(x)=\partial_x g(x,q)$.

%\begin{multline}
%\label{eq:gamma2-full}
%\frac{\gamma_2}{2\pi}=-\left(\frac{\chi_{\Delta}\Delta_1^2}{{a}/{l_T}-{3 l_T}/{(4a)}}\left(\frac{g(\tau-a)-g(\tau+a)}{4}\right) +\right. \\
%\left. \frac{2\chi_t t_2}{3} \left(\frac{g'(\tau)}{2}-\frac{g'(\tau-a)+g'(\tau+a)}{4}\right) \right) \frac{4eE_0 v_p \sqrt{n}}{L_y},
%\end{multline}

Comparing to the approximate solution in Eq. (\ref{eq:gamma2-full}), we find a strong quantitative disagreement even in the far-field limit.
The reason formeglio s this  disagreement can be traced back to the different approximation scheme used, and, specifically, to the different values of the qubit susceptibility to an homogeneous electric field $E$ calculated  in the two approaches.

In the perturbative approach of Sec. \ref{sec:PT-coup}, in fact, the susceptiblity is found by first projecting the double dot Hamiltonian $H_D$ (\ref{eq:qubit Hamiltonian-general}) onto the subspace of the  Hilbert space that is spanned by the $E$\textit{ field dependent} ground state wavefunctions. After this projection, the perturbation theory is formulated in the conventional way by computing the $E$ dependence of the matrix elements of the effective Hamiltonian (see Eqs. (\ref{eq:delta1})  and (\ref{eq:t2})) and by finding how much the eigenvectors are rotated by these terms.

In contrast, the Hartree integral approach of Sec. \ref{sec:Hartree-text} follows a different procedure: the perturbation theory is formulated starting  from the projection of the Hamiltonian $H_D$ onto the subspace spanned by ground state wavefunctions that \textit{do not} depend on $E$.
In other words, to the linear order in $E$, the  single-particle matrix elements $H_O^{ij}$ in Eq. (\ref{eq:second-quantized-Hd}) computed from the Hartree integral do not include the  terms
\begin{equation}
\label{eq:one-body-E-additional}
E \sum_{\alpha\beta} P_{i\alpha} \left( \frac{\partial}{\partial E}\langle \Psi_{00}^{\alpha}(E)|H_O(E=0)|\Psi_{00}^{\beta}(E)\rangle\right) P_{\beta j}.
\end{equation}

These additional terms do not affect the dependence of $\gamma$ on $\tau$, but they change the susceptibility to $E$. In fact, neglecting them, we obtain
\begin{equation}
\label{eq:hartree-suscept}
\delta J_{\text{ex}}^{{\text{Ha}}}=- 2\frac{ \chi_{\Delta}\Delta_1 eE_0 a}{\sqrt{1-s^2}}eE 
\end{equation}
instead of Eq. (\ref{eq:exchange-detunedefield}).
Eq. (\ref{eq:hartree-suscept}) is consistent with the  value of $\gamma_2$ shown in Eq. (\ref{eq:gamma2-full-Hartree}) in the same sense discussed for $\gamma_1$, i.e. if we interpret $E$ in (\ref{eq:hartree-suscept}) as the discrete derivative of the EMP voltage in Eq. (\ref{eq:gamma2-full-Hartree}). 

This interpretation suggests a possible ad-hoc  modification of Eq. (\ref{eq:gamma2-full-Hartree}) to include a posteriori the terms neglected in the Hartree approach.
Namely, we use the susceptibility from Eq. (\ref{eq:exchange-detunedefield}) instead of the one in Eq. (\ref{eq:hartree-suscept}). By performing the substitution
\begin{equation}
\label{eq:subst-susc-E}
\frac{\chi_{\Delta}\Delta_1 a}{\sqrt{1-s^2}}\rightarrow  -\left(\chi_t t_2+\chi_{\Delta} \Delta_1^2\right),
\end{equation} 
in Eq. (\ref{eq:gamma2-full-Hartree}) and setting $n=1$, we obtain Eq. (\ref{eq:gamma2-full}).\\

Note that if we were to consider the rotated configuration, with the two dots aligned to the resonator edge, in the long wavelength limit, Eq. (\ref{eq:coupling-strength-full}) is valid if we make the substitutions $a\rightarrow i a \sqrt{\beta}$ and $b_0\rightarrow 0$. In this case, as expected, an homogeneous electric field $E_0$ parallel to the resonator edge has no effect, since the term linear in $E_0$ is $\propto \mathrm{Im} (g(\tau+i a\sqrt{\beta}))=0$. 

%Also, the Hartree integral (\ref{eq:Hartree_int}) does not account for the changes in the qubit wavefunctions due to the electric field of the resonator. In fact, the interaction Hamiltonian also contains single particle terms depending on the projection
%\begin{equation}
%\langle \Psi_{00}^{\alpha}(E)|H_{O}(\textbf{r})|\Psi_{00}^{\beta}(E)\rangle,
%\end{equation}
%with $H_O$ being the orbital Hamiltonian (\ref{eq:orbital-hamilt}) and $\Psi_{00}^{\alpha}(E)$ being the single dot eigenfunction including the 

Finally, the effective Hamiltonian (\ref{eq:effective-H-SW}) captures the behavior of the system as long as the coupling between the computational and the non-computational subspaces of the double dot can be neglected, i.e. when
\begin{equation}
\label{eq:zeta-leakage}
\zeta\equiv \mathrm{max}\left|\frac{\left(M_S^{\dagger}\kappa_SM_S\right)_{3,i}}{\epsilon_0-\epsilon_i\pm \hbar \omega_R}\right|\ll 1;
\end{equation}
the value  $3$ in the index of the matrix product comes from the chosen ordering of the eigenenergies.
%We can also compute the amplitude of the first order corrections and verify that they are quantitatively small compared to $\gamma$. To the first order, we get

%\begin{multline}
%H_Q=\frac{J_{\text{ex}}+J_{\text{ex}}^{(1)}+J_{\text{ex}}^{\text{TL}}a^{\dagger}a}{2}\sigma_z+\left(\left(\alpha _q\right){}^{\dagger }+\alpha _q\right)+\alpha _q \left(\alpha _q\right){}^{\dagger } \left(\omega _{\text{TL}}^Q+\omega _{\text{TL}}\right)+\left(\left(\alpha _q\right){}^{\dagger }+\alpha _q\right) \left(\gamma _0+\sigma _z \gamma _{\text{S3}-\text{T3}}\right)+\Omega _2 \left(\sigma _z+1\right) \left(\left(\alpha _q\right){}^{\dagger } \left(\alpha _q\right){}^{\dagger }+\alpha _q \alpha _q\right)
%\end{multline}

\end{appendix}
\bibliography{lit}

\end{document}